\documentclass[aps,prb,twocolumn,showpacs,nobibnotes,superscriptaddress]{revtex4-1}

\usepackage{amsmath}
\usepackage{mathtools}
\usepackage{braket}
\usepackage[makeroom]{cancel}
\usepackage{graphicx}
\renewcommand{\vec}[1]{\boldsymbol{\mathbf{#1}}}
\newcommand{\matr}[1]{\boldsymbol{\mathbf{#1}}}

\usepackage{hyperref} 
\hypersetup{
            colorlinks = true, 
            urlcolor   = blue, 
            linkcolor  = blue, 
            citecolor  = blue, 
            }

\begin{document}
\title{Precise effective masses from density functional perturbation theory}

\author{J. Laflamme Janssen}
\affiliation{%
European Theoretical Spectroscopy Facility and Institute of Condensed Matter and Nanosciences, 
Universit\'e catholique de Louvain, Chemin des \'etoiles 8, bte L07.03.01, B-1348 Louvain-la-neuve, Belgium.
}%
\email[Corresponding author: ]{laflammejanssenjonathan@gmail.com}
\author{Y. Gillet}
\affiliation{%
European Theoretical Spectroscopy Facility and Institute of Condensed Matter and Nanosciences, 
Universit\'e catholique de Louvain, Chemin des \'etoiles 8, bte L07.03.01, B-1348 Louvain-la-neuve, Belgium.
}%
\author{S. Ponc\'e}
\affiliation{%
European Theoretical Spectroscopy Facility and Institute of Condensed Matter and Nanosciences, 
Universit\'e catholique de Louvain, Chemin des \'etoiles 8, bte L07.03.01, B-1348 Louvain-la-neuve, Belgium.
}%
\author{A. Martin}
\affiliation{%
CEA, DAM, DIF. F-91297 Arpajon, France
}%
\author{M. Torrent}
\affiliation{%
CEA, DAM, DIF. F-91297 Arpajon, France
}%
\author{X. Gonze}
\affiliation{%
European Theoretical Spectroscopy Facility and Institute of Condensed Matter and Nanosciences, 
Universit\'e catholique de Louvain, Chemin des \'etoiles 8, bte L07.03.01, B-1348 Louvain-la-neuve, Belgium.
}%

\date{\today}

\begin{abstract}

The knowledge of effective masses is a key ingredient to analyze numerous properties of semiconductors, like carrier mobilities, (magneto-)transport properties, or band extrema characteristics yielding carrier densities and density of states. 
Currently, these masses are usually calculated using finite-difference estimation of density functional theory (DFT) electronic band curvatures.
However, finite differences require an additional convergence study and are prone to numerical noise. 
Moreover, the concept of effective mass breaks down at degenerate band extrema.
We assess the former limitation by developing a method that allows to obtain the Hessian of DFT bands directly, using density functional perturbation theory (DFPT).
Then, we solve the latter issue by adapting the concept of `transport equivalent effective mass' to the $\vec{k} \cdot \hat{\vec{p}}$ framework.
The numerical noise inherent to finite-difference methods is thus eliminated, along with the associated convergence study. 
The resulting method is therefore more general, more robust and simpler to use, which makes it especially appropriate for high-throughput computing.
After validating the developed techniques, we apply them to the study of silicon, graphane, and arsenic.
The formalism is implemented into the ABINIT software and supports the norm-conserving pseudopotential approach, the projector augmented-wave method, and the inclusion of spin-orbit coupling.
The derived expressions also apply to the ultrasoft pseudopotential method.

\end{abstract}


\maketitle

\section{Introduction}

Accurate and precise \emph{ab initio} effective masses are desirable for the description of, e.g., transport properties, optoelectronic properties, cyclotron frequencies and Landau levels~\cite{Kittel,AshcroftMermin,Marder,YuCardona}.
In particular, this topic has recently seen renewed interest motivated by optoelectronics~\cite{Filip:2015iz,Hautier:2014dx,HautierTCO,Kim:2010cc,Kim:2009dj,Chantis:2006dh,Geller:2001it} and thermoelectric applications~\cite{Wang:2011fb,Hummer:2007hf}.

Effective masses enter the above quantities as a description of the band dispersion around an extrema. 
They appear within the context of the $\vec{k}\cdot\hat{\vec{p}}$ theory, which is a second-order perturbation expansion of eigenenergies with respect to the electronic wave vector $\vec{k}$, assuming an Hamiltonian with local potential $V(\vec{r},\vec{r}') = V(\vec{r}-\vec{r}')$.
In this framework, they are defined as the inverse second-order expansion coefficient. 
Thus, for non-degenerate bands, the effective mass are inverse curvatures of bands in one dimension and inverse Hessian of bands in three dimensions. 

In the first-principle context, these second-order derivatives of eigenenergies are usually obtained through finite-difference calculations~\cite{Filip:2015iz,Kim:2010cc,Kim:2009dj,Hummer:2007hf,Wang:2011fb} or integrations over the Brillouin zone~\cite{HautierTCO,Boltztrap} of DFT results. 
Such calculations require a convergence on the finite-difference parameter (or the $\vec{k}$-point grid density) and are prone to numerical noise.
These extra convergences lead to additional work and possible precision issues.
Moreover, since DFT eigenvalues show limited agreement with experimental eigenenergies (see e.g. \S 7.4 of Ref.~\onlinecite{Martin}), their use raises accuracy issues.
However, while these accuracy concerns have been investigated using many-body perturbation theory~\cite{Filip:2015iz,Kim:2010cc,Kim:2009dj,Hummer:2007hf}, the precision issues have yet to be addressed. 
Thus, within this work, we focus on the latter problem, i.e. the calculation of precise DFT effective masses.

We note that circumventing the use of finite differences is already possible using Wannier functions~\cite{Yates:2007er}.
However, the issues of additional work and precision remain to some extent.
Indeed, the Wannier function optimization procedure can get stuck in a local minima and, to prevent this, the user needs to choose the starting functions with some care. 
Thus, a method avoiding any such task and the associated precision issues remains desirable. 

Another difficulty in the calculation of effective masses is the treatment of degeneracies. 
Indeed, subtleties arise when one considers the $\vec{k}\cdot\hat{\vec{p}}$ framework in the context of degenerate perturbation theory~\cite{Luttinger55,Dresselhaus55,Kane56,Kane57}.
In one dimension, the perturbation coefficients become matrices (derivatives of Hamiltonian matrix elements within the degenerate subspace) instead of scalars.
Still, obtaining the effective masses remains a simple matter of diagonalizing the second-order matrix and attributing the inverse eigenvalues to the effective masses. 
However, in three dimensions, the second-order expansion coefficient becomes a matrix of tensors (i.e. the Hessian of the Hamiltonian in the degenerate subspace), as first noted by Luttinger and Kohn~\cite{Luttinger55}. 
Since it is not possible to diagonalize such a matrix of tensors, it would appear that simple individual quantities cannot be attributed to degenerate bands for describing their dispersion at second order. 

Dresselhaus, Kip, and Kittel~\cite{Dresselhaus55} assessed this issue for the top valence band of crystals with diamond structure.
Starting with symmetry group arguments to justify the results of Luttinger and Kohn~\cite{Luttinger55}, they added spin-orbit coupling and obtained a simple formula that describes degenerate band dispersions individually within the $\vec{k}\cdot\hat{\vec{p}}$ framework
\footnote{See Eq.~(63) of Ref.~\onlinecite{Dresselhaus55} or Eq.~\eqref{kittel_fit} of the present work.}.
While it successfully describes the complex directional dependence of the band curvature for these degenerate bands, it is limited to electronic states belonging to a specific symmetry group and to lattices with cubic symmetry. 
Moreover, the determination of e.g. transport properties from these parameters is less conventional than from effective mass tensors. 
A more general and convenient formalism would thus be welcome.

Mecholsky and coworkers~\cite{Mecholsky14} recently proposed such a formalism for the specialized case of transport tensor calculations. 
They first justify rigorously the breakdown of the concept of effective mass tensor for degenerate bands and argue that degenerate band curvatures should instead be described as a function of spherical angles $f(\theta,\phi)$. 
Then, they derive the relation between $f(\theta,\phi)$ and transport tensors, assuming a parabolic band extremum.
Finally, they define a `transport equivalent effective mass' tensor that generates the same contribution to transport tensors than $f(\theta,\phi)$.
This tensor has the benefit of being well defined for parabolic band extrema in any material and being straightforward to use for transport properties calculations; however it does not describe band dispersions anymore.

In this work, we first eliminate the necessity of carrying out finite-difference differentiation of eigenenergies to obtain effective masses. 
To do so, we derive analytical expressions for the second-order derivative of non-degenerate eigenenergies within the density functional perturbation theory (DFPT) framework~\cite{Baroni87,Gonze95}, which has already been used successfully to compute various derivatives with respect to the electronic wave vector $\vec{k}$ in the past~\cite{Gonze:1997vy,Gonze:1997ux,Ghosez:2000ct,Umari:2004hq}. 
More specifically, in Section~\ref{nd}, we first derive the relation between the effective mass tensor and the derivatives of the Hamiltonian. 
Then, we differentiate the relevant contributions of the Hamiltonian expressed in a plane wave basis set.
We start with the modified kinetic energy used to smooth the total energy dependence on primitive cell size~\cite{Bernasconi95,Gonze:2009aa}. 
Then, we proceed to the non-local potential involved in the norm-conserving pseudopotential (NCPP)~\cite{Hamann79}, projector augmented-wave (PAW)~\cite{Blochl}, and ultrasoft pseudopotential (USPP)~\cite{Vanderbilt:1990aa} methods. 
Finally, we consider the spin-orbit coupling contribution at the end of the Section. 

Also, since the present work targets particularly the high-throughput design of materials with optimized transport properties~\cite{HautierTCO,Wang:2011fb}, we adapt the `transport equivalent effective mass' formalism of Mecholsky and coworkers~\cite{Mecholsky14} to the DFPT context for the description of degenerate bands in Section~\ref{d}. 
To do so, we first generalize the formalism of Luttinger and Kohn~\cite{Luttinger55} for the description of degenerate states in the $\vec{k}\cdot\hat{\vec{p}}$ framework~\footnote{More specifically, we generalize their Eq.~(IV.9)} to the NCPP and PAW contexts. 
Then, we bridge the gap between this generalized formalism and the `transport equivalent effective mass' formalism. 

We validate in Section~\ref{results} our implementation within the ABINIT software~\cite{Gonze:2016te,Gonze:2009aa} by comparison with finite-difference calculations. 
Then, we apply it to a semiconductor (silicon), a 2D material (graphane) and a semimetal ($\alpha$-arsenic). 

Atomic units are used throughout.




\section{The non-degenerate case}
\label{nd}

\subsection{Effective mass in the DFPT framework}
\label{nd_eps}

Within density functional theory (DFT), Schr\"odinger's equation for periodic systems is
\begin{equation}
\label{schro}
\hat H \ket{\psi_{n\vec{k}}} 
= \left( \frac{\hat p^2}{2} + \hat V \right) \ket{\psi_{n\vec{k}}} 
= \varepsilon_{n\vec{k}} \ket{\psi_{n\vec{k}}},
\end{equation}
where $\hat H$ is the Hamiltonian, $\ket{\psi_{n\vec{k}}}$ and $\varepsilon_{n\vec{k}}$ are its eigenstates and eigenenergies, $n$ is the band index, $\vec{k}$ is Bloch's wave vector, $\hat{\vec{p}} = -i \hat{\vec{\nabla}}$ is the momentum operator and $\hat V$ is a local potential
\begin{equation}
\label{Vloc}
\bra{\vec{r}} \hat V \ket{\vec{r}'} = V(\vec{r}) \delta(\vec{r},\vec{r}').
\end{equation}
Using Bloch's theorem, the wavefunction can be expressed as the product of a crystal periodic function $\ket{u_{n\vec{k}}}$ and a phase
\begin{equation}
\label{bloch}
\ket{\psi_{n\vec{k}}} = e^{i \vec{k} \cdot \hat{\vec{r}}} \ket{u_{n\vec{k}}}.
\end{equation}
Eq.~\eqref{schro} then becomes
\begin{align}
\label{k-schro}
\hat H_{\vec{k}} \ket{u_{n\vec{k}}}
&= \left( \frac{k^2 + 2 \vec{k} \cdot \hat{\vec{p}} + \hat p^2}{2} + \hat V \right) \ket{u_{n\vec{k}}} \nonumber \\
&= \varepsilon_{n\vec{k}} \ket{u_{n\vec{k}}},
\end{align}
where we have defined $\vec{k}$-dependent operators $\hat O_{\vec{k}}$ as
\begin{equation}
\label{bloch_op}
\hat O_{\vec{k}} \triangleq e^{-i \vec{k} \cdot \hat{\vec{r}}} \hat O e^{i \vec{k} \cdot \hat{\vec{r}}},
\end{equation}
and where the eigenstates are orthonormalized
\begin{equation}
\label{orthonorm}
\braket{u_{n\vec{k}} | u_{n'\vec{k}}} = \delta_{nn'}.
\end{equation}

We now consider the situation where band $n$ is non-degenerate at $\vec{k}$ and where $\ket{u_{n\vec{k}}}$ and $\varepsilon_{n\vec{k}}$ are known. 
The Taylor expansion of the band dispersion around $\vec{k}$ yields
\begin{equation}
\label{taylor_eps}
\varepsilon_{n \vec{k} + \delta \vec{k}}
= \varepsilon_{n\vec{k}} 
+ \sum_\alpha \varepsilon^\alpha_{n\vec{k}} \delta k_\alpha 
+ \frac{1}{2} \sum_{\alpha \beta} \delta k_\alpha \varepsilon^{\alpha \beta}_{n\vec{k}} \delta k_\beta
+ O(\delta k^3),
\end{equation}
where Greek letters $\alpha,\beta,\gamma,...$ stand for Cartesian directions $\{x,y,z\}$ and where derivatives of any quantity $X$ with respect to a Cartesian component of the wave vector $\vec{k}$ are noted
\begin{equation}
X^\alpha \triangleq \frac{\partial X}{\partial k_\alpha}; \qquad 
X^{\alpha \beta} \triangleq \frac{\partial^2 X}{\partial k_\alpha \partial k_\beta}.
\end{equation}
Within perturbation theory, we wish to obtain derivatives of observables from derivatives of the Hamiltonian.
Thus, we first project Eq.~\eqref{k-schro} on $\bra{u_{n\vec{k}}}$
\begin{equation}
\varepsilon_{n \vec{k}} = \bra{u_{n\vec{k}}} \hat H_{\vec{k}} \ket{u_{n\vec{k}}},
\end{equation}
and then differentiate with respect to $k_\alpha$, which yields the first derivative appearing in Eq.~\eqref{taylor_eps}
\begin{align}
\varepsilon^{\alpha}_{n \vec{k}} 
&= \bra{u_{n\vec{k}}} \hat H^\alpha_{\vec{k}} \ket{u_{n\vec{k}}}
+\left( \bra{u^\alpha_{n\vec{k}}} \hat H_{\vec{k}} \ket{u_{n\vec{k}}} + \mathrm{c.c.} \right), \nonumber \\
&= \bra{u_{n\vec{k}}} \hat H^\alpha_{\vec{k}} \ket{u_{n\vec{k}}}  
+ \varepsilon_{n \vec{k}} \left( \braket{u^\alpha_{n\vec{k}} | u_{n\vec{k}}} + \mathrm{c.c.} \right), \nonumber \\
&= \bra{u_{n\vec{k}}} \hat H^\alpha_{\vec{k}} \ket{u_{n\vec{k}}}. \label{der1}
\end{align}
The last two relations have been obtained in the spirit of the Hellmann-Feynman force theorem~\cite{Feynman:1939bg,Martin}, using respectively Eq.~\eqref{k-schro} and the derivative of the normalization condition (Eq.~\eqref{orthonorm})
\begin{equation}
\label{dernorm}
(\braket{u^\alpha_{n\vec{k}} | u_{n\vec{k}}} + \mathrm{c.c.}) = 0,
\end{equation}
where $\mathrm{c.c.}$ stands for the complex conjugate of the previous term. 
The second-order derivative thus reads
\footnote{While Eq.~\eqref{der2} do not look symmetric with respect to $\alpha\beta$ at first glance, using Eqs.~\eqref{PQnd} and \eqref{SOS} along with the fact that $\ket{\hat{P}_{\vec{k}} u_{n\vec{k}}}$ does not contribute to $\varepsilon^{\alpha \beta}_{n \vec{k}}$ (see Eq.~\eqref{noPu} and associated discussion) reveals that the expression is indeed symmetric.}  
\begin{equation}
\label{der2}
\varepsilon^{\alpha \beta}_{n \vec{k}}
= \bra{u_{n\vec{k}}} \hat H^{\alpha \beta}_{\vec{k}} \ket{u_{n\vec{k}}}
+ \left( \bra{u^\beta_{n\vec{k}}} \hat H^\alpha_{\vec{k}} \ket{u_{n\vec{k}}} + \mathrm{c.c.} \right).
\end{equation}
Defining the following complementary projectors
\begin{equation}
\label{PQnd}
\hat{P}_{\vec{k}} \triangleq \ket{u_{n\vec{k}}} \bra{u_{n\vec{k}}}; \qquad \hat{Q}_{\vec{k}} \triangleq 1 - \hat{P}_{\vec{k}},
\end{equation}
we realize that $\ket{\hat{P}_{\vec{k}} u^\beta_{n\vec{k}}}$ does not contribute to $\varepsilon^{\alpha \beta}_{n \vec{k}}$ 
\begin{multline}
\label{noPu}
\varepsilon^{\alpha \beta}_{n \vec{k}}
=  \bra{u_{n\vec{k}}} \hat H^{\alpha \beta}_{\vec{k}} \ket{u_{n\vec{k}}} 
 + \left( \bra{\hat{Q}_{\vec{k}} u^\beta_{n\vec{k}}} \hat H^\alpha_{\vec{k}} \ket{u_{n\vec{k}}} + \mathrm{c.c.} \right) \\
 + \left( \braket{u^\beta_{n\vec{k}} | u_{n\vec{k}}} \bra{u_{n\vec{k}}} \hat H^\alpha_{\vec{k}} \ket{u_{n\vec{k}}} + \mathrm{c.c.} \right) \\
=  \bra{u_{n\vec{k}}} \hat H^{\alpha \beta}_{\vec{k}} \ket{u_{n\vec{k}}} 
 + \left( \bra{\hat{Q}_{\vec{k}} u^\beta_{n\vec{k}}} \hat H^\alpha_{\vec{k}} \ket{u_{n\vec{k}}} + \mathrm{c.c.} \right), 
\end{multline}
since $( \braket{u^\beta_{n\vec{k}} | u_{n\vec{k}}} \bra{u_{n\vec{k}}} \hat H^\alpha_{\vec{k}} \ket{u_{n\vec{k}}} + \mathrm{c.c.}) = \varepsilon^{\alpha}_{n \vec{k}} (\braket{u^\beta_{n\vec{k}} | u_{n\vec{k}}} + \mathrm{c.c.}) = 0$, following Eqs.~\eqref{der1} and \eqref{dernorm}.

Now, $\ket{u^\alpha_{n\vec{k}}}$ still has to be expressed in terms of derivatives of the Hamiltonian.
To do so, we differentiate Eq.~\eqref{k-schro} with respect to $\vec{k}$
\begin{align}
\hat H^\alpha_{\vec{k}} \ket{u_{n\vec{k}}} + \hat H_{\vec{k}} \ket{u^\alpha_{n\vec{k}}} 
&= \varepsilon^{\alpha}_{n \vec{k}} \ket{u_{n\vec{k}}} + \varepsilon_{n \vec{k}} \ket{u^\alpha_{n\vec{k}}}, \nonumber \\
\Rightarrow \left( \hat H_{\vec{k}} - \varepsilon_{n \vec{k}} \right) \ket{u^\alpha_{n\vec{k}}} 
&= - \left( \hat H^\alpha_{\vec{k}} - \varepsilon^{\alpha}_{n \vec{k}} \right) \ket{u_{n\vec{k}}}, \label{rawStern}
\end{align}
and apply $\hat{Q}_{\vec{k}}$ to the left
\begin{equation}
\left( \hat H_{\vec{k}} - \varepsilon_{n \vec{k}} \right) \ket{\hat{Q}_{\vec{k}} u^\alpha_{n\vec{k}}} 
= - \hat{Q}_{\vec{k}} \hat H^\alpha_{\vec{k}} \ket{u_{n\vec{k}}}, \label{stern} 
\end{equation}
which allows to deduce $\ket{\hat{Q}_{\vec{k}} u^\alpha_{n\vec{k}}}$ from $\hat H^\alpha_{\vec{k}}$ and unperturbed quantities. 

Directly solving Eq.~\eqref{stern} using linear algebra techniques such as conjugate gradients~\cite{Hestenes52} can be unstable because the left-hand side operator $\hat H_{\vec{k}} - \varepsilon_{n \vec{k}}$ is not, in general, positive definite~\cite{Golub}. 
It is more practical to invert the operator $\hat H_{\vec{k}} - \varepsilon_{n \vec{k}}$ in Eq.~\eqref{stern} then use Eqs.~\eqref{k-schro} and \eqref{PQnd} to obtain the usual sum-over-state expression
\begin{equation}
\label{SOS}
\ket{\hat{Q}_{\vec{k}} u^\alpha_{n\vec{k}}} = \sum_{n' \neq n} \ket{u_{n'\vec{k}}} \frac{\bra{u_{n'\vec{k}}} \hat H^\alpha_{\vec{k}} \ket{u_{n\vec{k}}}}{\varepsilon_{n \vec{k}} - \varepsilon_{n' \vec{k}}}.
\end{equation}
Substituting Eq.~\eqref{SOS} in Eq.~\eqref{noPu} gives
\begin{multline}
\label{eps2SOS}
\varepsilon^{\alpha \beta}_{n \vec{k}}
= \bra{u_{n\vec{k}}} \hat H^{\alpha \beta}_{\vec{k}} \ket{u_{n\vec{k}}} \\
+ \bigg(\sum_{n' \neq n} \frac{\bra{u_{n\vec{k}}} \hat H^\beta_{\vec{k}} \ket{u_{n'\vec{k}}} \bra{u_{n'\vec{k}}} \hat H^\alpha_{\vec{k}} \ket{u_{n\vec{k}}}}{\varepsilon_{n \vec{k}} - \varepsilon_{n' \vec{k}}}
+ \mathrm{c.c.} \bigg).
\end{multline}

While the above expression is numerically easier to handle than Eq.~\eqref{stern}, it has the downside being much less efficient.
Indeed, Eq.~\eqref{SOS} exhibits a notoriously slow convergence with the number of states included in the summation~\cite{Gonze:2011ej}.
However, it is possible to combine the technical ease of Eq.~\eqref{SOS} with the efficiency of Eq.~\eqref{stern} by using the former to obtain the contribution of the active space (i.e. bands up to the highest one $N$ for which eigenenergies derivatives are desired) to $\ket{\hat{Q}_{\vec{k}} u^\alpha_{n\vec{k}}}$ and the latter to obtain the contribution of the complementary subspace (band index above $N$) to $\ket{\hat{Q}_{\vec{k}} u^\alpha_{n\vec{k}}}$. 
Indeed, this strategy guarantees the left-hand side operator of Eq.~\eqref{stern} to be positive definite (thus allowing the use of, e.g., conjugated gradients), while minimizing the number of bands treated using Eq.~\eqref{SOS}. 

Defining the projector
\begin{equation}
\label{PN}
\hat{Q}_{N \vec{k}} \triangleq \sum_{n' > N} \ket{u_{n'\vec{k}}} \bra{u_{n'\vec{k}}},
\end{equation}
we can write the second-order eigenenergies in the form described above
\begin{multline}
\label{eps_Pstern}
\varepsilon^{\alpha \beta}_{n \vec{k}}
= \bra{u_{n\vec{k}}} \hat H^{\alpha \beta}_{\vec{k}} \ket{u_{n\vec{k}}}
+ \left( \bra{\hat{Q}_{N \vec{k}} u^\beta_{n\vec{k}}} \hat H^\alpha_{\vec{k}} \ket{u_{n\vec{k}}} + \mathrm{c.c.} \right) \\
+ \bigg(\sum_{n' \neq n}^N \frac{\bra{u_{n\vec{k}}} \hat H^\beta_{\vec{k}} \ket{u_{n'\vec{k}}} \bra{u_{n'\vec{k}}} \hat H^\alpha_{\vec{k}} \ket{u_{n\vec{k}}}}{\varepsilon_{n \vec{k}} - \varepsilon_{n' \vec{k}}}
+ \mathrm{c.c.}\bigg),
\end{multline}
with $\hat{Q}_{N \vec{k}} \ket{u^\alpha_{n\vec{k}}}$ given by the projection of Eq.~\eqref{rawStern} on bands above $N$
\begin{equation}
\label{Pstern}
\left( \hat H_{\vec{k}} - \varepsilon_{n \vec{k}} \right) \ket{\hat{Q}_{N \vec{k}} u^\alpha_{n\vec{k}}} 
= - \hat{Q}_{N \vec{k}} \hat H^\alpha_{\vec{k}} \ket{u_{n\vec{k}}}.
\end{equation}

Once $\varepsilon^{\alpha \beta}_{n \vec{k}}$ is known, one can obtain the effective mass from the usual expression~\cite{Kittel,AshcroftMermin,Marder}
\begin{equation}
\label{efmas_nd}
[\matr{M}_{n\vec{k}}^{-1}]_{\alpha \beta} \triangleq \varepsilon^{\alpha \beta}_{n \vec{k}},
\end{equation}
for non-degenerate $\varepsilon_{n \vec{k}}$.

In the case of an Hamiltonian with a local potential, as described in Eqs.~\eqref{Vloc} and \eqref{k-schro}, the perturbed Hamiltonian reduces to 
\begin{equation}
\label{der_mfH}
\hat H^\alpha_{\vec{k}} = (\vec{k} + \hat{\vec{p}})_\alpha; \qquad \hat H^{\alpha \beta}_{\vec{k}} = \delta_{\alpha \beta}.
\end{equation}

For practical reasons, reduced coordinates are often used internally by DFT codes instead of Cartesian coordinates. 
We therefore provide in Appendix~\ref{red_coords} the relation between derivatives with respect to $\vec{k}$ in both coordinate systems.

\subsection{Derivatives of kinetic energy operator with cutoff smearing}
\label{kin_sec}

To avoid discontinuities in the total energy with respect to primitive cell size, one can modify the kinetic energy to ensure that the number of available degrees of freedom for energy minimization varies continuously with cell size~\cite{Bernasconi95} (see Appendix~\ref{ecutsm-sec}).
In the ABINIT software, this kinetic energy reads (see Eqs.~\eqref{Kin} and \eqref{Kinx})
\begin{equation}
\label{Kin}
\bra{\vec{G}} \hat T_{\vec{k}} \ket{\vec{G}'} = \frac{1}{2}(\vec{k}+\vec{G})^2 \delta_{\vec{G}\vec{G}'} \, p(x),
\end{equation}
where $p(x)$ is a function that is one for most of the plane waves, but diverges when $\frac{1}{2}(\vec{k}+\vec{G})^2$ becomes close to the plane wave kinetic energy cut-off $E_c$. 
Its accurate formulation is given in Appendix~\ref{ecutsm-sec}. 
The quantity x is equal to
\begin{equation}
\label{Kinx}
x \triangleq \frac{E_c - \frac{1}{2}(\vec{k}+\vec{G})^2}{E_s},
\end{equation}
where the parameter $E_s$ is the energy range around the cutoff energy $E_c$ where the occupations start to be forced towards 0, i.e. $E_s$ can be interpreted as a smearing of the cutoff energy.  

The present method implements the derivatives of this modified kinetic energy
\begin{multline}
\label{Kin1}
\bra{\vec{G}} \hat T^\alpha_{\vec{k}} \ket{\vec{G}'} 
= \left( p(x) - \frac{1}{2}(\vec{k}+\vec{G})^2 \frac{p'(x)}{E_s} \right) \\
  (\vec{k}+\vec{G})_\alpha \delta_{\vec{G}\vec{G}'},
\end{multline}
\begin{multline}
\label{Kin2}
\bra{\vec{G}} \hat T^{\alpha \beta}_{\vec{k}} \ket{\vec{G}'} 
= \left( p(x) - \frac{1}{2}(\vec{k}+\vec{G})^2 \frac{p'(x)}{E_s} \right) \delta_{\alpha \beta} \delta_{\vec{G}\vec{G}'} \\
+ \left( \frac{1}{2}(\vec{k}+\vec{G})^2 \frac{p''(x)}{E_s^2} - 2 \frac{p'(x)}{E_s} \right) \\
 (\vec{k}+\vec{G})_\alpha (\vec{k}+\vec{G})_\beta \delta_{\vec{G}\vec{G}'},
\end{multline}
where $p'(x)$ stands for the derivative of $p(x)$ with respect to $x$.
Their reduced coordinate version can then be obtained by using the reverse of Eqs.~\eqref{der1red} and \eqref{der2red}. 

\subsection{Pseudopotentials and derivatives of associated non-local operators}
\label{nl}

The potential $V(\vec{r})$ appearing in a DFT Hamiltonian involves the Coulomb potential generated by the nuclei of the simulated system. 
It therefore has sharp features, which are cumbersome to represent accurately with a plane wave basis set. 
Numerous methods have been developed to alleviate this problem, 
among which two are supported in the present implementation: the norm-conserving pseudopotential (NCPP)~\cite{Hamann79} and projector-augmented wave (PAW)~\cite{Blochl, Audouze08} methods. 
We derive the relevant expressions in the PAW framework, since it generalizes the NCPP framework~\cite{Blochl}. 
We then obtain the NCPP expressions by carrying out the appropriate simplifications.

Since the relationship between all-electrons wavefunctions $\ket{\psi_{n\vec{k}}}$ and pseudo wavefunctions $\ket{\tilde \psi_{n\vec{k}}}$ (Eq.~\eqref{tr_PAW}) has the same form in the ultrasoft pseudopotential formalism (USPP)~\cite{Vanderbilt:1990aa,Audouze08,Umari:2004hq,Miwa:2011ez,dalCorso:2001aa} and the PAW formalism, it results that the present Section applies to both PAW and USPP (see Ref.~\onlinecite{Umari:2004hq} for a more detailed discussion of this).
This also allows us to build upon existing DFPT developments within the USPP framework~\cite{Audouze08,Umari:2004hq,Miwa:2011ez,dalCorso:2001aa}.

We offer a short PAW reminder in Appendix~\ref{PAW-reminder} to put Eqs.~\eqref{PAW-up}-\eqref{PAW-down} below into context.

In PAW, Eq.~\eqref{k-schro} becomes (see Eq.~\eqref{Schro-PAW})
\begin{equation}
\hat{\tilde{H}}_{\vec{k}} \ket{\tilde{u}_{n\vec{k}}} = \varepsilon_{n\vec{k}} \hat{\tilde{1}}_{\vec{k}} \ket{\tilde{u}_{n\vec{k}}} \label{PAW-up}, 
\end{equation}
where (see Eqs.~\eqref{U-sca}, \eqref{U-op}, and \eqref{S-sca})
\begin{align}
\hat{\tilde{H}}_{\vec{k}} & = \hat H_{\vec{k}} + \hat D_{\vec{k}} \nonumber \\
& = \hat H_{\vec{k}} + \sum_{\vec{R}ij} e^{-i \vec{k} \cdot \hat{\vec{r}}} \ket{\tilde{p}_{\vec{R}i}} D_{\vec{R}ij} \bra{\tilde{p}_{\vec{R}j}} e^{i \vec{k} \cdot \hat{\vec{r}}}, \label{PAW-H} \\
\hat{\tilde{1}}_{\vec{k}} & = 1 + \sum_{\vec{R}ij} e^{-i \vec{k} \cdot \hat{\vec{r}}} \ket{\tilde{p}_{\vec{R}i}} S_{\vec{R}ij} \bra{\tilde{p}_{\vec{R}j}} e^{i \vec{k} \cdot \hat{\vec{r}}},
\end{align}
and where $\bra{\tilde{p}_{\vec{R}i}}$, $D_{\vec{R}ij}$, and $S_{\vec{R}ij}$ are defined in Eqs.~\eqref{PAW-p}, \eqref{PAW-U}, and \eqref{PAW-S}, respectively. 

The relations required to carry out the differentiation of the non-local part of the Hamiltonian $\hat D_{\vec{k}}$ are (see Eqs.~\eqref{UGG-pbar}, \eqref{ptildeK}, and \eqref{ptildes})
\begin{align}
\bra{\vec{G}} \hat D_{\vec{k}} \ket{\vec{G}'} 
&= \sum_{\vec{R}ij} \braket{\vec{K} | \bar p_{\vec{R}i}} D_{\vec{R}ij} \braket{\bar p_{\vec{R}j} | \vec{K}'}, \\
\braket{\vec{K} | \bar p_{\vec{R}i}} 
&= 4 \pi i^{l_i} Y_{l_i m_i}(\hat{\vec{K}}) \tilde{P}_{\vec{R}i}(K) e^{- i \vec{G} \cdot \vec{R}}, \\
\tilde{P}_{\vec{R}i}(K)
&= \int_0^{s_c} ds \, s \, \tilde{\mathcal{P}}_{\vec{R}i}(s) j_{l_i}(Ks), \label{PAW-down}
\end{align}
where $\vec{K}$, $\vec{s}$, $Y_{lm}$, $j_l(ks)$, $s_c$, $\tilde{\mathcal{P}}_{\vec{R}i}(s)$, $\tilde{P}_{\vec{R}i}(K)$, and $\braket{\vec{K} | \bar p_{\vec{R}i}}$ are defined in Appendix~\ref{PAW-reminder}
\footnote{For convenience, we still summarize the definitions here: $\vec{K} \triangleq \vec{k}+\vec{G}$, $\hat{\vec{K}}$ is the unit vector in the direction of $\vec{K}$, $\vec{s} \triangleq \vec{r}-\vec{R}$, $Y_{lm}$ are the spherical harmonics, $j_l(ks)$ are spherical Bessel functions, $s_c$ is the radius of the PAW augmentation regions (see Eq.~\eqref{EqB17}), $\tilde{\mathcal{P}}_{\vec{R}i}(s)$ is defined in Eq.~\eqref{PAW-Ps} (see also Eq.~\eqref{ptildes}), $\tilde{P}_{\vec{R}i}(K)$ is defined in Eq.~\eqref{ptildeK}, and $\braket{\vec{K} | \bar p_{\vec{R}i}}$ differs from $\braket{\vec{K} | \tilde p_{\vec{R}i}}$ only trough the substitution $\vec{K} \to \vec{G}$ in Eq.~\eqref{ptildes} (see Eqs.~\eqref{ptildes}-\eqref{PAW-pbar}).}.

Since the $D_{\vec{R}ij}$ have no dependence on $\vec{k}$, the derivatives of $\braket{\vec{K} | \bar p_{\vec{R}i}}$ suffice to obtain those of $\bra{\vec{G}} \hat D_{\vec{k}} \ket{\vec{G}'}$. 
Thus,
\begin{multline}
\label{UGG_der1}
\bra{\vec{G}} \hat D^\alpha_{\vec{k}} \ket{\vec{G}'}
= \sum_{\vec{R}ij} \braket{\vec{K} | \bar p_{\vec{R}i}}^\alpha D_{\vec{R}ij} \braket{\bar p_{\vec{R}j} | \vec{K}'} \\
+ \braket{\vec{K} | \bar p_{\vec{R}i}} D_{\vec{R}ij} \braket{\bar p_{\vec{R}j} | \vec{K}'}^\alpha,
\end{multline}
\begin{multline}
\label{UGG_der2}
\bra{\vec{G}} \hat D^{\alpha \beta}_{\vec{k}} \ket{\vec{G}'}
= \sum_{\vec{R}ij} \braket{\vec{K} | \bar p_{\vec{R}i}}^{\alpha \beta} D_{\vec{R}ij} \braket{\bar p_{\vec{R}j} | \vec{K}'} \\
+ \braket{\vec{K} | \bar p_{\vec{R}i}} D_{\vec{R}ij} \braket{\bar p_{\vec{R}j} | \vec{K}'}^{\alpha \beta} \\
+ \Big( \braket{\vec{K} | \bar p_{\vec{R}i}}^{\alpha} D_{\vec{R}ij} \braket{\bar p_{\vec{R}j} | \vec{K}'}^\beta \\
+ \alpha \leftrightarrow \beta \Big),
\end{multline}
where $\alpha \leftrightarrow \beta$ stands for the transpose of the previous term (with respect to $\alpha$ and $\beta$), where 
\begin{align}
\label{pbar_der1}
\braket{\vec{K} | \bar p_{\vec{R}i}}^{\alpha} 
=& 4 \pi i^{l_i} e^{- i \vec{G} \cdot \vec{R}} \Big( Y^{\alpha}_{l_i m_i}(\hat{\vec{K}}) \tilde{P}_{\vec{R}i}(K) \nonumber\\
& + Y_{l_i m_i}(\hat{\vec{K}}) \tilde{P}^{\alpha}_{\vec{R}i}(K) \Big),
\end{align}
\begin{multline}
\label{pbar_der2}
\braket{\vec{K} | \bar p_{\vec{R}i}}^{\alpha \beta} 
= 4 \pi i^{l_i} e^{- i \vec{G} \cdot \vec{R}} \\
\Big( Y^{\alpha \beta}_{l_i m_i}(\hat{\vec{K}}) \tilde{P}_{\vec{R}i}(K) + Y_{l_i m_i}(\hat{\vec{K}}) \tilde{P}^{\alpha \beta}_{\vec{R}i}(K) \\
+ \big( Y^{\alpha}_{l_i m_i}(\hat{\vec{K}}) \tilde{P}^{\beta}_{\vec{R}i}(K) + \alpha \leftrightarrow \beta \big) \Big),
\end{multline}
with $Y^\alpha_{lm}(\hat{\vec{K}})$ ($Y^{\alpha \beta}_{lm}(\hat{\vec{K}})$) the Cartesian component $\alpha$ ($\alpha \beta$) of the gradient (Hessian) of spherical harmonics $Y_{lm}$, and where
\begin{equation}
\label{ptilde_der1}
\tilde{P}^{\alpha}_{\vec{R}i}(K)
= \int_0^{s_c} ds \, s \, \tilde{\mathcal{P}}_{\vec{R}i}(s) j'_{l_i}(Ks) \, s \frac{K_\alpha}{K}, 
\end{equation}
\begin{multline}
\label{ptilde_der2}
\tilde{P}^{\alpha \beta}_{\vec{R}i}(K)
= \int_0^{s_c} ds \, s \, \tilde{\mathcal{P}}_{\vec{R}i}(s) \bigg( j''_{l_i}(Ks) \, s^2 \frac{K_\alpha}{K} \frac{K_\beta}{K} \\
+ \frac{j'_{l_i}(Ks)}{K} \, s \Big( \delta_{\alpha \beta} - \frac{K_\alpha}{K} \frac{K_\beta}{K} \Big) \bigg), 
\end{multline}
with $j'_l(x)$ and $j''_l(x)$ the first and second derivatives of spherical Bessel functions with respect to their argument $x$, respectively.
The calculation of the derivatives of $\bra{\vec{G}} \hat{\tilde{1}}_{\vec{k}} \ket{\vec{G}'}$ is conceptually identical to that of $\bra{\vec{G}} \hat D_{\vec{k}} \ket{\vec{G}'}$ and the final result is identical to Eqs.~\eqref{UGG_der1} and \eqref{UGG_der2}, with the substitution $D_{\vec{R}ij} \to S_{\vec{R}ij}$.
Also, to obtain the NCPP version of this section, one simply has to substitute $\hat{\tilde{1}}_{\vec{k}} \to 1$ ($S_{\vec{R}ij} \to 0$).

The implementation of the first-order perturbed quantities $\hat{\tilde{H}}_{\vec{k}}^\alpha$ and $\hat{\tilde{1}}_{\vec{k}}^\alpha$ (as per Eqs.~\eqref{UGG_der1}, \eqref{pbar_der1}, \eqref{ptilde_der1} and their analogs for $\hat{\tilde{1}}_{\vec{k}}$) in ABINIT was already done by C. Audouze and co-workers~\cite{Audouze08}. 
Therefore, we only had to implement the second-order quantities  $\hat{\tilde{H}}_{\vec{k}}^{\alpha \beta}$ and $\hat{\tilde{1}}_{\vec{k}}^{\alpha \beta}$ (as per Eqs.~\eqref{UGG_der2}, \eqref{pbar_der2}, \eqref{ptilde_der2} and their analogs for $\hat{\tilde{1}}_{\vec{k}}$) in the code.

\subsection{Spin-orbit coupling}
\label{SO}

A simple, approximate way to take into account spin-orbit coupling (SOC) within the PAW framework is to calculate the coupling only in the PAW augmentation regions, which is what is done in the ABINIT software~\cite{Gonze:2016te,Gonze:2009aa}. 
The hypotheses underlying this approximation, and its practical validity, have been discussed in Ref.~\onlinecite{Corso:2012aa}.
Note that since it involves the PAW augmentation regions, it cannot be applied to the NCPP and USPP methods.

Within this approximation, the PAW Hamiltonian of Eq.~\eqref{PAW-H} becomes 
\begin{multline}
\label{SO-H}
\hat{\tilde{H}}_{\vec{k}} = 
\hat H_{\vec{k}} + 
\sum_{\vec{R}ij\sigma\sigma'} 
e^{-i \vec{k} \cdot \hat{\vec{r}}} \ket{\tilde{p}_{\vec{R}i\sigma}} 
\big( D_{\vec{R}ij} \delta_{\sigma\sigma'} \\
+ D_{\vec{R}ij\sigma\sigma'}^{\rm SO} \big)
\bra{\tilde{p}_{\vec{R}j\sigma'}} e^{i \vec{k} \cdot \hat{\vec{r}}}, 
\end{multline}
where $\sigma$ denotes the spin component of the wavefunction acted upon by the projector and where $D_{\vec{R}ij\sigma\sigma'}^{\rm SO}$ are the matrix elements of the spin-orbit Hamiltonian between all electron partial waves $\ket{\phi_{\vec{R}i\sigma}}$~\cite{Corso:2012aa}.

We note that Eq.~\eqref{SO-H} has the same $\vec{k}$-dependence than the non-local operators studied in Section~\ref{nl}. 
Therefore, since we already implemented the corresponding derivatives (Eqs.~\eqref{UGG_der1}-\eqref{ptilde_der2}) and since the ground-state spin-orbit Hamiltonian (i.e. $D_{\vec{R}ij\sigma\sigma'}^{\rm SO}$) was already available in the ABINIT software, we added SOC support to our effective mass implementation by introducing spinors $\ket{\tilde{\psi}_{n\vec{k}}} \to \ket{\tilde{\psi}_{n\vec{k}\sigma}}$ in Section~\ref{nd_eps}-\ref{nl} and substituting $D_{\vec{R}ij} \delta_{\sigma\sigma'} \to D_{\vec{R}ij} \delta_{\sigma\sigma'} + D_{\vec{R}ij\sigma\sigma'}^{\rm SO}$.
For NCPP, the equations that allow to treat the SOC within DFPT have been presented in Ref.~\onlinecite{Verstraete:2008aa} and applied in the case of phonons. 
The present formalism might easily be generalized to this case, although this has not been part of the present work.

\section{The degenerate case and transport equivalent effective masses}
\label{d}

\subsection{Effective mass tensor and degeneracy}
\label{d_eps}

In degenerate perturbation theory, the corrections to observables $\hat{O}$ at successive orders $(i)$ with respect to a variable $x$ are calculated $O_{nn' \vec{k}}^{(i)} = \partial^i \bra{u_{n'\vec{k}}} \hat{O}_{\vec{k}} \ket{u_{n\vec{k}}} / \partial x^i$ until one finds an order at which the degenerescence is lifted $O_{nn' \vec{k}}^{(i)} \neq C_{\vec{k}}^{(i)} \, \delta_{nn'}$, $n$ and $n'$ labelling states within the degenerate subspace and $C_{\vec{k}}^{(i)}$ being a proportionality constant. 
Thus, Section~\ref{nd_eps} applies if there is no degenerescence at second order, i.e. if the degeneracy is lifted at 0th or 1st order. 
We now consider the case where the degeneracy is maintained at 0th and 1st order (e.g. a degenerate band extrema), which is a case often encountered in important technological materials, like the III-V or II-VI semiconductors. 
Eq.~\eqref{taylor_eps} can then be generalized to (see Eq.~(IV.9) of Ref.~\onlinecite{Luttinger55})
\begin{multline}
\varepsilon_{nn' \vec{k} + \delta \vec{k}}
= \varepsilon_{\{d\}\vec{k}} \delta_{nn'}
+ \sum_\alpha \varepsilon^\alpha_{\{d\} \vec{k}} \delta_{nn'} \delta k_\alpha \\
+ \frac{1}{2} \sum_{\alpha \beta} \delta k_\alpha \varepsilon^{\alpha \beta}_{nn' \vec{k}} \delta k_\beta 
+ O(\delta k^3);
\quad \ n,n' \in \{d\}
\end{multline}
where $\{d\}$ represents the degenerate subspace and with $\varepsilon_{n\vec{k}} = \varepsilon_{\{d\}\vec{k}}$ and $\varepsilon^\alpha_{n\vec{k}} = \varepsilon^\alpha_{\{d\}\vec{k}} \ \forall \ n \in \{d\}$.
In such cases, after obtaining the $\varepsilon^{\alpha \beta}_{nn' \vec{k}}$ matrix within the degenerate subspace, one must diagonalize it to find the relevant eigenstates, eigenenergies, and the associated effective masses. 

The first step is therefore to generalize $\varepsilon^{\alpha \beta}_{n \vec{k}}$ to $\varepsilon^{\alpha \beta}_{nn' \vec{k}}$. 
We will also consider in this subsection the more general case of PAW (where $\hat{\tilde{1}}_{\vec{k}}$ is present, see Subsection~\ref{nl}), from which the norm-conserving expressions can be obtained by substituting $\hat{\tilde{1}}_{\vec{k}} \to 1$. 
Moreover, as discussed at the beginning of Section~\ref{nl}, the equations derived in this Section for PAW also apply to the USPP method within the parallel gauge (see Refs.~\onlinecite{Gonze95,Audouze08,Audouze06} for more details on this gauge).
This Section thus generalizes Eqs.~\eqref{eps_Pstern} and \eqref{Pstern} to both the degenerate case and the PAW (USPP) formalism(s).

We will systematically express $\varepsilon_{\{d\}\vec{k}}$ as $\frac{\varepsilon_{n\vec{k}}+\varepsilon_{n'\vec{k}}}{2}$ and $\varepsilon_{\{d\}\vec{k}}^\alpha$ as $\frac{\varepsilon_{n\vec{k}}^\alpha+\varepsilon_{n'\vec{k}}^\alpha}{2}$ in order to be explicitly symmetric in $n,n'$ and to facilitate the comparison with other PAW expressions within the parallel gauge (see e.g. Eq.~(78) of Ref.~\onlinecite{Audouze08}).

We obtain analytically the derivatives of $\varepsilon_{nn' \vec{k}}$ from the derivatives of the Hamiltonian and overlap operator, starting from 
\begin{equation}
\label{PAW_eps}
\varepsilon_{nn' \vec{k}}
= \bra{\tilde{u}_{n' \vec{k}}} \hat{\tilde{H}}_{\vec{k}} \ket{\tilde{u}_{n\vec{k}}} 
\end{equation}
and differentiating with respect to $k_\alpha$
\begin{multline}
\label{eps1_raw}
\varepsilon^\alpha_{nn' \vec{k}}
= \bra{\tilde{u}^\alpha_{n' \vec{k}}} \hat{\tilde{H}}_{\vec{k}} \ket{\tilde{u}_{n\vec{k}}} 
+ \bra{\tilde{u}_{n'\vec{k}}} \hat{\tilde{H}}^\alpha_{\vec{k}} \ket{\tilde{u}_{n\vec{k}}} \\
+ \bra{\tilde{u}_{n' \vec{k}}} \hat{\tilde{H}}_{\vec{k}} \ket{\tilde{u}^\alpha_{n\vec{k}}}. 
\end{multline}
Using Eq.~\eqref{PAW-up} and evaluating at the degenerate point $\vec{k}$ yields 
\begin{multline}
\label{eps1_PAW_b}
\varepsilon^\alpha_{nn' \vec{k}}
= \frac{\varepsilon_{n \vec{k}} + \varepsilon_{n' \vec{k}}}{2} 
\Big( \bra{\tilde{u}^\alpha_{n' \vec{k}}} \hat{\tilde{1}}_{\vec{k}} \ket{\tilde{u}_{n\vec{k}}} \\
+ \bra{\tilde{u}_{n' \vec{k}}} \hat{\tilde{1}}_{\vec{k}} \ket{\tilde{u}^\alpha_{n\vec{k}}} \Big) 
+ \bra{\tilde{u}_{n'\vec{k}}} \hat{\tilde{H}}^\alpha_{\vec{k}} \ket{\tilde{u}_{n\vec{k}}}. 
\end{multline}
Using the derivative of the PAW version of the orthonormalization condition (from Eqs.~\eqref{tr_PAW}, \eqref{overlap_op}, \eqref{bloch}, and \eqref{bloch_op})
\begin{equation}
\label{orthonorm-PAW}
\delta_{nn'}
= \braket{\psi_{n' \vec{k}} | \psi_{n\vec{k}}} 
= \bra{\tilde{u}_{n' \vec{k}}} \hat{\tilde{1}}_{\vec{k}} \ket{\tilde{u}_{n\vec{k}}} 
\end{equation}
\begin{multline}
\label{orthonorm1}
\Rightarrow 0 = 
\bra{\tilde{u}^\alpha_{n' \vec{k}}} \hat{\tilde{1}}_{\vec{k}} \ket{\tilde{u}_{n\vec{k}}} 
+ \bra{\tilde{u}_{n' \vec{k}}} \hat{\tilde{1}}^\alpha_{\vec{k}} \ket{\tilde{u}_{n\vec{k}}} \\
+ \bra{\tilde{u}_{n' \vec{k}}} \hat{\tilde{1}}_{\vec{k}} \ket{\tilde{u}^\alpha_{n\vec{k}}}, 
\end{multline}
Eq.~\eqref{eps1_PAW_b} becomes
\begin{equation}
\label{eps1}
\varepsilon^\alpha_{nn' \vec{k}}
= \bra{\tilde{u}_{n'\vec{k}}} 
    \hat{\tilde{H}}^\alpha_{\vec{k}} 
    - \frac{\varepsilon_{n \vec{k}} + \varepsilon_{n' \vec{k}}}{2} \hat{\tilde{1}}^\alpha_{\vec{k}} 
  \ket{\tilde{u}_{n\vec{k}}}. 
\end{equation}

We now proceed to the second-order derivative, starting from Eq.~\eqref{eps1_raw}.
Differentiating with respect to $k_\beta$ yields
\begin{multline}
\varepsilon_{nn' \vec{k}}^{\alpha\beta}
=          \bra{\tilde{u}_{n' \vec{k}}              } \hat{\tilde{H}}_{\vec{k}}^{\alpha\beta} \ket{\tilde{u}_{n\vec{k}}              }         \\
  + \left( \bra{\tilde{u}_{n' \vec{k}}^{\alpha\beta}} \hat{\tilde{H}}_{\vec{k}}               \ket{\tilde{u}_{n\vec{k}}              }
  +        \bra{\tilde{u}_{n' \vec{k}}              } \hat{\tilde{H}}_{\vec{k}}               \ket{\tilde{u}_{n\vec{k}}^{\alpha\beta}} \right) \\
  + \Big ( \bra{\tilde{u}_{n' \vec{k}}^\alpha       } \hat{\tilde{H}}_{\vec{k}}^\beta         \ket{\tilde{u}_{n\vec{k}}              } 
  +        \bra{\tilde{u}_{n' \vec{k}}^\alpha       } \hat{\tilde{H}}_{\vec{k}}               \ket{\tilde{u}_{n\vec{k}}^\beta        }          \\
  +        \bra{\tilde{u}_{n' \vec{k}}              } \hat{\tilde{H}}_{\vec{k}}^\alpha        \ket{\tilde{u}_{n\vec{k}}^\beta        } \Big  )
  + \alpha \leftrightarrow \beta. 
\end{multline}
Using Eq.~\eqref{PAW-up} and the second-order derivative of the orthonormalization condition (Eq.~\eqref{orthonorm-PAW}) 
\begin{multline}
\label{orthonorm2}
           \bra{\tilde{u}_{n' \vec{k}}              } \hat{\tilde{1}}_{\vec{k}}^{\alpha\beta} \ket{\tilde{u}_{n\vec{k}}              }         
 +  \left( \bra{\tilde{u}_{n' \vec{k}}^{\alpha\beta}} \hat{\tilde{1}}_{\vec{k}}               \ket{\tilde{u}_{n\vec{k}}              }
 +         \bra{\tilde{u}_{n' \vec{k}}              } \hat{\tilde{1}}_{\vec{k}}               \ket{\tilde{u}_{n\vec{k}}^{\alpha\beta}} \right) \\
 +  \Big ( \bra{\tilde{u}_{n' \vec{k}}^\alpha       } \hat{\tilde{1}}_{\vec{k}}^\beta         \ket{\tilde{u}_{n\vec{k}}              } 
 +         \bra{\tilde{u}_{n' \vec{k}}^\alpha       } \hat{\tilde{1}}_{\vec{k}}               \ket{\tilde{u}_{n\vec{k}}^\beta        }          
 +         \bra{\tilde{u}_{n' \vec{k}}              } \hat{\tilde{1}}_{\vec{k}}^\alpha        \ket{\tilde{u}_{n\vec{k}}^\beta        } \Big  )  \\
 +  \alpha \leftrightarrow \beta 
= 0
\end{multline}
as well as evaluating at the degenerate point $\vec{k}$ leads to
\begin{multline}
\label{eps2a}
\varepsilon_{nn' \vec{k}}^{\alpha\beta}
=         \bra{\tilde{u}_{n' \vec{k}}              } \hat{\tilde{H}}_{\vec{k}}^{\alpha\beta} - \frac{\varepsilon_{n \vec{k}} + \varepsilon_{n' \vec{k}}}{2} \hat{\tilde{1}}_{\vec{k}}^{\alpha\beta} \ket{\tilde{u}_{n\vec{k}}              }         \\
 + \Big ( \bra{\tilde{u}_{n' \vec{k}}^\alpha       } \hat{\tilde{H}}_{\vec{k}}^\beta         - \frac{\varepsilon_{n \vec{k}} + \varepsilon_{n' \vec{k}}}{2} \hat{\tilde{1}}_{\vec{k}}^\beta         \ket{\tilde{u}_{n\vec{k}}              }          \\ 
 +        \bra{\tilde{u}_{n' \vec{k}}^\alpha       } \hat{\tilde{H}}_{\vec{k}}               - \frac{\varepsilon_{n \vec{k}} + \varepsilon_{n' \vec{k}}}{2} \hat{\tilde{1}}_{\vec{k}}               \ket{\tilde{u}_{n\vec{k}}^\beta        }          \\
 +        \bra{\tilde{u}_{n' \vec{k}}              } \hat{\tilde{H}}_{\vec{k}}^\alpha        - \frac{\varepsilon_{n \vec{k}} + \varepsilon_{n' \vec{k}}}{2} \hat{\tilde{1}}_{\vec{k}}^\alpha        \ket{\tilde{u}_{n\vec{k}}^\beta        } \Big  )  \\
 + \alpha \leftrightarrow \beta .
\end{multline}

Since the analytical derivatives of $\hat{\tilde{H}}_{\vec{k}}$ and $\hat{\tilde{1}}_{\vec{k}}$ are available (see Section~\ref{nl}), the only task left is to find $\ket{\tilde{u}_{n\vec{k}}^\alpha}$ or, more precisely, to generalize Eqs.~\eqref{eps_Pstern} and \eqref{Pstern} to the degenerate case and the PAW formalism. 
To handle the degeneracy, we separate the components of $\ket{\tilde{u}_{n\vec{k}}^\alpha}$ lying in the degenerate subspace $\{{\rm d}\}$ from the rest
\begin{equation}
\label{psi1_dnd}
\ket{\tilde{u}_{n\vec{k}}^\alpha} = \hat{\tilde{P}}_{\vec{k}} \ket{\tilde{u}_{n\vec{k}}^\alpha} + \hat{\tilde{Q}}_{\vec{k}} \ket{\tilde{u}_{n\vec{k}}^\alpha},
\end{equation}
with
\begin{align}
\label{Pdeg}
\hat{\tilde{P}}_{\vec{k}} &\triangleq \sum_{n' \in \{{\rm d}\}} \ket{\tilde{u}_{n'\vec{k}}} \bra{\tilde{u}_{n'\vec{k}}} \hat{\tilde{1}}_{\vec{k}},\\
\hat{\tilde{Q}}_{\vec{k}} &\triangleq \sum_{n \notin \{{\rm d}\}} \ket{\tilde{u}_{n\vec{k}}} \bra{\tilde{u}_{n\vec{k}}} \hat{\tilde{1}}_{\vec{k}}.
\end{align}
Substituting Eq.~\eqref{psi1_dnd} in Eq.~\eqref{eps2a} and carrying out some algebra presented in Appendix~\ref{PAW-alg}, we obtain (see Eq.~\eqref{eps2c-A}
\begin{widetext}
\begin{multline}
\label{eps2c}
\varepsilon_{nn' \vec{k}}^{\alpha\beta}
= \bra{\tilde{u}_{n' \vec{k}}} \hat{\tilde{H}}_{\vec{k}}^{\alpha\beta} - \frac{\varepsilon_{n \vec{k}} + \varepsilon_{n' \vec{k}}}{2} \hat{\tilde{1}}_{\vec{k}}^{\alpha\beta} \ket{\tilde{u}_{n\vec{k}}} \\
 + \Big ( \bra{\hat{\tilde{Q}}_{\vec{k}} \tilde{u}_{n' \vec{k}}^\alpha - \frac{1}{2} \delta \tilde{u}_{n\vec{k}}^\alpha}   
          \hat{\tilde{H}}_{\vec{k}}^\beta - \frac{\varepsilon_{n \vec{k}} + \varepsilon_{n' \vec{k}}}{2} \hat{\tilde{1}}_{\vec{k}}^\beta 
          \ket{\tilde{u}_{n\vec{k}}}  
 + \bra{\tilde{u}_{n' \vec{k}}} 
   \hat{\tilde{H}}_{\vec{k}}^\alpha - \frac{\varepsilon_{n \vec{k}} + \varepsilon_{n' \vec{k}}}{2} \hat{\tilde{1}}_{\vec{k}}^\alpha    
   \ket{\hat{\tilde{Q}}_{\vec{k}} \tilde{u}_{n\vec{k}}^\beta - \frac{1}{2} \delta \tilde{u}_{n\vec{k}}^\beta} \\
 + \bra{\hat{\tilde{Q}}_{\vec{k}} \tilde{u}_{n' \vec{k}}^\alpha - \frac{1}{2} \delta \tilde{u}_{n\vec{k}}^\alpha} 
   \hat{\tilde{H}}_{\vec{k}}    
   - \frac{\varepsilon_{n \vec{k}} + \varepsilon_{n' \vec{k}}}{2} \hat{\tilde{1}}_{\vec{k}} 
   \ket{\hat{\tilde{Q}}_{\vec{k}} \tilde{u}_{n\vec{k}}^\beta - \frac{1}{2} \tilde{u}_{n\vec{k}}^\beta} \Big )   
 + \alpha \leftrightarrow \beta,
\end{multline}
\end{widetext}
where we have defined (see Eq.~\eqref{dpsi1a-A})
\begin{equation}
\label{dpsi1a}
 \ket{\delta \tilde{u}_{n\vec{k}}^\alpha} \triangleq \sum_{n' \in \{d\}} \ket{\tilde{u}_{n'\vec{k}}} \bra{\tilde{u}_{n'\vec{k}}} \hat{\tilde{1}}_{\vec{k}}^\alpha \ket{\tilde{u}_{n\vec{k}}}.  
\end{equation}

We are now left with the task of finding $\ket{\hat{\tilde{Q}}_{\vec{k}} \tilde{u}_{n\vec{k}}^\alpha}$ within PAW.
To do so, we first differentiate Eq.~\eqref{PAW-up} 
\begin{multline}
 \hat{\tilde{H}}_{\vec{k}}^\alpha \ket{\tilde{u}_{n\vec{k}}} 
+ \hat{\tilde{H}}_{\vec{k}} \ket{\tilde{u}_{n\vec{k}}^\alpha} \\
=\varepsilon_{n\vec{k}}^\alpha \hat{\tilde{1}}_{\vec{k}} \ket{\tilde{u}_{n\vec{k}}}
+ \varepsilon_{n\vec{k}} \hat{\tilde{1}}_{\vec{k}}^\alpha \ket{\tilde{u}_{n\vec{k}}}
+ \varepsilon_{n\vec{k}} \hat{\tilde{1}}_{\vec{k}} \ket{\tilde{u}_{n\vec{k}}^\alpha}.
\end{multline}
Re-arranging the terms and applying $\sum_{n' \notin \{{\rm d}\}} \ket{\tilde{u}_{n'\vec{k}}} \bra{\tilde{u}_{n'\vec{k}}}$ on both sides yields
\begin{multline}
\sum_{n' \notin \{{\rm d}\}} \ket{\tilde{u}_{n'\vec{k}}} \bra{\tilde{u}_{n'\vec{k}}} 
\hat{\tilde{H}}_{\vec{k}} - \varepsilon_{n\vec{k}} \hat{\tilde{1}}_{\vec{k}} \ket{\tilde{u}_{n\vec{k}}^\alpha} \\
= - \sum_{n' \notin \{{\rm d}\}} \ket{\tilde{u}_{n'\vec{k}}} \bra{\tilde{u}_{n'\vec{k}}} 
\hat{\tilde{H}}_{\vec{k}}^\alpha - \varepsilon_{n\vec{k}} \hat{\tilde{1}}_{\vec{k}}^\alpha \ket{\tilde{u}_{n\vec{k}}},
\end{multline}
where we have used the orthonormality condition of Eq.~\eqref{orthonorm-PAW}.
Finally, using Eq.~\eqref{PAW-up}, dividing by $\varepsilon_{n'\vec{k}} - \varepsilon_{n\vec{k}}$, and using Eq.~\eqref{Pdeg} yields
\begin{multline}
\label{QuSOS}
\hat{\tilde{Q}}_{\vec{k}} \ket{\tilde{u}_{n\vec{k}}^\alpha} = \\
 - \sum_{n' \notin \{{\rm d}\}} \frac{\ket{\tilde{u}_{n'\vec{k}}} \bra{\tilde{u}_{n'\vec{k}}} \hat{\tilde{H}}_{\vec{k}}^\alpha - \varepsilon_{n\vec{k}} \hat{\tilde{1}}_{\vec{k}}^\alpha \ket{\tilde{u}_{n\vec{k}}}}{\varepsilon_{n'\vec{k}} - \varepsilon_{n\vec{k}}}.
\end{multline}
For reasons that have already been mentioned in the paragraph after Eq.~\eqref{eps2SOS}, we calculate the contribution to $\hat{\tilde{Q}}_{\vec{k}} \ket{\tilde{u}_{n\vec{k}}^\alpha}$ stemming from the subspace generated by the bands explicitly treated in the calculation (the active subspace $n \leq N$) using a sum-over-states approach while we solve a linear equation to obtain the contribution stemming from the complementary subspace 
\begin{equation}
\label{QN-PAW}
\hat{\tilde{Q}}_{N\vec{k}} = \sum_{n > N} \ket{\tilde{u}_{n\vec{k}}} \bra{\tilde{u}_{n\vec{k}}} \hat{\tilde{1}}_{\vec{k}}.
\end{equation}
We thus split $\hat{\tilde{Q}}_{\vec{k}} \ket{\tilde{u}_{n\vec{k}}^\alpha}$ into these two contributions and add the $- \frac{1}{2} \ket{\delta \tilde{u}_{n\vec{k}}^\alpha}$ contribution required to calculate Eq.~\eqref{eps2c}.
Also, since it is more convenient to calculate and store the symmetrized matrix elements of $\hat{\tilde{H}}_{\vec{k}}^\alpha - \varepsilon_{n\vec{k}} \hat{\tilde{1}}_{\vec{k}}^\alpha$ (i.e. $\bra{\tilde{u}_{n'\vec{k}}} \hat{\tilde{H}}_{\vec{k}}^\alpha - \frac{\varepsilon_{n\vec{k}} + \varepsilon_{n'\vec{k}}}{2} \hat{\tilde{1}}_{\vec{k}}^\alpha \ket{\tilde{u}_{n\vec{k}}}$), we re-express the part of Eq.~\eqref{QuSOS} that stems from the active subspace ($n \leq N$) so that it uses these symmetric matrix elements.
We obtain
\begin{multline}
\label{eps2c-ket}
\hat{\tilde{Q}}_{\vec{k}}  \ket{\tilde{u}_{n\vec{k}}^\alpha} - \frac{1}{2} \ket{\delta \tilde{u}_{n\vec{k}}^\alpha} \\
=- \sum_{n' \notin \{{\rm d}\}}^N \frac{\ket{\tilde{u}_{n'\vec{k}}} \bra{\tilde{u}_{n'\vec{k}}}
\hat{\tilde{H}}_{\vec{k}}^\alpha - \frac{\varepsilon_{n\vec{k}} + \varepsilon_{n'\vec{k}}}{2} \hat{\tilde{1}}_{\vec{k}}^\alpha \ket{\tilde{u}_{n\vec{k}}}}{\varepsilon_{n'\vec{k}} - \varepsilon_{n\vec{k}}} \\
 + \ket{\hat{\tilde{Q}}_{N\vec{k}} \tilde{u}_{n\vec{k}}^\alpha}
  - \frac{1}{2} \ket{\delta \tilde{u}_{Nn\vec{k}}^\alpha},
\end{multline}
where we have defined 
\begin{equation}
\label{dpsi1}
 \ket{\delta \tilde{u}_{Nn\vec{k}}^\alpha} \triangleq \sum_{n'}^N \ket{\tilde{u}_{n'\vec{k}}} \bra{\tilde{u}_{n'\vec{k}}} \hat{\tilde{1}}_{\vec{k}}^\alpha \ket{\tilde{u}_{n\vec{k}}}  
\end{equation}
(following the notation introduced in Eq.~(42) of Ref.~\onlinecite{Audouze08}, with an additional $N$ index) and 
\begin{multline}
\label{PNpsi1SOS}
\ket{\hat{\tilde{Q}}_{N\vec{k}} \tilde{u}_{n\vec{k}}^\alpha} =\\
- \sum_{n'>N} \frac{\ket{\tilde{u}_{n'\vec{k}}} \bra{\tilde{u}_{n'\vec{k}}}
\hat{\tilde{H}}_{\vec{k}}^\alpha - \varepsilon_{n\vec{k}} \hat{\tilde{1}}_{\vec{k}}^\alpha \ket{\tilde{u}_{n\vec{k}}}} {\varepsilon_{n'\vec{k}} - \varepsilon_{n\vec{k}}}.  
\end{multline}
To obtain the last contribution from a linear equation problem, we start from Eq.~\eqref{PNpsi1SOS}, use Eq.~\eqref{QN-PAW}, multiply by $\varepsilon_{n'\vec{k}} - \varepsilon_{n\vec{k}}$, then use Eq.~\eqref{PAW-up} and the conjugate transpose of Eq.~\eqref{QN-PAW}
\begin{equation}
\label{PNpsi1}
\hat{\tilde{Q}}_{N\vec{k}}^\dagger 
\Big( \hat{\tilde{H}}_{\vec{k}} - \varepsilon_{n\vec{k}} \hat{\tilde{1}}_{\vec{k}} \Big) 
\ket{\hat{\tilde{Q}}_{N\vec{k}} \tilde{u}_{n\vec{k}}^\alpha}
= - \hat{\tilde{Q}}_{N\vec{k}}^\dagger 
\Big( \hat{\tilde{H}}_{\vec{k}}^\alpha - \varepsilon_{n\vec{k}} \hat{\tilde{1}}_{\vec{k}}^\alpha \Big) 
\ket{\tilde{u}_{n\vec{k}}}.  
\end{equation}

Now, substituting Eq.~\eqref{eps2c-ket} into Eq.~\eqref{eps2c} and simplifying using Eqs.~\eqref{QN-PAW}, \eqref{dpsi1}, \eqref{PAW-up}, and \eqref{orthonorm-PAW} yields 
\begin{widetext}
\begin{multline}
\label{eps2}
\varepsilon_{nn' \vec{k}}^{\alpha\beta} = 
   \bra{\tilde{u}_{n' \vec{k}}} \hat{\tilde{H}}_{\vec{k}}^{\alpha\beta} - \frac{\varepsilon_{n \vec{k}} + \varepsilon_{n' \vec{k}}}{2} \hat{\tilde{1}}_{\vec{k}}^{\alpha\beta} \ket{\tilde{u}_{n\vec{k}}}   
  + \Big ( \bra{\hat{\tilde{Q}}_{N\vec{k}} \tilde{u}_{n'\vec{k}}^\alpha + \delta \tilde{u}_{Nn'\vec{k}}^\alpha}
           \hat{\tilde{H}}_{\vec{k}} - \frac{\varepsilon_{n \vec{k}} + \varepsilon_{n' \vec{k}}}{2} \hat{\tilde{1}}_{\vec{k}}   
           \ket{\hat{\tilde{Q}}_{N\vec{k}} \tilde{u}_{n\vec{k}}^\beta + \delta \tilde{u}_{Nn\vec{k}}^\beta} \\
  +        \bra{\hat{\tilde{Q}}_{N\vec{k}} \tilde{u}_{n'\vec{k}}^\alpha + \delta \tilde{u}_{Nn'\vec{k}}^\alpha}
           \hat{\tilde{H}}_{\vec{k}}^\beta - \frac{\varepsilon_{n \vec{k}} + \varepsilon_{n' \vec{k}}}{2} \hat{\tilde{1}}_{\vec{k}}^\beta 
           \ket{\tilde{u}_{n\vec{k}}}    
  +        \bra{\tilde{u}_{n' \vec{k}}} 
           \hat{\tilde{H}}_{\vec{k}}^\alpha - \frac{\varepsilon_{n \vec{k}} + \varepsilon_{n' \vec{k}}}{2} \hat{\tilde{1}}_{\vec{k}}^\alpha 
           \ket{\hat{\tilde{Q}}_{N\vec{k}} \tilde{u}_{n\vec{k}}^\beta + \delta \tilde{u}_{Nn\vec{k}}^\beta} \\
  +        \sum_{n'' \notin \{{\rm d}\}}^N 
           \bra{\tilde{u}_{n'\vec{k}}} 
           \hat{\tilde{H}}_{\vec{k}}^\alpha - \frac{\varepsilon_{n'\vec{k}} + \varepsilon_{n''\vec{k}}}{2} \hat{\tilde{1}}_{\vec{k}}^\alpha 
           \ket{\tilde{u}_{n''\vec{k}}}   
           \frac{1}{\varepsilon_{n \vec{k}} - \varepsilon_{n''\vec{k}}} \bra{\tilde{u}_{n''\vec{k}}}
           \hat{\tilde{H}}_{\vec{k}}^\beta - \frac{\varepsilon_{n''\vec{k}} + \varepsilon_{n\vec{k}}}{2} \hat{\tilde{1}}_{\vec{k}}^\beta
           \ket{\tilde{u}_{n\vec{k}}} \Big ) + \alpha \leftrightarrow \beta.
\end{multline}
\end{widetext}
Eqs.~\eqref{eps2} and \eqref{PNpsi1} are the extension of Eqs.~\eqref{eps_Pstern} and \eqref{Pstern} to degeneracy and PAW (USPP) that we were seeking. 

By substituting $\hat{\tilde{1}}_{\vec{k}} \to 1$ in Eq.~\eqref{eps2}, we recover the NCPP expression for degenerate states
\begin{multline}
\label{eps2-NCPP}
\varepsilon_{nn' \vec{k}}^{\alpha\beta} = 
   \bra{u_{n' \vec{k}}} \hat{H}_{\vec{k}}^{\alpha\beta} \ket{u_{n\vec{k}}} \\  
  + \Big ( \bra{\hat{Q}_{N\vec{k}} u_{n'\vec{k}}^\alpha}
           \hat{H}_{\vec{k}} - \frac{\varepsilon_{n \vec{k}} + \varepsilon_{n' \vec{k}}}{2}   
           \ket{\hat{Q}_{N\vec{k}} u_{n\vec{k}}^\beta} \\
  +        \bra{\hat{Q}_{N\vec{k}} u_{n'\vec{k}}^\alpha}
           \hat{H}_{\vec{k}}^\beta  
           \ket{u_{n\vec{k}}}  
  +        \bra{u_{n' \vec{k}}} 
           \hat{H}_{\vec{k}}^\alpha 
           \ket{\hat{Q}_{N\vec{k}} u_{n\vec{k}}^\beta} \\
  +        \sum_{n'' \notin \{{\rm d}\}}^N 
           \bra{u_{n'\vec{k}}} 
           \hat{H}_{\vec{k}}^\alpha  
           \frac{\ket{u_{n''\vec{k}}} \bra{u_{n''\vec{k}}}}{\varepsilon_{n \vec{k}} - \varepsilon_{n''\vec{k}}} 
           \hat{H}_{\vec{k}}^\beta 
           \ket{u_{n\vec{k}}} \Big ) \\
  +        \alpha \leftrightarrow \beta,
\end{multline}
where $\ket{\hat{Q}_{N\vec{k}} u_{n\vec{k}}^\alpha}$ is given by Eq.~\eqref{Pstern}.
The preceding expression can be simplified using Eqs.~\eqref{PNpsi1SOS} and \eqref{k-schro} to yield
\begin{multline}
\varepsilon_{nn' \vec{k}}^{\alpha\beta} = 
   \bra{u_{n' \vec{k}}} \hat{H}_{\vec{k}}^{\alpha\beta} \ket{u_{n\vec{k}}} \\  
  +        \bra{u_{n' \vec{k}}} 
           \hat{H}_{\vec{k}}^\alpha 
           \ket{\hat{Q}_{N\vec{k}} u_{n\vec{k}}^\beta}   
  +        \bra{\hat{Q}_{N\vec{k}} u_{n'\vec{k}}^\beta} 
           \hat{H}_{\vec{k}}^\alpha 
           \ket{u_{n\vec{k}}} \\
  + \Bigg( \sum_{n'' \notin \{{\rm d}\}}^N 
    \frac{ \bra{u_{n'\vec{k}}} 
           \hat{H}_{\vec{k}}^\alpha  
           \ket{u_{n''\vec{k}}} \bra{u_{n''\vec{k}}} 
           \hat{H}_{\vec{k}}^\beta 
           \ket{u_{n\vec{k}}}}{\varepsilon_{n \vec{k}} - \varepsilon_{n''\vec{k}}} \\
  +        \alpha \leftrightarrow \beta \Bigg),
\end{multline}
which simplifies to Eq.~\eqref{eps_Pstern} when degeneracy is lifted at 0th order.

\subsection{Transport equivalent effective mass from degenerate perturbation theory}
\label{TEefmas}

However, deducing transport properties from $\varepsilon_{nn' \vec{k}}^{\alpha\beta}$ as per Eqs.~\eqref{PNpsi1} and \eqref{eps2} is more involved than in the non-degenerate case, where we simply had $[\matr{M}_{n\vec{k}}^{-1}]_{\alpha \beta} \triangleq \varepsilon_{n\vec{k}}^{\alpha\beta}$ (Eq.~\eqref{efmas_nd}).
It is indeed not possible to obtain effective mass tensors that describe the dispersion of the individual bands: this would involve diagonalizing $\varepsilon_{nn' \vec{k}}^{\alpha\beta}$ with respect to $nn'$, but each matrix element in this case is a tensor $\matr{\varepsilon}_{nn' \vec{k}}$ (with $[\matr{\varepsilon}_{nn' \vec{k}}]_{\alpha\beta} \triangleq \varepsilon_{nn' \vec{k}}^{\alpha\beta}$) and it is not in general possible to diagonalize a matrix of tensors. 
It is however still possible to circumvent this complication by associating $\varepsilon_{nn' \vec{k}}^{\alpha\beta}$ to a set of $N_{\rm deg}$ `transport equivalent mass' tensors~\cite{Mecholsky14} $\bar{\matr{M}}_{n\vec{k}}$, which generate the same contribution to transport properties as the less intuitive $\varepsilon_{nn' \vec{k}}^{\alpha\beta}$ does. 
To make this association, we begin by transforming the tensorial matrix elements into scalar quantities. 
This is easily done by picking a direction $(\theta,\phi)$ in spherical coordinates along which we describe the bands dispersion
\begin{align}
\varepsilon_{nn'\vec{k}+\vec{q}} 
&= \varepsilon_{\{d\} \vec{k}} + \frac{1}{2} \sum_{\alpha \beta} \vec{q}_\alpha \varepsilon_{nn' \vec{k}}^{\alpha\beta} \vec{q}_\beta \nonumber \\
&= \varepsilon_{\{d\} \vec{k}} + \frac{q^2}{2} \sum_{\alpha \beta} \hat{\vec{q}}_\alpha (\theta,\phi) \varepsilon_{nn' \vec{k}}^{\alpha\beta} \hat{\vec{q}}_\beta (\theta,\phi) \label{f_d}\\
&\triangleq \varepsilon_{\{d\} \vec{k}} + \frac{q^2}{2} f_{nn'\vec{k}}(\theta,\phi), \nonumber
\end{align}
where we have supposed that the degenerate point $\vec{k}$ is also a band extrema $\varepsilon_{nn' \vec{k}}^{\alpha} = 0$, where $\vec{q}$ represents a position in the Brillouin zone with respect to the band extrema $\vec{k}$, and where we have defined the curvature $f_{nn'\vec{k}}(\theta,\phi)$ of the matrix elements $\varepsilon_{nn'\vec{k}}$ at the band extrema along the direction $(\theta,\phi)$. 
We now deal with an angular-dependent matrix of scalars $f_{nn'\vec{k}}(\theta,\phi)$, which can be diagonalized for each combination of $(\theta,\phi)$ considered in the calculation
\begin{multline}
\label{f_def}
\sum_{n'} f_{nn'\vec{k}}(\theta,\phi) \ket{\nu_{n'\vec{k}}(\theta,\phi)} = \\
f_{n\vec{k}}(\theta,\phi) \ket{\nu_{n\vec{k}}(\theta,\phi)}.
\end{multline}
The resulting eigenvalues $f_{n\vec{k}}(\theta,\phi)$ and eigenvectors $\ket{\nu_{n\vec{k}}(\theta,\phi)}$ yield the diagonalized version of Eq.~\eqref{f_d}
\begin{align}
\varepsilon_{n\vec{k}+\vec{q}} &= \varepsilon_{\{d\} \vec{k}} + \frac{q^2}{2} f_{n\vec{k}}(\theta,\phi) \\
&\triangleq \varepsilon_{\{d\} \vec{k}} + \frac{q^2}{2} \frac{1}{m_{n\vec{k}}(\theta,\phi)}, \label{m_th_ph}
\end{align}
where we have defined the angular-dependent effective mass $m_{n\vec{k}}(\theta,\phi)$. 

Once these angular dependent quantities are available for each band in the degenerate set, it becomes possible possible to apply the idea of Ref.~\onlinecite{Mecholsky14}. 
The latter associates to $f_{n\vec{k}}(\theta,\phi)$ a `transport equivalent mass' tensors $\bar{\matr{M}}_{n\vec{k}}$ that generate the same contribution to transport properties. 
This association supposes that the relaxation time approximation to Boltzmann's transport equation~\cite{AshcroftMermin,Marder} holds.
It furthermore supposes that the relaxation time depends on the energy only and that this dependence can be factored out of the tensor (i.e. a relaxation time of the form $\matr U_n \tau_{n\vec{k}}(\varepsilon)$).
It also requires that $\bar{\matr{M}}_{n\vec{k}}$ be calculated at a band extrema ($\varepsilon_{nn' \vec{k}}^{\alpha} = 0$), which is why we have already supposed that the degenerate point $\vec{k}$ is also a band extrema in Eq.~\eqref{f_d}.
For convenience, we summarize in Appendix~\ref{Meqv} the derivation of this association, generalized to an energy dependent relaxation time tensor $\matr \tau_{n\vec{k}}(\varepsilon)$ and specialized to the case of conductivity $\matr \sigma$ for concision. 
Since the 2D case presents some peculiarities with respect to the 3D case, we also extend the concept of `transport equivalent effective mass' to the 2D case in Appendix~\ref{Meqv-2D}.

The prescription to obtain $\bar{\matr{M}}_{n\vec{k}}$ from $f_{n\vec{k}}(\theta,\phi)$ begins with the calculation of the angular dependence of the electronic velocity (see Eq.~(29) of Ref.~\onlinecite{Mecholsky14} or Eq.~\eqref{speed-nu} of present work)
\begin{widetext}
\begin{equation}
\label{speed-nu-2}
\bar{\vec{v}}_{n\vec{k}}(\hat{\vec{q}}) \triangleq 
\begin{pmatrix}
2 f_{n\vec{k}}(\theta,\phi) \sin(\theta) \cos(\phi) + \frac{\partial f_{n\vec{k}}}{\partial \theta} \cos(\theta) \cos(\phi) - \frac{\partial f_{n\vec{k}}}{\partial \phi} \frac{\sin(\phi)}{\sin(\theta)} \\
2 f_{n\vec{k}}(\theta,\phi) \sin(\theta) \sin(\phi) + \frac{\partial f_{n\vec{k}}}{\partial \theta} \cos(\theta) \sin(\phi) + \frac{\partial f_{n\vec{k}}}{\partial \phi} \frac{\cos(\phi)}{\sin(\theta)} \\
2 f_{n\vec{k}}(\theta,\phi) \cos(\theta)            - \frac{\partial f_{n\vec{k}}}{\partial \theta} \sin(\theta) 
\end{pmatrix}.
\end{equation}
\end{widetext}
Then, one can deduce from $\bar{\vec{v}}_{n\vec{k}}(\hat{\vec{q}})$ and $f_{n\vec{k}}(\theta,\phi)$ a tensorial quantity $\matr{C}_{n\vec{k}}$ representing the angular contribution of extrema $\vec{k}$ of band $n$ to transport properties (e.g. $\matr{\sigma}$) (see Eqs.~\eqref{sigma}, \eqref{Kn} and \eqref{Cn}, which are analogous to Eqs.~(34), (35) and (36) of Ref.~\onlinecite{Mecholsky14})
\begin{equation}
\label{C_tensor}
\matr{C}_{n\vec{k}} = \int_0^{2\pi} d\phi \int_0^\pi d\theta \sin(\theta) \frac{\bar{\vec{v}}_{n\vec{k}}(\theta,\phi) \bar{\vec{v}}_{n\vec{k}}^T(\theta,\phi)}{2 |f_{n\vec{k}}(\theta,\phi)|^{5/2}}.
\end{equation}
In the present implementation, we use Gauss-Legendre quadrature to numerically integrate with respect to $\theta$ and $\phi$.
Finally, one has to rotate the Cartesian axes to diagonalize this tensor
\begin{equation}
\matr{U}_{n\vec{k}}^T \matr{C}_{n\vec{k}} \matr{U}_{n\vec{k}} \triangleq 
\begin{pmatrix}
C_{n\vec{k} x} & 0 & 0 \\
0 & C_{n\vec{k} y} & 0 \\
0 & 0 & C_{n\vec{k} z} 
\end{pmatrix},
\end{equation}
where $\matr{U}_{n\vec{k}}$ is the desired rotation, then deduce the components of $\bar{\matr{M}}_{n\vec{k}}$ in the rotated system (using Eq.~(38) of Ref.~\onlinecite{Mecholsky14} or Eq.~\eqref{m_ni} of present work) and rotate back to the original Cartesian axes 
\begin{multline}
\bar{\matr{M}}_{n\vec{k}} = \left( \frac{3}{8\pi} \right)^2 \matr{U}_{n\vec{k}} \\
\begin{pmatrix}
C_{n\vec{k} y} C_{n\vec{k} z} & 0 & 0 \\
0 & C_{n\vec{k} x} C_{n\vec{k} z} & 0 \\
0 & 0 & C_{n\vec{k} x} C_{n\vec{k} y} 
\end{pmatrix} \matr{U}_{n\vec{k}}^T .
\end{multline}

Note that the partial derivatives with respect to spherical angles of $f_{n\vec{k}}(\theta,\phi)$ can easily be obtained analytically 
\begin{align}
&f_{n\vec{k}}(\theta,\phi) \nonumber \\
&= \sum_{n'n'' \in \{{\rm d}\}} \braket{\nu_{n\vec{k}}(\theta,\phi) | u_{n'\vec{k}}} f_{n'n''\vec{k}}(\theta,\phi) \braket{u_{n''\vec{k}} | \nu_{n\vec{k}}(\theta,\phi)}, \nonumber \\
& \triangleq \bra{\nu_{n\vec{k}}(\theta,\phi)} \hat{f}_{\vec{k}}(\theta,\phi) \ket{\nu_{n\vec{k}}(\theta,\phi)}, \\
&\Rightarrow \frac{\partial f_{n\vec{k}}}{\partial \theta} \nonumber \\
&= \frac{\partial \bra{\nu_{n\vec{k}}}}{\partial \theta} \hat{f}_{\vec{k}} \ket{\nu_{n\vec{k}}}
 + \bra{\nu_{n\vec{k}}} \frac{\partial \hat{f}_{\vec{k}}}{\partial \theta} \ket{\nu_{n\vec{k}}}
 + \bra{\nu_{n\vec{k}}} \hat{f}_{\vec{k}} \frac{\partial \ket{\nu_{n\vec{k}}}}{\partial \theta}, \nonumber \\
&= \bra{\nu_{n\vec{k}}} \frac{\partial \hat{f}_{\vec{k}}}{\partial \theta} \ket{\nu_{n\vec{k}}} 
 + f_{n\vec{k}} \cancel{\frac{\partial \braket{\nu_{n\vec{k}} | \nu_{n\vec{k}}}}{\partial \theta}}, \nonumber \\
&= \sum_{n'n'' \in \{{\rm d}\}} \braket{\nu_{n\vec{k}}(\theta,\phi) | u_{n'\vec{k}}} \frac{\partial f_{n'n''\vec{k}}(\theta,\phi)}{\partial \theta} \braket{u_{n''\vec{k}} | \nu_{n\vec{k}}(\theta,\phi)},
\end{align}
with (see Eq.~\eqref{f_d}) 
\begin{multline}
\frac{\partial f_{nn'\vec{k}}(\theta,\phi)}{\partial \theta} 
= \sum_{\alpha \beta} \frac{\partial \hat{\vec{q}}_\alpha (\theta,\phi)}{\partial \theta} \varepsilon_{nn' \vec{k}}^{\alpha\beta} \hat{\vec{q}}_\beta (\theta,\phi) \\
+ \hat{\vec{q}}_\alpha (\theta,\phi) \varepsilon_{nn' \vec{k}}^{\alpha\beta} \frac{\partial \hat{\vec{q}}_\beta (\theta,\phi)}{\partial \theta},
\end{multline}
and with $\hat{\vec{q}}^T = (\sin \theta \cos \phi, \sin \theta \sin \phi, \cos \theta)$.
The analogous expressions for $\phi$ derivatives are easily obtained by substituting $\partial \theta \to \partial \phi$. 

It is also interesting to note that Eq.~\eqref{C_tensor} diverges in a 2D system, i.e. a system where $f_{n\vec{k}}(\theta,\phi) \to 0$ for some directions. 
This problem is assessed in Appendix~\ref{Meqv-2D}.

\section{Results}
\label{results}

\subsection{Numerical validation by direct comparison of DFPT against finite-difference effective masses in the case of silicon}
\label{DFPT-FD}

To check our implementation, we assessed its agreement with finite-difference calculations.
Finite-difference derivatives of the form $\varepsilon_{n \vec{k}}^{\alpha\alpha}$ were computed using second-order derivative formulae of order 2 and 8 calculated on a regularly spaced grid from Ref.~\onlinecite{Fornberg:1988en}
\begin{equation}
\frac{\partial^2 X}{\partial h^2} \Big{|}_{h_i} =  
\frac{X_{h_{i+1}} - 2X_{h_{i}} + X_{h_{i-1}}}{\Delta h^2} + O(\Delta h^2)
\end{equation}
and
\begin{multline}
\frac{\partial^2 X}{\partial h^2} \Big{|}_{h_i} = \\ 
\Big( \tfrac{-1}{560} X_{h_{i+4}} + \tfrac{8}{315} X_{h_{i+3}} - \tfrac{1}{5} X_{h_{i+2}} + \tfrac{8}{5} X_{h_{i+1}} - \tfrac{205}{72} X_{h_{i}} \\
+ \tfrac{8}{5} X_{h_{i-1}} - \tfrac{1}{5} X_{h_{i-2}} + \tfrac{8}{315} X_{h_{i-3}} - \tfrac{1}{560} X_{h_{i-4}} \Big) \big/ \Delta h^2 \\
+ O(\Delta h^8),
\end{multline}
where $X$ stands for any function of the independent variable $h$ computed on a regular grid of spacing $\Delta h$, with its points labeled by the index $i$ and where $O(\Delta h^n)$ denotes the order $n$ of the error. 
Also, derivatives of the form $\varepsilon_{n \vec{k}}^{\alpha\beta}$ where $\alpha \neq \beta$ were obtained by substituting the expression for the first-order derivative along the first direction $\alpha$ into the expression for the first-order derivative along the second direction $\beta$. 
The first-order derivative formulae used were those of Ref.~\onlinecite{Fornberg:1988en} for precision order 2 and 8 with a regularly spaced grid  
\begin{equation}
\frac{\partial X}{\partial h} \Big{|}_{h_i} =  
\frac{\tfrac{1}{2} X_{h_{i+1}} - \tfrac{1}{2} X_{h_{i-1}}}{\Delta h} + O(\Delta h^2)
\end{equation}
and
\begin{multline}
\frac{\partial X}{\partial h} \Big{|}_{h_i} =  
\Big( \tfrac{-1}{280} X_{h_{i+4}} + \tfrac{4}{105} X_{h_{i+3}} - \tfrac{1}{5} X_{h_{i+2}} + \tfrac{4}{5} X_{h_{i+1}} \\
- \tfrac{4}{5} X_{h_{i-1}} + \tfrac{1}{5} X_{h_{i-2}} - \tfrac{4}{105} X_{h_{i-3}} + \tfrac{1}{280} X_{h_{i-4}} \Big) \big/ \Delta h \\
+ O(\Delta h^8).
\end{multline}
The resulting expressions for $\varepsilon_{n \vec{k}_i}^{\alpha\beta}$ (where $\alpha \neq \beta$) at precision order 2 and 8 require 4 and 64 evaluations of $\varepsilon_{n \vec{k}_i}$, respectively.

In both DFPT and finite-difference calculations, the PAW formalism was used along with spin-orbit coupling within the PAW augmentation regions~\cite{Corso:2012aa,Gonze:2009aa} (see Section~\ref{SO}).
A cutoff energy of 20 Ha for the plane wave basis and 40 Ha for the PAW double grid along with a 6 6 6 Monkhorst-Pack grid for $\vec{k}$-point integrations were used to ensure that the calculated effective masses were converged to four significant digits. 
The local-density approximation of Perdew and Wang~\cite{Perdew:1992zza} was used, for coherence with available effective-mass calculations in the literature.
The cell parameter was fully relaxed, yielding $5.4015$ \AA.

Direct comparison of tensorial effective masses can only be done for non-degenerate bands, since the tensorial formalism becomes inappropriate for degenerate states (see Section~\ref{d}).
Thus, we compare in Table~\ref{FDvsRF} the values provided by the two methods for the effective mass of the first three non-degenerate bands of silicon at $\vec{\Gamma}$ ($\vec{\Gamma}_{1v}$, split-off $\vec{\Gamma}_{25v}'$, and split-off $\vec{\Gamma}_{15c}$).
Since the $\vec{\Gamma}$ point exhibit cubic symmetry, the effective mass tensors become proportional to the identity at this point, so that only one scalar needs to be reported in this case.

Also, for degenerate bands, it is still possible to obtain scalar effective masses along specific directions (see Section~\ref{TEefmas}). 
Thus, in Table~\ref{FDvsRFd}, we compare the values provided by the two methods for the scalar effective masses of the top valence bands of silicon at $\vec{\Gamma}$ in the Cartesian directions $(100)$, $(111)$, and $(110)$.

Finally, to assess the quality of the finite differences, we also compare the results obtained using order 2 and order 8 finite differences. 

\begin{table}
\caption{
Comparison of effective masses for the two non-degenerate valence bands and the first conduction band of silicon at $\vec{\Gamma}$. 
Due to the cubic symmetry, the effective masses tensors are proportional to the identity, so that only one scalar is reported.
We use order 2 and order 8 finite differences as well as DFPT to compute the masses.
The agreement between the finite-difference methods and DFPT is provided (DFPT - FD 2 and DFPT - FD 8, respectively).
Results are provided in atomic units ($m_e=1$).}
\label{FDvsRF}
\begin{ruledtabular}
\begin{tabular}{lccc}
Band         & $\vec{\Gamma}_{1v}$ & split-off $\vec{\Gamma}_{25v}'$ & split-off $\vec{\Gamma}_{15c}$ \\
\hline
FD order 2   & 1.161 530 {\bf65}   & -0.222 58{\bf9 10}    & 0.385 38{\bf8 13} \\
FD order 8   & 1.161 530 {\bf63}   & -0.222 58{\bf8 25}    & 0.385 38{\bf7 92} \\
DFPT         & 1.161 530 {\bf54}   & -0.222 58{\bf8 29}    & 0.385 38{\bf7 87} \\
DFPT - FD 2  & -1E-7               &  8E-7                 & -3E-7              \\
DFPT - FD 8  & -9E-8               & -5E-8                 & -5E-8              \\
\end{tabular}
\end{ruledtabular}
\end{table}

\begin{table}
\caption{
Comparison of effective masses for the two degenerate valence bands of silicon at $\vec{\Gamma}$ ($\vec{\Gamma}_{25v}'$), usually referred to as `light hole' and `heavy hole' bands. 
We provide scalar effective masses along the Cartesian directions $(100)$, $(111)$, and $(110)$ since the concept of effective mass tensor is not suitable for degenerate bands.
We use order 2 and order 8 finite differences as well as DFPT to compute the masses.
The agreement between the finite-difference methods and DFPT is provided (DFPT - FD 2 and DFPT - FD 8, respectively).
Results are provided in atomic units ($m_e=1$).}
\label{FDvsRFd}
\begin{ruledtabular}
\begin{tabular}{lccc}
Direction    & $(100)$            & $(111)$            & $(110)$            \\
\hline
{\bf light hole} &                    &                    &                    \\
FD order 2   & -0.188 25{\bf4 38} & -0.129 7{\bf44 50} & -0.136 67{\bf7 68} \\
FD order 8   & -0.188 25{\bf2 32} & -0.129 7{\bf39 62} & -0.136 67{\bf2 64} \\
DFPT         & -0.188 25{\bf2 26} & -0.129 7{\bf39 40} & -0.136 67{\bf2 54} \\
DFPT - FD 2  &  2E-6              &  5E-6              &  5E-6              \\
DFPT - FD 8  &  6E-8              &  2E-7              &  1E-7              \\
\hline                                                                      
{\bf heavy hole} &                    &                    &                    \\
FD order 2   & -0.253 93{\bf4 62} & -0.648 381 {\bf27} & -0.517 2{\bf32 33} \\
FD order 8   & -0.253 93{\bf3 45} & -0.648 381 {\bf48} & -0.517 2{\bf48 27} \\
DFPT         & -0.253 93{\bf3 47} & -0.648 381 {\bf69} & -0.517 2{\bf49 81} \\
DFPT - FD 2  &  1E-6              & -4E-7              & -2E-5              \\
DFPT - FD 8  & -2E-8              & -2E-7              & -2E-6              \\
\end{tabular}
\end{ruledtabular}
\end{table}

The results in Tables~\ref{FDvsRF} and \ref{FDvsRFd} show good agreement between DFPT and order 8 finite differences (agreement in the range $10^{-8} - 10^{-6}$). 
Furthermore, we observe that the finite-difference results converge (with increasing order) towards the DFPT results.
This observation suggests that the difference between DFPT and order 8 finite differences can be attributed to the numerical precision of the latter method. 

\begin{figure}
\includegraphics[width=1.0 \linewidth]{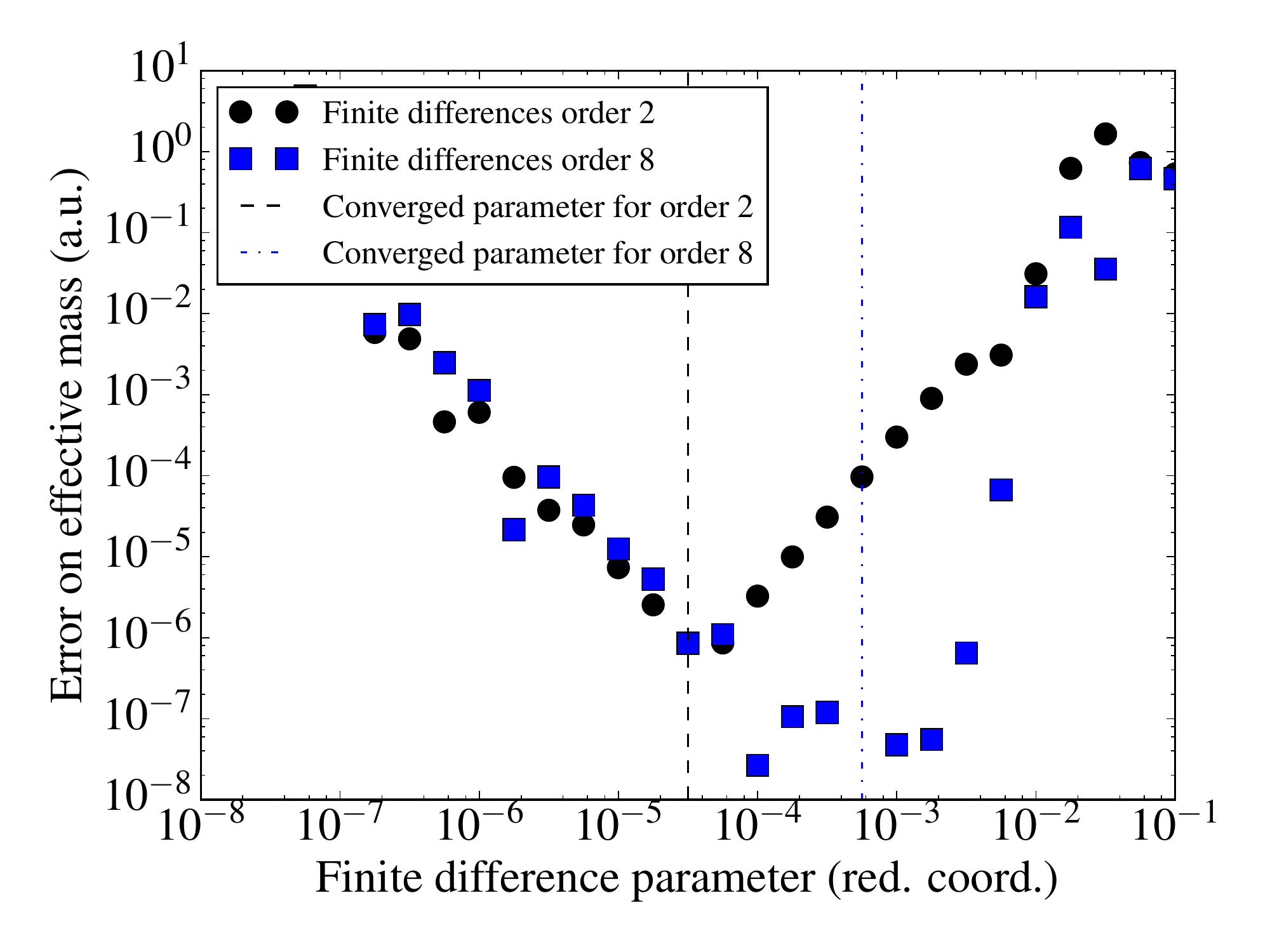}
\caption{\label{FDgraph} 
(Color online) 
Convergence study with respect to the finite-difference parameter for the effective mass of the first conduction band of silicon at $\vec{\Gamma}$.
The converged value $m_i$ is taken to be the one where $|m_{i+1}-m_i|+|m_i-m_{i-1}|$ is smallest, the index $i$ labelling the successive calculations in the convergence study.
Only the absolute difference between the calculated values and this converged value is shown for clarity.
Note that this places the converged value outside the logarithmic scale ($0=10^{-\infty}$).
}
\end{figure}

To further support this attribution, it is interesting to study the convergence behavior of the finite-difference calculations with respect to the finite-difference parameter. 
We show such a convergence study in Figure~\ref{FDgraph} for the first conduction band of silicon at $\vec{\Gamma}$.

We first note that the order 8 finite-difference results around the converged value vary by an amount similar to the agreement between finite difference and DFPT, which further supports the attribution of the discrepancies to the finite-difference method. 

Also, we note the characteristic V-shape of the convergence. 
Indeed, there is an optimal value of the finite-difference parameter that yields the best compromise between numerical noise and sampling of the band close to the extrema (where it is parabolic). 
Thus, finite-difference calculations of effective masses require this supplementary convergence study.
In contrast, DFPT is devoid of a parameter analogous to the finite-difference parameter and thus does not require such a convergence study.

\subsection{Effective mass for selected materials}
\label{results_mater}

\subsubsection{Silicon}
\label{Sec_Si}

Now that our implementation has been validated, we compare our results for silicon to those available in the literature.
Our calculations use the parameters specified in the first paragraph of Section~\ref{DFPT-FD}, along with 500 points Gauss-Legendre quadrature integration for the computation of the transport effective mass tensor (see Eq.~\eqref{C_tensor}) and the spherically averaged effective mass.
The conduction band minima was determined to be at 0.42252 (0.42253) on the $\vec{\Gamma}-\vec{X}$ path with (without) SOC and the effective mass was calculated at this point. 

For the non-degenerate bands, we simply compare the effective mass tensor with other results in the literature. 
In the degenerate case, the literature tends to focus on scalar effective mass along the standard Cartesian directions $(100) \parallel \vec{\Gamma}-\vec{X}$, $(111)\parallel \vec{\Gamma}-\vec{L}$, and $(110) \parallel \vec{\Gamma}-\vec{K}$; the comparison is therefore based on these results.

However, these comparisons do not validate the implementation of the transport equivalent formalism (see Section~\ref{TEefmas}).
Thus, we also provide results for this quantity and compare them to the only others results currently available (Ref.~\onlinecite{Mecholsky14}).

Finally, we also assess the importance of taking into account the SOC in effective-mass calculations by also providing PAW results without SOC. 
The results are provided in Table~\ref{Sidirt}.

\begin{table}
\caption{Effective masses of silicon at the valence band maximum (VBM) ($\vec{\Gamma}_{25v}'$) and the conduction band minimum (CBM) (located at $0.42252 \vec{X}$ with SOC and $0.42253 \vec{X}$ without SOC). 
When degenerate, transport equivalent and spherically averaged effective masses are given, along with effective masses along standard Cartesian directions.
When non-degenerate, the effective mass tensor is given.
At $\vec{\Gamma}$, the cubic symmetry makes the tensor proportional to the identity, so that only one scalar needs to be reported.
On the $\vec{\Gamma}-\vec{X}$ path, Cartesian coordinates diagonalize the tensor and the 2 directions perpendicular to the path are identical, so that only two scalars need to be reported (the diagonal components of the tensor $\parallel$ and $\perp$ to $\vec{\Gamma}-\vec{X}$).
Results are provided in atomic units ($m_e=1$).}
\label{Sidirt}
\begin{ruledtabular}
\begin{tabular}{lcccccc}
Band                      & \multicolumn{2}{c}{This work} & \multicolumn{2}{c}{Theory} & \multicolumn{2}{c}{Experiment} \\
                          & no SOC    & SOC & [\onlinecite{Mecholsky14}] & [\onlinecite{Ramos:2001cy}] & [\onlinecite{Dexter:1954}] & [\onlinecite{Hensel:1965ht}] \\
\hline
{\bf split-off}           &           &           &        &       &           &           \\
Eff. mass                 & -         & 0.2226    & -      & 0.22  & -         & -         \\
Trans. eqv.               & 0.1336    & 0.2226    & -      & -     & -         & -         \\
Sph. avg.                 & 0.1101    & 0.2226    & -      & -     & -         & -         \\
$(1 0 0)$                 & 0.1671    & 0.2226    & -      & -     & -         & -         \\
$(1 1 1)$                 & 0.0938    & 0.2226    & -      & -     & -         & -         \\
$(1 1 0)$                 & 0.1054    & 0.2226    & -      & -     & -         & -         \\
\hline                                
{\bf light hole}          &           &           &        &       &           &           \\
Trans. eqv.               & 0.5715    & 0.1559    & 0.1731 & -     & -         & -         \\
Sph. avg.                 & 0.3026    & 0.1431    & 0.1531 & -     & -         & -         \\
$(1 0 0)$                 & 0.2578    & 0.1882    & -      & 0.18  & 0.171     & -         \\
$(1 1 1)$                 & 0.6495    & 0.1297    & -      & 0.13  & 0.160     & -         \\
$(1 1 0)$                 & 0.2578    & 0.1367    & -      & 0.14  & 0.163     & -         \\
\hline                                
{\bf heavy hole}          &           &           &        &       &           &           \\
Trans. eqv.               & 10.60     & 0.7294    & 1.1567 & -     & -         & -         \\
Sph. avg.                 & 0.7405    & 0.4416    & 0.5322 & -     & -         & -         \\
$(1 0 0)$                 & 0.2578    & 0.2540    & -      & 0.26  & 0.46      & -         \\
$(1 1 1)$                 & 0.6495    & 0.6484    & -      & 0.67  & 0.56      & -         \\
$(1 1 0)$                 & 2.703     & 0.5173    & -      & 0.54  & 0.53      & -         \\
\hline                                
{\bf CBM}                             &           &           &        &       &           &           \\
$\parallel$ to $\vec{\Gamma}-\vec{X}$ & 0.9455    & 0.9455    & -      & 0.96  & -         &  0.9163   \\
$\perp$ to $\vec{\Gamma}-\vec{X}$     & 0.1875    & 0.1876    & -      & 0.16  & -         &  0.1905   \\
\end{tabular}
\end{ruledtabular}
\end{table}

Let us first assess the impact of SOC. 
As expected, the conduction band is almost not influenced by the inclusion of SOC.
However, this is not the case for the valence bands. 
Close examination of the split-off band immediately suggests that this qualitative change of behavior is related to the degeneracy of the bands.
Indeed, the split-off band is not degenerate when SOC is included, which allows its description by an effective mass tensor.
The cubic symmetry of the $\vec{\Gamma}$ point then makes this tensor proportional to the identity, which makes the split-off effective mass spherically symmetric. 

In contrast, neglecting SOC makes the split-off band degenerate with the two other valence bands, which prevents its description by an effective mass tensor and thus requires the formalism of Section~\ref{TEefmas}.
Moreover, increasing the degenerate subspace dimension from 2 to 3 adds two off-diagonal coupling terms in $f_{nn'\vec{k}}$ (see Eq.~\eqref{f_def}), which breaks the spherical symmetry of the mass and cause the qualitative changes observed in Table~\ref{Sidirt}. 

This attribution can be more formally proven by (unphysically) treating the split-off band in a SOC calculation as degenerate with the two other valence bands within the formalism of Section~\ref{TEefmas}.
Since our implementation decides whether two bands are degenerate or not using a numerical threshold on their eigenenergy difference, we carried out the preceding test by increasing this threshold to a value above the spin-orbit splitting. 
We thus obtained transport equivalent effective masses of 0.1335, 0.5708, and 10.58, in very good agreement with the values obtained without SOC (respectively 0.1336, 0.5715, and 10.60; see Table~\ref{Sidirt}), which demonstrates the validity of our interpretation.

Yet, this analysis does not settle whether both results are correct and really reflect the curvature of the valence bands extrema. 
For the results with SOC, we have already demonstrated in Table~\ref{FDvsRFd} that the DFPT results were indeed correct (i.e. that they agreed very well with finite-difference calculations). 
Carrying out the same test for the valence bands without SOC (see Table~\ref{FDvsRFdnoSOC}) reveals the same agreement, which proves that the qualitative change of behavior of the effective masses predicted by our implementation is correct.

\begin{table}
\caption{
Comparison of effective masses for the three degenerate valence bands of silicon (without SOC) at $\vec{\Gamma}$. 
We provide scalar effective masses along the Cartesian directions $(100)$, $(111)$, and $(110)$ since the concept of effective mass tensor is not suitable for degenerate bands.
We use order 8 finite differences and DFPT to compute the masses.
The agreement between the finite-difference methods and DFPT is provided.
Results are provided in atomic units ($m_e=1$).}
\label{FDvsRFdnoSOC}
\begin{ruledtabular}
\begin{tabular}{lccc}
Direction    & $(100)$            & $(111)$            & $(110)$            \\
\hline
{\bf Band 2} &                    &                    &                    \\
FD order 8   & -0.167 180 4{\bf4} & -0.093 807 8{\bf2} & -0.105 36{\bf9 00} \\
DFPT         & -0.167 180 4{\bf0} & -0.093 807 8{\bf0} & -0.105 36{\bf8 95} \\
DFPT - FD 8  & -4E-8              & -3E-8              & -5E-8              \\
\hline                                                                      
{\bf Band 3} &                    &                    &                    \\
FD order 8   & -0.257 803 {\bf79} & -0.649 497 {\bf22} & -0.257 803 8{\bf0} \\
DFPT         & -0.257 803 {\bf82} & -0.649 497 {\bf35} & -0.257 803 8{\bf2} \\
DFPT - FD 8  &  3E-8              &  1E-7              &  2E-8              \\
\hline                                                                      
{\bf Band 4} &                    &                    &                    \\
FD order 8   & -0.257 803 82      & -0.649 497 {\bf16} & -2.702 563 {\bf30} \\
DFPT         & -0.257 803 82      & -0.649 497 {\bf35} & -2.702 563 {\bf84} \\
DFPT - FD 8  &  4E-9              &  2E-7              &  5E-7              \\
\end{tabular}
\end{ruledtabular}
\end{table}

\begin{figure}
\includegraphics[width=1.0 \linewidth]{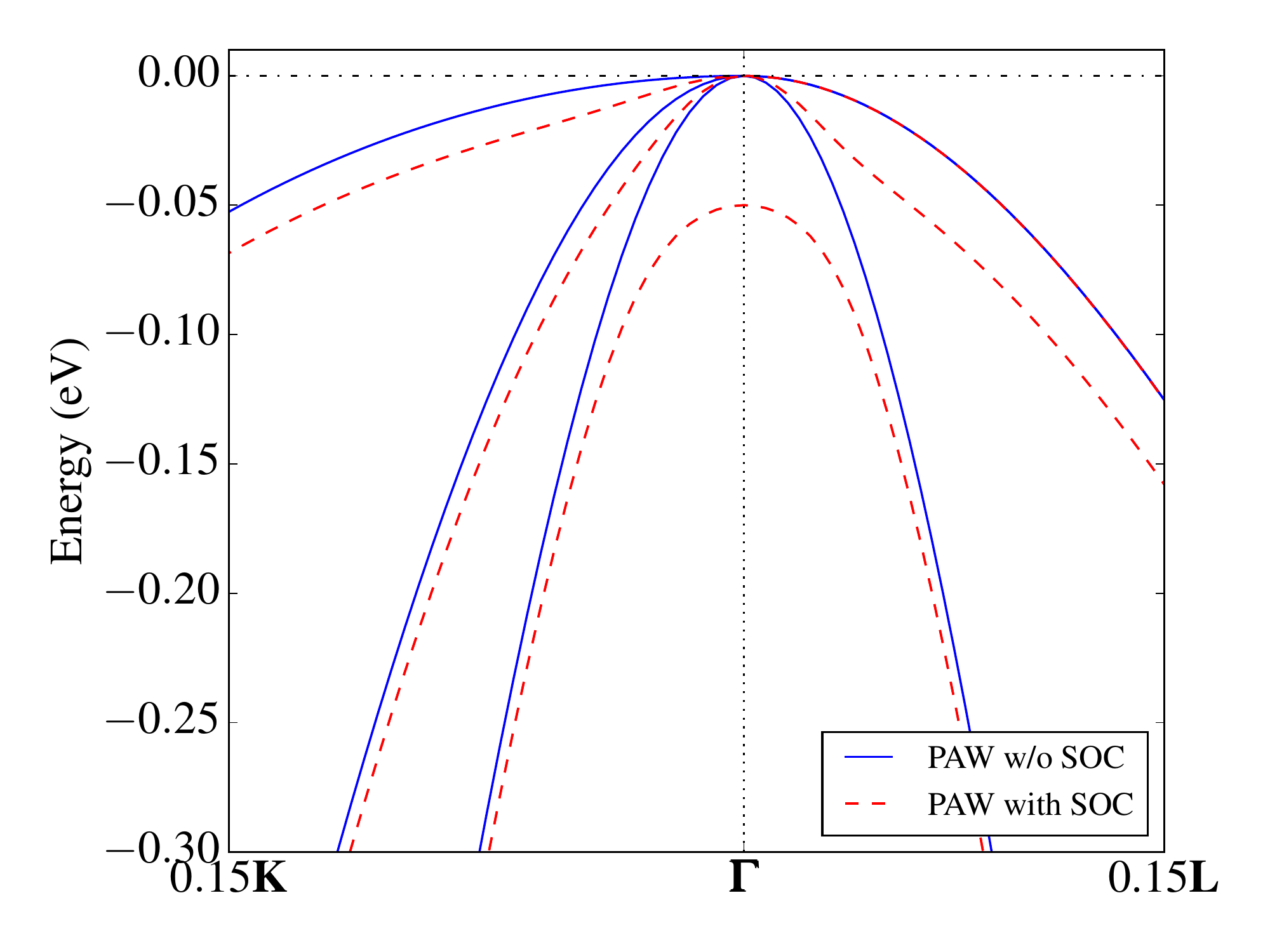}
\caption{\label{band_Si} 
(Color online)
PAW calculation of the valence bands of silicon around $\vec{\Gamma}$ in the directions $(110) \parallel \vec{K}$ and $(111) \parallel \vec{L}$, with and without SOC.
For the light hole band (middle band) in direction $\vec{L}$ with SOC, we see that the effective mass has a change of regime around the spin-orbit energy splitting.
After this change of regime, the effective mass of the band with SOC tends to the mass of the band without SOC.
}
\end{figure}

To explain this fact, we plot the valence bands with and without SOC around the $\vec{\Gamma}$ point in the $(111)$ and the $(110)$ directions in Figure~\ref{band_Si}. 
Examining e.g. the light hole band (middle one) with SOC in the $(111)$ direction reveals a change of regime in the effective mass around the SOC energy splitting. 
Indeed, it can be seen that the curvature is relatively high at $\vec{\Gamma}$, but then changes rather rapidly and settles to a lower value around $0.10\vec{L}-0.15\vec{L}$. 
This suggests that the effective mass evolve from the SOC value at low energy to the non-SOC value at higher energies (with respect to the SOC splitting).
This behavior is not surprising, since one would expect that treating the split-off band as degenerate with the two other valence bands would become a good approximation at energies where the SOC splitting becomes negligible. 
However, in the case of silicon, this occurs relatively far from the $\vec{\Gamma}$ point, so that the validity of the quadratic expansion of the eigenenergies becomes questionable. 
Thus, we postpone further analysis of this question to the next section, where a material with much smaller SOC splitting of the valence bands is studied (graphane) (see in particular Figure~\ref{Grsm}). 

Since we have established the critical role of SOC in the proper description of the effective masses at degenerate band extrema, we now compare our results including SOC with results from the literature (Table~\ref{Sidirt}), which all include SOC. 
As could be expected, the agreement with Ref.~\onlinecite{Ramos:2001cy} (DFT results) is good and globally better than with Ref.~\onlinecite{Dexter:1954,Hensel:1965ht} (experimental results). 
However, the agreement of the `transport equivalent effective mass' with Ref.~\onlinecite{Mecholsky14} is surprisingly poor.

To assess this issue, we first replaced $f_{n\vec{k}}(\theta,\phi)$ (see Eq.~\eqref{f_def}) in our implementation by the fit used in Ref.~\onlinecite{Mecholsky14} (see their Eq.~(16), or Eq.~(63) of Ref.~\onlinecite{Dresselhaus55} for the original version)
\begin{multline}
\label{kittel_fit}
f_{n\vec{k}}(\theta,\phi) = A_{n\vec{k}} \pm \\
\sqrt{B_{n\vec{k}}^2 + C_{n\vec{k}}^2 \sin^2(\theta) \Big(\cos^2(\theta) + \sin^2(\theta) \sin^2(\phi) \cos^2(\phi)\Big)},
\end{multline}
where the parameters $A_{n\vec{k}}$, $B_{n\vec{k}}$, and $C_{n\vec{k}}$ are independent of $n$ for a given pair of degenerate bands (it is only the $\pm$ sign that distinguishes the two bands). 
Using the values obtained by Ref.~\onlinecite{Mecholsky14} for the parameters of the top valence band of silicon at $\vec{\Gamma}$ ($A=−4.20449$, $B=0.378191$, and $C=5.309$) yields, to all significant digits provided, a transport equivalent effective mass identical to theirs. 
We also checked that analytic and finite-difference differentiation of $f_{n\vec{k}}(\theta,\phi)$ (in Eq.~\eqref{speed-nu-2}) yielded the same results, both for our implementation's $f_{n\vec{k}}(\theta,\phi)$ (see Eq.~\eqref{f_def}) and Eq.~\eqref{kittel_fit}'s. 
Therefore, the discrepancy does not stem from the implementation of the transport effective mass formalism (from Eq.~\eqref{f_def} to the end of Section~\ref{TEefmas}).

Thus, two possibilities remain: either Eq.~\eqref{kittel_fit} does not fit well $f_{n\vec{k}}(\theta,\phi)$ or the underlying results for $f_{n\vec{k}}(\theta,\phi)$ differ.
To discriminate between the two, we fitted Eq.~\eqref{kittel_fit} to our own results and calculated the transport equivalent effective mass tensor from this fit. 
The parameters we obtained ($A=-4.62503$, $B=0.686991$, and $C=5.20517$) yield a tensor identical ($10^{-5}$ difference) to the one we obtained directly with our implementation.
Thus, the discrepancy between the results of Ref.~\onlinecite{Mecholsky14} and the present study can be traced back to the numerical results obtained for $f_{n\vec{k}}(\theta,\phi)$.

This is in line with the explanation given in Ref.~\onlinecite{Mecholsky14} for the poor agreement between their fit and their DFT values for $f_{n\vec{k}}(\theta,\phi)$, i.e. the nonparabolicity of the bands introduced errors in their finite-difference calculations of $f_{n\vec{k}}(\theta,\phi)$.
Indeed, it is known that coupling between bands (which cause the band warping of degenerate states) can also cause strong nonparabolicity of the bands~\cite{Kane56,Willatzen:2009iv}. 
Provided that such strong nonparabolicity occur for a substantial portion of the directions $(\theta,\phi)$, finite-difference methods would become unsuitable for computing $f_{n\vec{k}}(\theta,\phi)$ in a calculation of $\bar{\matr{M}}_{n\vec{k}}$. 
This would explain the poor agreement between our results and Ref.~\onlinecite{Mecholsky14}'s as well as why Eq.~\eqref{kittel_fit} fit perfectly our results while it is not the case for Ref.~\onlinecite{Mecholsky14}'s.
It also illustrates the convenience and reliability of direct (DFPT) calculations of $f_{n\vec{k}}(\theta,\phi)$ with respect to finite-difference computations.

\subsubsection{Graphane}

Graphane has emerged in recent years as a new two-dimensional material with promising properties~\cite{Vishnyakova:2016bn,Inagaki:2014ig,Zhou:2014dg}. 
However, the effective masses in this material have received little attention. 
Indeed, to the author's knowledge, only the effective mass of the conduction band has been roughly assessed~\cite{Restrepo:2014jj}.
We thus decided to investigate this topic from our first-principles framework.

The grafting of one hydrogen per carbon atom on graphene to produce graphane can be done following many different patterns~\cite{Zhou:2014dg}.
However, we focus here on the most stable one: the so-called `chair' configuration, where two hydrogen atoms attached to neighbouring carbon atoms are located on opposite sides of the graphene sheet. 
This form of graphane has the same primitive cell as graphene, barring the addition of the two hydrogen atoms and a slight distortion of the carbon-carbon bonds.
Thus, it also features the same Brillouin zone. 
However, in the case of graphane, the valence band maximum and the conduction band minimum occur at $\vec{\Gamma}$. 
Moreover, the valence band is doubly degenerate.

In our simulations, we used both PAW with SOC (as presented in Section~\ref{SO}) and NCPP without SOC (to validate it, since it is not used elsewhere in this work).
To ensure convergence (effective masses precise to 3 significant digits), the NCPP simulations were carried out using a cutoff energy of 40~Ha for the plane wave basis while, for the PAW case, a cutoff energy of 20~Ha for the plane wave basis and 40~Ha for the PAW double grid were used.
Also, in both cases, a 8 8 1 $\vec{k}$-point grid, an interlayer spacing of 22.5 \AA, and the local-density approximation of Perdew and Wang~\cite{Perdew:1992zza} were used.
Moreover, in each case, the structure was fully relaxed.

The hexagonal structure of graphane is described by 2 primitive vectors of equal lengths with a 120$^\circ$ angle between them, with the atoms placed at positions $(0,0)$ (one carbon and one hydrogen) and $(\tfrac{1}{3},\tfrac{2}{3})$ (the other carbon and hydrogen).
Moreover, by symmetry, all the C-C bonds and C-H bonds have the same length. 
This leaves only three quantities to be reported to define the structure of graphane: the primitive vector length $a$, the C-C bond length and the C-H bond length. 
Our NCPP structural relaxation yielded $a$=2.495 \AA, C-C=1.509 \AA, and C-H=1.110 \AA~while PAW yielded $a$=2.504 \AA, C-C=1.515 \AA, and C-H=1.117 \AA.
The latter result compares very well with the other PAW LDA result reported in the literature~\cite{Zhou:2014dg,He:2012br} ($a$=2.504 \AA, C-C=1.537 \AA, and C-H=1.110 \AA).

Since graphane is a 2D material, the 3D formalism for the `transport equivalent' effective mass presented in Section~\ref{TEefmas} and Appendix~\ref{Meqv} becomes numerically unstable. 
We discuss this issue and adapt the formalism of Appendix~\ref{Meqv} to the 2D case in Appendix~\ref{Meqv-2D}. 
Thus, it is with the latter formalism that we handled the degenerate valence band extremum in graphane. 

Since the $\vec{\Gamma}$ point exhibits hexagonal symmetry, the tensors at this point are proportional to the identity, which leaves only one quantity per band to be reported.
We summarize these results in Table~\ref{Grt}.
Also, for the degenerate 2D case, a scaling factor of the transport tensor $c_{nk}$ needs to be taken into account (see Eq.~\eqref{scaling_factor}).
However, since it is found to be $1.000$ for the valence band of graphane, it is omitted from Table~\ref{Grt}. 

\begin{table}
\caption{Effective masses for the conduction band minimum (CBM) and the two-fold degenerate valence band maximum (VBM) of `chair' graphane.
The `light hole' (lh) and `heavy hole' (hh) bands designate respectively the lower and upper bands in the inset of Figure~\ref{GrBands}.
Results were obtained using NCPP calculations without SOC (labelled NCPP) as well as PAW calculations with SOC (labelled PAW).
To explain the large difference between these calculations, we treated the valence bands as degenerate in a PAW calculation with SOC (labelled PAW+deg).
We also carried out finite-difference calculations (order 8) and obtained their difference with respect to DFPT calculations (labelled $\Delta$@NCPP and $\Delta$@PAW) using the methodology of Section~\ref{DFPT-FD}.
Also, since the $\vec{\Gamma}$ point exhibits hexagonal symmetry, the tensors at this point are proportional to the identity, so that effective masses can be reported using a single number.
Finally, degenerate bands are handled using the formalism of Appendix~\ref{Meqv-2D} since the formalism of Appendix~\ref{Meqv} becomes unstable for 2D materials. 
Results are provided in atomic units ($m_e=1$).}
\label{Grt}
\begin{ruledtabular}
\begin{tabular}{lccccc}
Band           & NCPP   &$\Delta$@NCPP& PAW    & $\Delta$@PAW & PAW+deg \\ 
\hline                                                                    
{VBM, lh}  & -0.271 & -4E-7  & -0.373 &  8E-8  & -0.267  \\ 
{VBM, hh}  & -0.616 &  1E-6  & -0.372 &  8E-8  & -0.613  \\ 
{CBM}      &  1.012 &  2E-7  &  1.012 &  5E-8  &  1.012  \\ 
\end{tabular}
\end{ruledtabular}
\end{table}

We observe good agreement between all results for the conduction effective mass.
Moreover, our results agree well with Ref.~\onlinecite{Restrepo:2014jj}, which reports an effective mass of $1$ for the conduction band of graphane using first-principles calculations. 
However, the comparison is more complex in the case of the valence bands.

\begin{figure}
\includegraphics[width=1.0 \linewidth]{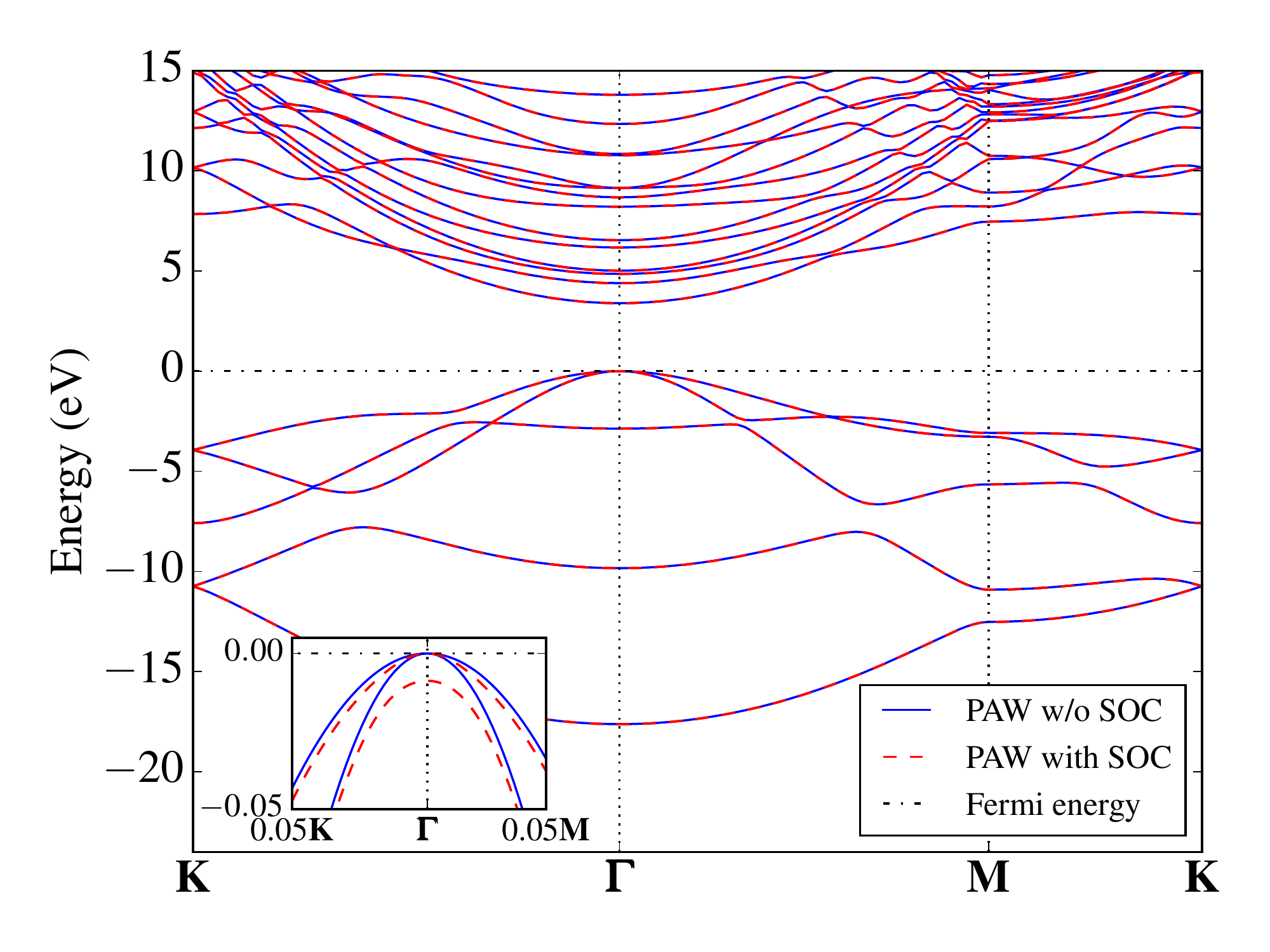}
\caption{\label{GrBands} 
(Color online)
Band structure of graphane within the PAW formalism, with and without SOC.
The inset shows the 8.71~meV spin-orbit splitting of the valence bands.
}
\end{figure}

The first issue is that valence bands are degenerate in the NCPP calculations, while the PAW calculations with SOC lift the degeneracy by 8.71~meV (see Figure~\ref{GrBands}).
Thus, the nature of the effective masses is not the same for NCPP and PAW with SOC: the former are `transport equivalent' effective masses while the latter are plain effective masses.
Still, examination of the angular dependent effective masses $m_{n\vec{k}}(\theta,\phi)$ (see Eq.~\eqref{m_th_ph}) for both degenerate valence bands in the NCPP calculations reveals that they exhibit spherical symmetry (i.e. there is no warping).
In this case, `transport equivalent' effective masses coincide with plain effective masses, so that direct comparison between NCPP and PAW with SOC results is possible.

Yet, as in the case of silicon, the results with and without SOC strongly differ.
In Table~\ref{Grt}, we prove that this difference can be attributed to the extra coupling term present in $f_{nn'\vec{k}}$ (see Eq.~\eqref{f_def}) when the bands are considered degenerate.
We do so by increasing the numerical threshold for degeneracy to a value above the SOC splitting in the PAW (with SOC) calculation (see the PAW+deg column).
As expected, the results obtained are in very good agreement with the NCPP ones.

We then prove that both results (NCPP and PAW with SOC) really reflects the curvature of the bands at $\vec{\Gamma}$ by comparing them with finite-difference calculations, using the methodology of Section~\ref{DFPT-FD}. 
For clarity, only the difference between DFPT and finite-difference results is reported in Table~\ref{Grt}. 
The good agreement confirms both the accuracy of NCPP and PAW with SOC effective masses (and validate our NCPP DFPT implementation).

\begin{figure}
\includegraphics[width=1.0 \linewidth]{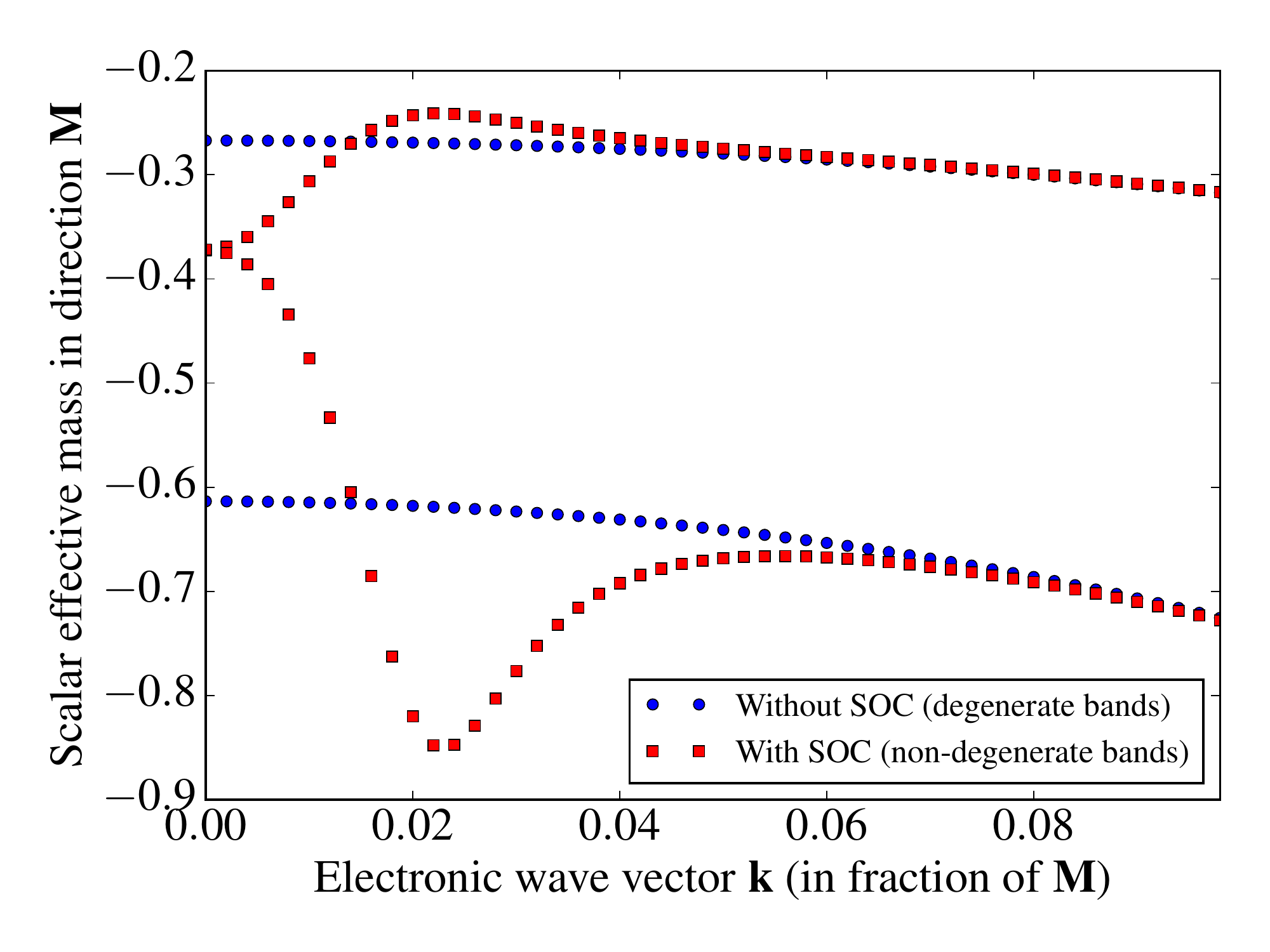}
\caption{\label{Grsm} 
(Color online)
PAW calculations of scalar effective masses of graphane in the $\vec{\Gamma}-\vec{M}$ direction at different point on the $\vec{\Gamma}-\vec{M}$ line, with (red squares) and without (blue dots) SOC. 
}
\end{figure}

In the case of silicon, observation of the band structure suggested that the effective masses with SOC evolve towards the values without SOC when moving away from the band extrema. 
However, such an observation is more difficult to make in the case of graphane.
Indeed, direct examination of the band structure (Figure~\ref{GrBands}) does not allow to see much besides a rigid shift of a valence band when SOC in included. 
Thus, to more precisely investigate the issue, we calculated the scalar effective masses in the $\vec{\Gamma}-\vec{M}$ direction at different points along the $\vec{\Gamma}-\vec{M}$ line, with and without SOC.
The results are presented in Figure~\ref{Grsm}.

The results confirm what was suggested in the case of silicon: the masses with SOC evolve towards the masses without SOC, with a transition that occurs around $\vec{k}=0.02\vec{M}$, which coincides with an energy scale that is comparable to the SOC energy splitting.
Indeed, for PAW calculations with SOC, $\varepsilon_{\mathrm{lh}\vec{k}}-\varepsilon_{\mathrm{lh}\vec{\Gamma}}=9.5$~meV and $\varepsilon_{\mathrm{hh}\vec{k}}-\varepsilon_{\mathrm{hh}\vec{\Gamma}}=7.6$~meV, where lh and hh stand for light hole and heavy hole band, respectively (i.e. the lower and upper bands in the inset of Figure~\ref{GrBands}).
This observation suggests that it would be pertinent to take into account a relevant magnitude of $\vec{k}$ or $\varepsilon_{n\vec{k}}$ around the band extrema (e.g. the doping level, gate voltage, ...) when computing effective masses. 
A possible solution would be to merge the formalism of Kane models~\cite{Kane56} into the formalism of Section~\ref{d}.

\subsubsection{Arsenic}

$\alpha$-arsenic (or grey arsenic) is a semimetal with the A7 crystal structure~\cite{Wyckoff:1960vk}. 
The latter structure has a trigonal primitive cell with 2 atoms per cell. 
The Brillouin zone of $\alpha$-As is described in Section~II and Appendix~A of Ref.~\onlinecite{Falicov:1965go}.
It features a non-degenerate ellipsoidal electron pocket at $\vec{L}$ and a non-degenerate hole pocket of complex shape around $\vec{T}$~\cite{Lin:1966hy,Silas:2013bz}. 

The fact that an effective mass at a non-degenerate band extrema is a tensor (see, e.g., Eq.~\eqref{efmas_nd}) implies that the angle-dependent effective mass $m_{n\vec{k}}(\theta,\phi)$ should be an ellipsoid. 
Consequently, the effective mass formalism is inappropriate for the description of the hole pockets in $\alpha$-As, since they are far from an ellipsoidal shape. 
We therefore focus on the electron pockets, which originate from the $\vec{L}_1$ band. 

In the literature on $\alpha$-As effective masses~\cite{Falicov:1965go,Lin:1966hy,Priestley:1967aa,Ih:1970aa,Cooper:1971aa}, Cartesian directions in the reciprocal space are conventionally chosen to be the `trigonal' axis ((1 1 1) in reduced coordinates) for the $z$ direction and the `binary' axis ((0 1 -1) in reduced coordinates) for the $x$ axis, which leaves $y$ (the `bisectrix' axis) to be $\hat{\vec{z}} \times \hat{\vec{x}}$.

One of the principal axes of the effective mass ellipsoid $m_1$ lies along the `binary' axis $x$, which leaves the two other $m_2$ and $m_3$ to be in the $yz$ plane~\cite{Falicov:1965go,Lin:1966hy,Priestley:1967aa,Ih:1970aa,Cooper:1971aa}. 
The orientation of the largest of these two principal effective masses $m_3$ is conventionally described in term of the `tilt' angle it forms with the $z$ axis. 
Positive angles denote a rotation from the $z$ axis towards the $y$ axis. 
We use the same convention here.

In our calculations, the PAW formalism was used with and without SOC (see Section~\ref{SO}).
A cutoff energy of 30 Ha for the plane wave basis and 60 Ha for the PAW double grid along with a 30 30 30 $\vec{k}$-point grid were used to ensure fully converged calculations (effective masses precise to 3 significant digits). 
Also, the local-density approximation of Perdew and Wang~\cite{Perdew:1992zza} was used and the structure was fully relaxed.
The trigonal primitive vector were found to be 3.9783 \AA~long with 56.441$^\circ$ angles between each of them and the atomic positions were found to be $0.2305\cdot(1,1,1)$ and $0.76947\cdot(1,1,1)$ in reduced coordinates. 

We compare our results for $m_1$, $m_2$, $m_3$, and the tilt angle with results from the literature in Table~\ref{As-exp}.
To further assess the nonparabolicity of the band, we also compute three scalar effective masses (with SOC) from the band extrema and the points of the Fermi surface lying on the effective mass principal axes. 

\begin{table}
\caption{DFPT effective masses for the electron pocket of $\alpha$-As at $\vec{L}=(0.5,0,0)$ from the present implementation and comparison with the literature.
PAW results with and without SOC are provided (labeled SOC and no SOC, respectively).
To assess the nonparabolicity of the band extrema generating the electron pockets, we also compute effective masses (with SOC) along the effective mass principal axes by fitting a parabola to the band extremum and the point of the Fermi surface lying on the axis (the resulting masses are labeled FD@$\varepsilon_F$).
The convention for the tilt angle is the same as in Refs.~\onlinecite{Lin:1966hy,Priestley:1967aa,Ih:1970aa,Cooper:1971aa}.
Results are provided in atomic units ($m_e=1$).
}
\label{As-exp}
\begin{ruledtabular}
\begin{tabular}{lccccccc}
       & \multicolumn{3}{c}{This work}                     & \multicolumn{1}{c}{Theory}   & \multicolumn{3}{c}{Experiment} \\
       &       no SOC  &      SOC     & FD@$\varepsilon_F$ & [\onlinecite{Lin:1966hy}] & [\onlinecite{Cooper:1971aa}] & [\onlinecite{Ih:1970aa}] & [\onlinecite{Priestley:1967aa}] \\ 
\hline                                             
$m_1$  & -0.0448       & -0.0709      & 0.0867     & 0.11        & 0.134        & 0.121        & 0.163        \\ 
$m_2$  &  0.00462      & 0.00421      & 0.0311     & 0.038       & 0.140        & 0.138        & 0.105        \\ 
$m_3$  &  1.30         & 1.10         & 1.40       & 0.94        & 1.350        & 1.18         & 2.11         \\ 
Tilt   &  84.4$^\circ$ & 84.9$^\circ$ & -          & ~80$^\circ$ & 84.5$^\circ$ & 83.6$^\circ$ & 86.4$^\circ$ \\ 
\end{tabular}
\end{ruledtabular}
\end{table}

\begin{figure}
\includegraphics[width=1.0 \linewidth]{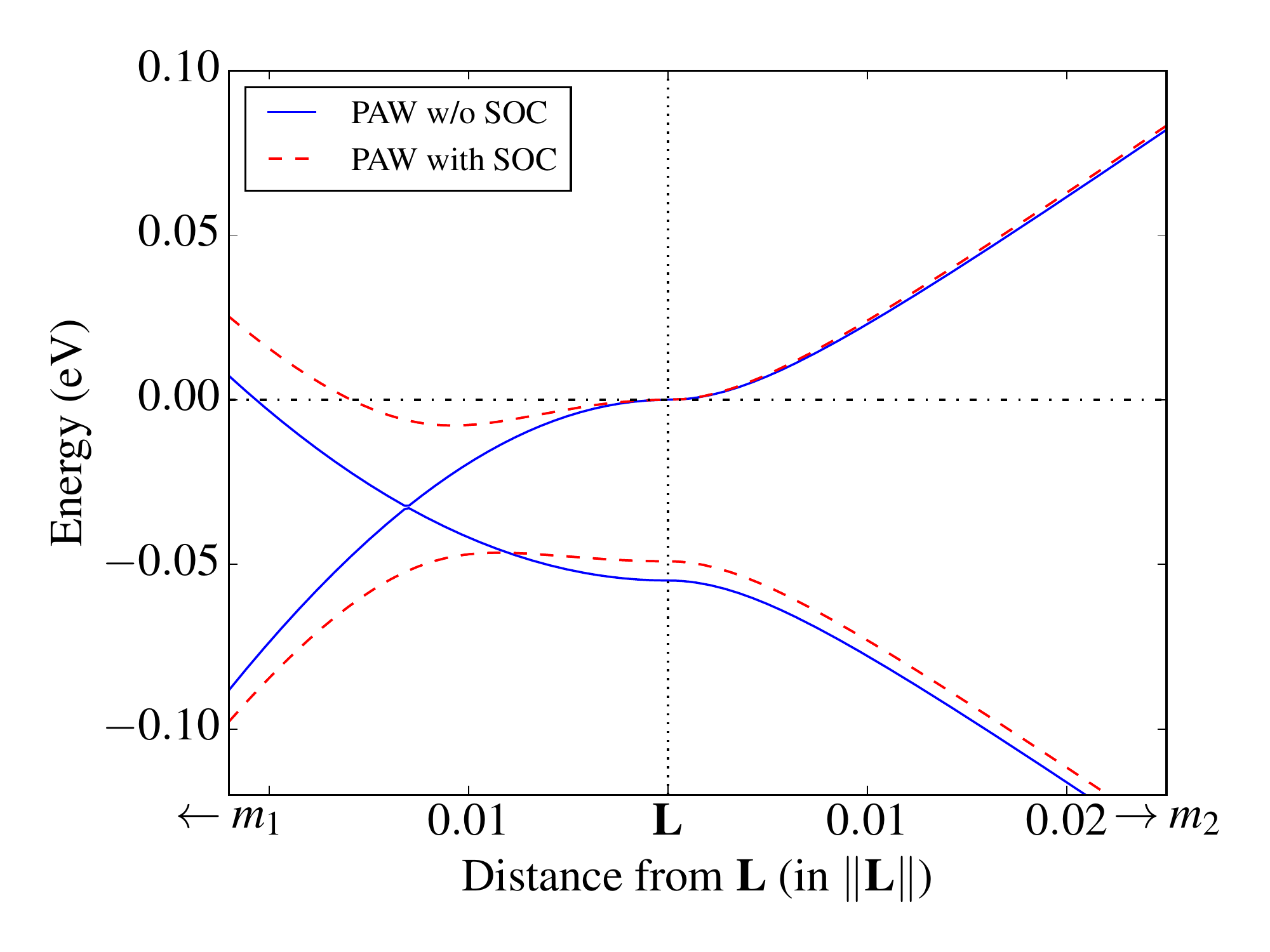}
\caption{\label{band_As} 
(Color online)
Band structure of $\alpha$-As with and without SOC around $\vec{L}$ in the directions of the two DFPT effective mass principal axes that exhibit strong nonparabolicity ($m_1$ and $m_2$). 
We observe a strong nonparabolicity in the $m_2$ direction.
Also, the $m_1$ direction looks reasonably parabolic without SOC, but changes sign due to a band crossing at finite wave vector.
When SOC is included, the $m_1$ direction exhibits strong nonparabolicity due to an avoided crossing.
}
\end{figure}

We immediately note the disagreement between the DFPT and the other results for $m_1$ and $m_2$. 
In particular, the disagreement between our DFPT results and our scalar effective masses (FD@$\varepsilon_F$) directs suspicion towards band nonparabolicity. 
To assess this, we plot the band structure with and without SOC around $\vec{L}$ along these directions in Figure~\ref{band_As}. 
Indeed, both band structure are clearly non-parabolic along these two directions and, in particular, have a curvature along $m_1$ that change sign at a finite wave vector $\vec{k}$.
This is in line with the strong nonparabolicity also observed by Ref.~\onlinecite{Falicov:1965go}. 

We also note that, for our scalar effective masses (FD@$\varepsilon_F$), the agreement with other theoretical results (Ref.~\onlinecite{Lin:1966hy}) is reasonably good. 
Indeed, the agreement is as good as could be expected, given that Ref.~\onlinecite{Lin:1966hy} uses an empirical pseudopotential approach, that our DFPT effective mass principal axes are likely to be slightly different from the Fermi surface principal axes and that the Fermi surface deviates slightly from an ellipsoid~\cite{Gonze:1990je,Cooper:1971aa}.

On the other hand, we observe that DFPT agrees with the scalar effective mass results (FD@$\varepsilon_F$) for $m_3$, which suggests that the effective mass is parabolic in this direction. 
In this case (as well as for the tilt angle, which is incidentally defined with respect to $m_3$), we also observe good overall agreement with experimental~\cite{Cooper:1971aa,Ih:1970aa,Priestley:1967aa} and theoretical~\cite{Lin:1966hy} results.

We conclude from this that effective masses (as defined in Eq.~\eqref{efmas_nd}, i.e. in the $\vec{k}\cdot\hat{\vec{p}}$ and DFPT sense) are a dangerous concept to use in metals, even if they have as little charge carriers as $\alpha$-As.
Indeed, in the present case, the range where a quadratic expansion reliably describes the band dispersion is much smaller than the electron pockets.
However, it is possible that some coupling between the bands in Figure~\ref{band_As} would explain the strong nonparabolicities observed. 
Thus, including the formalism of Kane models~\cite{Kane56} into the formalism of Section~\ref{d} may enhance the description of $\alpha$-As (and semimetals in general).

\section{Conclusion}

Up to now, effective masses were usually calculated using finite-difference estimation of density functional theory (DFT) electronic band curvatures~\cite{HautierTCO,Boltztrap,Filip:2015iz,Kim:2010cc,Kim:2009dj,Hummer:2007hf,Wang:2011fb}.
The only option available to circumvent their use relies on Wannier functions~\cite{Yates:2007er}.
However, finite differences require additional convergence studies and can lead to precision issues while Wannier functions require careful selection of the starting functions.
In contrast, the present DFPT-based method allows to obtain DFT effective masses with high precision without any such additional work. 
It is therefore more suitable for e.g. high-throughput material design.

Moreover, it is known that the concept of effective mass breaks down at degenerate band extrema due to the non-analyticity of the band structure at such a point. 
While this issue is usually addressed with a fitting procedure that aims at accurately describing the band structure at this point~\cite{Dresselhaus55}, it would be more convenient to directly determine a metric of the performance of the material in a design context. 
Since the most appropriate metric is usually some transport tensor, the concept of `transport equivalent mass tensor'~\cite{Mecholsky14} becomes quite suitable for our objective and has been integrated in our DFPT-based method.
This concept allows one to compute an effective mass-like tensor which gives the right contribution to transport tensors when used, even if the band extremum cannot be described by a tensor.
This makes our method more general and even simpler to use.

The developed techniques were validated by comparison with finite-difference calculations and excellent agreement was observed. 
Then, applying our method to the study of silicon, graphane, and arsenic, we found results coherent with previous studies, thus further validating our method.
Actually, the agreement with Ref.~\onlinecite{Mecholsky14} was, at first sight, not very good, as seen in Section~\ref{Sec_Si}, but a careful analysis pointed to the superiority of the DFPT versus finite-difference approach to extract a `transport equivalent effective mass'.

Still, our simulations (especially in the case of graphane) suggest that neither the non-degenerate formalism of Section~\ref{nd} nor the degenerate formalism of Section~\ref{d} is suitable when the energy scale relevant to the problem (e.g. doping level, gate voltage, ...) is comparable to the energy separation between the band of interest and its closest neighbour. 
Thus, it would be interesting to merge the formalism developed by Kane~\cite{Kane56} into the formalism of Section~\ref{d} in future developments.

Also, a substantial difference between DFT and experimental results remains. 
Provided that recent studies have exposed the substantial impact of electron-electron interactions on the calculated effective masses~\cite{Kim:2010cc,Filip:2015iz}, it becomes interesting to investigate approximate schemes to include these interactions in the calculations. 
Furthermore, it would be interesting to investigate the impact of electron-phonon interaction not only on the band gap~\cite{Ponce:2015jc,Antonius:2015fj,Ponce14,Antonius:2014im,Ponce:2014cm,Gonze:2011ej,Allen:1976dh,Allen:1981cq}, but also on the effective masses. 

\section{Acknowledgments}
This work was supported by the FRQNT through a postdoctoral research fellowship (J.L.J.) as well as the FRS-FNRS through Scientific Stay grant No. 2014/V~6/5/010-IB (J.L.J.), a FRIA fellowship (S.P.), and a FNRS fellowship (Y.G.).
Also, we would like to thank Yann Pouillon and Jean-Michel Beuken for their valuable technical support and help with the test and build system of ABINIT.
Computational resources have been provided by the supercomputing facilities of the Universit\'e catholique de Louvain (CISM/UCL) and the Consortium des \'Equipements de Calcul Intensif en F\'ed\'eration Wallonie Bruxelles (CECI) funded by the Fonds de la Recherche Scientifique de Belgique (FRS-FNRS) under Grant No.~2.5020.11.

\appendix
\section{Derivatives in reduced coordinates}
\label{red_coords}

Defining $\{\vec{a}_1, \vec{a}_2, \vec{a}_3\}$ as the real space and $\{\vec{b}_1, \vec{b}_2, \vec{b}_3\}$ as the reciprocal space primitive vectors, with $\vec{a}_\alpha \cdot \vec{b}_\beta = 2\pi \delta_{\alpha\beta}$, noting vectors in reduced coordinates as $\breve{\vec{v}}$, and defining
\begin{equation}
\label{transfo_matr}
 [\matr{A}]_{\alpha \beta} \triangleq [\vec{a}_\beta]_\alpha; \qquad [\matr{B}]_{\alpha \beta} \triangleq [\vec{b}_\beta]_\alpha, 
\end{equation}
we have
\begin{equation}
\vec{r}= \matr{A} \breve{\vec{r}}; \qquad
\vec{k}= \matr{B} \breve{\vec{k}}; \qquad 
\matr{A}^T \matr{B} = 2 \pi \matr{1},
\end{equation}
from which we deduce the inverse transformation
\begin{equation}
\label{kred}
\breve{\vec{k}}= \frac{1}{2 \pi} \matr{A}^T \vec{k}.
\end{equation}
We can now deduce using Eq.~\eqref{kred} the transformation that links derivatives with respect to Cartesian components of $\vec{k}$ (noted $X^\alpha$) to derivatives with respect to reduced coordinates (noted $\breve{X}^\alpha$) 
\begin{equation}
\delta X  = \sum_\alpha X^\alpha [\delta \vec{k}]_\alpha = \sum_\alpha \breve{X}^\alpha [\delta \breve{\vec{k}}]_\alpha,
\end{equation}
\begin{equation}
\label{der1red}
\Rightarrow X^\alpha = \sum_\beta \frac{[\matr{A}]_{\alpha \beta}}{2 \pi} \breve{X}^\beta.
\end{equation}
Similarly for second-order derivatives, we obtain
\begin{equation}
\label{der2red}
X^{\alpha \beta} = \sum_{\gamma \delta} \frac{[\matr{A}]_{\alpha \gamma}}{2 \pi} \breve{X}^{\gamma \delta} \frac{[\matr{A}^T]_{\delta \beta}}{2 \pi},
\end{equation}
and can thus retrieve e.g. $\varepsilon^{\alpha \beta}_{n \vec{k}}$ from $\breve{\varepsilon}^{\alpha \beta}_{n \vec{k}}$. 


\section{Kinetic energy with cutoff smearing}
\label{ecutsm-sec}

A known issue when optimizing the primitive cell size in a plane wave implementation of DFT are the spurious discontinuous drops in total energy that occur when increasing the cell size.
Indeed, dilating the real space lattice contracts the reciprocal space one.
This contraction increases discontinuously the number of plane waves located inside a sphere defined by the cutoff energy $E_c$.
This discontinuous increase of the size of the plane wave basis set translates into a discontinuous increase of the number of degrees of freedom available for the minimization of the total energy, which causes discontinuous drops when increasing the cell size.

A possible solution to this problem is to force the effective number of degrees of freedom to increase continuously as the cell size increases.
A numerical way to achieve this was first proposed in Ref.~\onlinecite{Bernasconi95} and involves modifying the kinetic energy close to $E_c$ so that it smoothly becomes large 
\begin{multline}
\bra{\vec{G}} \hat T_{\vec{k}} \ket{\vec{G}'} 
= \delta_{\vec{G}\vec{G}'} \bigg[ \frac{1}{2}(\vec{k}+\vec{G})^2 \\
+ A \bigg( 1 + {\rm erf} \Big( \frac{1/2(\vec{k}+\vec{G})^2 - E_c}{E_s} \Big) \bigg) \bigg],
\end{multline}
where ${\rm erf}(x)$ is the error function and where $E_s$ and $A$ are adjustable parameters. 
Thus, the weights of the plane waves become smoothly small close to $E_c$ and the change in the number of degrees of freedom with varying cell size can be made more continuous. 
This idea not only makes the total energy smoother with respect to cell size but also provides, as a side effect, a very good approximation of Pulay stress~\cite{Dacosta:1986aa} within the kinetic energy term~\cite{Bernasconi95}. 

\begin{figure}
\includegraphics[width=1.0 \linewidth]{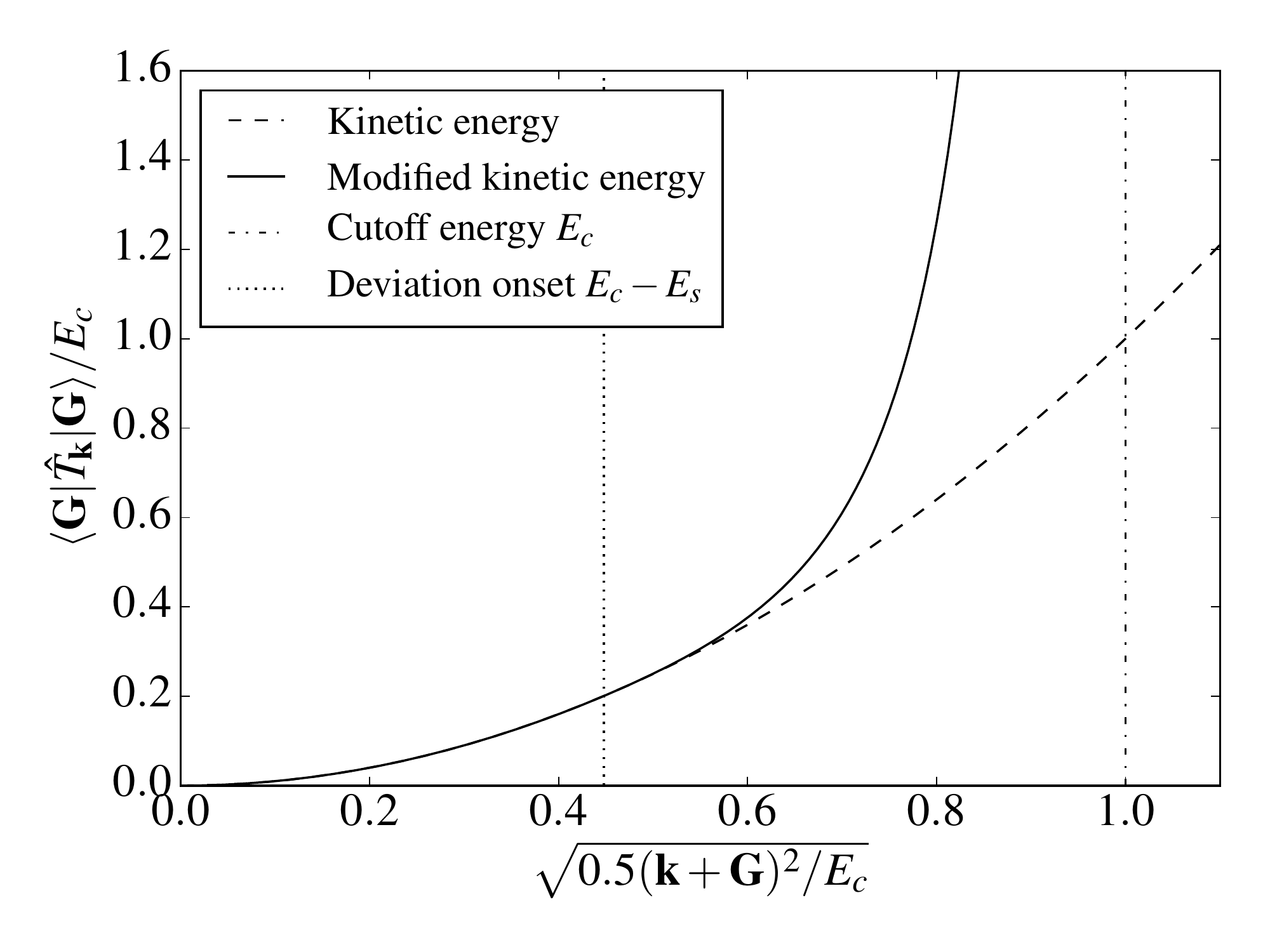} 
\caption{\label{ecutsm-graph} 
Within ABINIT, the kinetic energy rises asymptotically to infinity so that the change in the number of degrees of freedom with varying cell size is continuous.
The onset of the deviation from the physical expression $\frac{1}{2}(\vec{k}+\vec{G})^2$ occurs at the energy $E_c-E_s$.
The first and second-order derivatives of the kinetic energy remain continuous at this onset.}
\end{figure}

Within ABINIT, the implementation of this idea  takes a slightly different form.
Rather than becoming smoothly large at $E_c$, the kinetic energy rises asymptotically to infinity (see Figure~\ref{ecutsm-graph}), so that the change in the number of degrees of freedom with varying cell size becomes completely continuous.
The mathematical formulation of this idea takes the form
\begin{equation}
\label{Kin}
\bra{\vec{G}} \hat T_{\vec{k}} \ket{\vec{G}'} = \frac{1}{2}(\vec{k}+\vec{G})^2 \delta_{\vec{G}\vec{G}'} \, p(x),
\end{equation}
where 
\begin{equation}
\label{Kinx}
x \triangleq \frac{E_c - \frac{1}{2}(\vec{k}+\vec{G})^2}{E_s},
\end{equation}
and where $p(x) \to +\infty$ as $x \to 0^+$. 
The deviation from the physical expression $\frac{1}{2}(\vec{k}+\vec{G})^2$ starts at $E_c-E_s$, i.e. $p(x)$ starts deviating from $1$ below $x=1$.
The parameter $E_s$ is therefore the energy range around the cutoff energy $E_c$ where the occupations start to be forced towards 0, i.e. $E_s$ can be interpreted as a smearing of the cutoff energy.  

To avoid numerical issues, $p(x)$ is chosen so the first and second-order derivatives of the kinetic energy remain continuous at the onset of this deviation, that is, just above the plane wave kinetic energy $E_c-E_s$.
Moreover, its inverse approaches quadratically zero at $E_c$.
Among the possible functions with this behaviour, the following specific form for $p(x)$ was chosen and implemented within ABINIT
\begin{equation}
\label{Kinp}
p(x) \triangleq
\begin{cases}
 1 & \mathrm{if} \quad 1 < x, \\
 \frac{1}{x^2(3+x(1+x(-6+3x)))} & \mathrm{if} \quad 0 < x \leq 1. \\
\end{cases}
\end{equation}

Together, Eqs.~\eqref{Kin}, \eqref{Kinx}, and \eqref{Kinp} define the modified kinetic energy implemented within ABINIT and form the starting point from which its derivatives (Eqs.~\eqref{Kin1} and \eqref{Kin2}) are obtained.

\section{PAW reminder}
\label{PAW-reminder}

The basic idea of PAW is to transform the eigenfunctions $\ket{\psi_{n\vec{k}}}$ of the Hamiltonian $\hat H$ into `pseudo' wavefunctions $\ket{\tilde \psi_{n\vec{k}}}$, which are smoother and thus accurately described by a smaller plane wave basis. 
Since the sharp features of the wavefunction usually occur near the nucleus, the transformation only needs to differ from identity within an `augmentation' region $\Omega_{\vec{R}}$ around each nucleus position $\vec{R}$ within the primitive cell. 
Within these augmentation regions, an orthonormal basis $\{\ket{\phi_{\vec{R}i}}\}$ that contains the sharp components of the wavefunctions $\ket{\psi_{n\vec{k}}}$ is chosen, where the index $i$ labels the functions $\ket{\phi_{\vec{R}i}}$ belonging to a given atomic site $\vec{R}$. 
The transformation can then be defined as the replacement of these components $\{\ket{\phi_{\vec{R}i}}\}$ of the wavefunctions $\ket{\psi_{n\vec{k}}}$ by those of another orthonormal basis $\{\ket{\tilde \phi_{\vec{R}i}}\}$, which are smoother inside the augmentation region $\Omega_{\vec{R}}$ and identical outside.
More formally, this idea translates into
\begin{equation}
\label{itr_NC}
\ket{\tilde \psi_{n\vec{k}}} = \ket{\psi_{n\vec{k}}} + \sum_{\vec{R}i} \left( \ket{\tilde \phi_{\vec{R}i}} - \ket{\phi_{\vec{R}i}} \right) \braket{\phi_{\vec{R}i} | \psi_{n\vec{k}}}.
\end{equation}
At the end of the calculation, we can recover the full wavefunctions with the inverse transformation
\begin{equation}
\label{tr_NC}
\ket{\psi_{n\vec{k}}} = \ket{\tilde \psi_{n\vec{k}}} + \sum_{\vec{R}i} \left( \ket{\phi_{\vec{R}i}} - \ket{\tilde \phi_{\vec{R}i}} \right) \braket{\tilde \phi_{\vec{R}i} | \tilde \psi_{n\vec{k}}}.
\end{equation}

In the PAW method, the orthonormality constraint imposed to the bases $\{\ket{\phi_{\vec{R}i}}\}$ and $\{\ket{\tilde \phi_{\vec{R}i}}\}$ is relaxed into a completeness constraint.
This gives additional degrees of freedom to make the basis $\{\ket{\tilde \phi_{\vec{R}i}}\}$ even smoother.
However, this implies that $\braket{\tilde \phi_{\vec{R}i} | \tilde \phi_{\vec{R}j}} = I_{\vec{R}ij}$ is not in general the identity matrix $\delta_{ij}$ and thus $\sum_i \ket{\tilde \phi_{\vec{R}i}} \bra{\tilde \phi_{\vec{R}i}}$ is not the identity operator. 

To solve this issue, we introduce the projectors $\ket{\tilde{p}_{\vec{R}i}}$, defined by
\begin{equation}
\label{PAW-p}
\braket{\tilde{p}_{\vec{R}i} | \tilde \phi_{\vec{R}j}} \triangleq \delta_{ij},
\end{equation}
which allows to express the identity as $\sum_i \ket{\tilde \phi_{\vec{R}i}} \bra{\tilde{p}_{\vec{R}i}}$ and thus Eq.~\eqref{tr_NC} becomes
\begin{align}
\ket{\psi_{n\vec{k}}} &= \ket{\tilde \psi_{n\vec{k}}} + \sum_{\vec{R}i} \left( \ket{\phi_{\vec{R}i}} - \ket{\tilde \phi_{\vec{R}i}} \right) \braket{\tilde{p}_{\vec{R}i} | \tilde \psi_{n\vec{k}}}, \\
& \triangleq \hat \tau \ket{\tilde \psi_{n\vec{k}}}, \label{tr_PAW}
\end{align}
where $\hat \tau$ is a linear, but not unitary, transformation. 

For PAW to be computationally advantageous, the $\ket{\tilde \psi_{n\vec{k}}}$ (and not the $\ket{\psi_{n\vec{k}}}$) must be the variational parameter used in the calculation.
Ideally, one should avoid using the $\ket{\psi_{n\vec{k}}}$ whenever possible, and thus directly deduce the desired observables $O$ from the $\ket{\tilde \psi_{n\vec{k}}}$ 
\begin{align}
O_{n\vec{k}} &= \bra{\psi_{n\vec{k}}} \hat O \ket{\psi_{n\vec{k}}}, \nonumber \\
& = \bra{\tilde \psi_{n\vec{k}}} \hat \tau^\dagger \hat O \hat \tau \ket{\tilde \psi_{n\vec{k}}}, \nonumber \\
& \triangleq \bra{\tilde \psi_{n\vec{k}}} \hat{\tilde{O}} \ket{\tilde \psi_{n\vec{k}}}, \label{PAW_op}
\end{align}
where the transformed operator $\hat{\tilde{O}}$ takes the form 
\begin{multline}
\hat{\tilde{O}} 
= \left( 1 + \sum_{\vec{R}i} \ket{\tilde{p}_{\vec{R}i}} \left( \bra{\phi_{\vec{R}i}} - \bra{\tilde \phi_{\vec{R}i}} \right) \right) 
\hat O \\
\left( 1 + \sum_{\vec{R}'j} \left( \ket{\phi_{\vec{R}'j}} - \ket{\tilde \phi_{\vec{R}'j}} \right) \bra{\tilde{p}_{\vec{R}'j}} \right), \\
= \hat O + \sum_{\vec{R}ij} \ket{\tilde{p}_{\vec{R}i}} \left( \bra{\phi_{\vec{R}i}} \hat O \ket{\phi_{\vec{R}j}} - \bra{\tilde \phi_{\vec{R}i}} \hat O \ket{\tilde \phi_{\vec{R}j}} \right) \bra{\tilde{p}_{\vec{R}j}} \label{local_PAW_O}. 
\end{multline}
The last relation is valid only for semi-local operators $\hat O$ (that is, operators which depend only locally on the value and derivatives with respect to $\vec{r}$ of the wavefunctions).
Moreover, it supposes that the bases $\{\ket{\phi_{\vec{R}i}}\}$ and $\{\ket{\tilde \phi_{\vec{R}i}}\}$ are complete and that the augmentation regions $\Omega_{\vec{R}}$ do not overlap. 
However, in practical calculations, small bases usually suffice to achieve good accuracy and a small overlap of the augmentation regions can be tolerated without substantial loss of precision.

In the case of periodic solids, we wish to use the $\ket{\tilde{u}_{n\vec{k}}}$ (instead of the $\ket{\tilde \psi_{n\vec{k}}}$) directly as the variational parameter.
To do so, we must transform Sch\"odinger's equation (Eq.~\eqref{k-schro}) to its PAW version.
Starting from Eq.~\eqref{schro}, then using Eqs.~\eqref{tr_PAW}, \eqref{PAW_op} and, finally, applying Eqs.~\eqref{bloch}, \eqref{bloch_op} to the PAW quantities, we obtain
\begin{align}
\hat H \ket{\psi_{n\vec{k}}} & = \varepsilon_{n\vec{k}} \ket{\psi_{n\vec{k}}}, \nonumber \\
\Rightarrow \hat{\tilde{H}} \ket{\tilde \psi_{n\vec{k}}} & = \varepsilon_{n\vec{k}} \hat{\tilde{1}} \ket{\tilde \psi_{n\vec{k}}}, \nonumber \\
\Rightarrow \hat{\tilde{H}}_{\vec{k}} \ket{\tilde{u}_{n\vec{k}}} & = \varepsilon_{n\vec{k}} \hat{\tilde{1}}_{\vec{k}} \ket{\tilde{u}_{n\vec{k}}}, \label{Schro-PAW}
\end{align}
where we have defined the overlap operator 
\begin{equation}
\label{overlap_op}
\hat{\tilde{1}}  \triangleq \hat \tau^\dagger  \hat \tau, 
\end{equation}
and where the $\vec{k}$-dependent operators $\hat{\tilde{H}}_{\vec{k}}$ and $\hat{\tilde{1}}_{\vec{k}}$ have the following form, following Eq.~\eqref{local_PAW_O}
\begin{align}
\hat{\tilde{H}}_{\vec{k}} & = \hat H_{\vec{k}} + \sum_{\vec{R}ij} e^{-i \vec{k} \cdot \hat{\vec{r}}} \ket{\tilde{p}_{\vec{R}i}} D_{\vec{R}ij} \bra{\tilde{p}_{\vec{R}j}} e^{i \vec{k} \cdot \hat{\vec{r}}} \label{U-sca} \\
& \triangleq \hat H_{\vec{k}} + \hat D_{\vec{k}}, \label{U-op} \\
\hat{\tilde{1}}_{\vec{k}} & = 1 + \sum_{\vec{R}ij} e^{-i \vec{k} \cdot \hat{\vec{r}}} \ket{\tilde{p}_{\vec{R}i}} S_{\vec{R}ij} \bra{\tilde{p}_{\vec{R}j}} e^{i \vec{k} \cdot \hat{\vec{r}}}, \label{S-sca}
\end{align}
with
\begin{align}
D_{\vec{R}ij} & \triangleq \left( \bra{\phi_{\vec{R}i}} \hat H \ket{\phi_{\vec{R}j}} - \bra{\tilde \phi_{\vec{R}i}} \hat H \ket{\tilde \phi_{\vec{R}j}} \right), \label{PAW-U} \\
S_{\vec{R}ij} & \triangleq \left( \braket{\phi_{\vec{R}i} | \phi_{\vec{R}j}} - \braket{\tilde \phi_{\vec{R}i} | \tilde \phi_{\vec{R}j}} \right). \label{PAW-S}
\end{align}

As stated by Eqs.~\eqref{eps_Pstern} and \eqref{Pstern} for the norm-conserving case and as we demonstrate in Section~\ref{d} (see Eqs.~\eqref{eps2} and \eqref{PNpsi1}) for the PAW case, we need the first and second-order derivatives of $\hat{\tilde{H}}_{\vec{k}}$ and $\hat{\tilde{1}}_{\vec{k}}$ to compute the effective masses. 
From Eqs.~\eqref{PAW-U}, \eqref{PAW-S}, and the fact that we use a plane wave basis set, this means that we need expressions for $\bra{\vec{G}} \hat D_{\vec{k}} \ket{\vec{G}'}$ and $\bra{\vec{G}} \hat{\tilde{1}}_{\vec{k}} \ket{\vec{G}'}$. 
We obtain
\begin{align}
\bra{\vec{G}} \hat D_{\vec{k}} \ket{\vec{G}'} 
=& \sum_{\vec{R}ij} 
\bra{\vec{G}} e^{-i \vec{k} \cdot \hat{\vec{r}}} \ket{\tilde{p}_{\vec{R}i}} 
D_{\vec{R}ij} 
\bra{\tilde{p}_{\vec{R}j}} e^{i \vec{k} \cdot \hat{\vec{r}}} \ket{\vec{G}'} \nonumber \\
= & \sum_{\vec{R}ij} 
\braket{\vec{k}+\vec{G} | \tilde{p}_{\vec{R}i}}
D_{\vec{R}ij}
\braket{\tilde{p}_{\vec{R}j} | \vec{k}+\vec{G}'}. \label{UGG}
\end{align}
Since the projectors $\ket{\tilde{p}_{\vec{R}i}}$ stem from Schr\"odinger's equation with a spherical potential~\cite{Blochl}, they can be expressed as the product of their radial part times spherical harmonics~\cite{Audouze08} 
\begin{equation}
\label{PAW-Ps}
\braket{\vec{r} | \tilde{p}_{\vec{R}i}} \triangleq \frac{\tilde{\mathcal{P}}_{\vec{R}i}(s)}{s} Y_{l_i m_i}(\hat{\vec{s}}), 
\end{equation}
where we have defined $\tilde{\mathcal{P}}_{\vec{R}i}(s)$, where $\vec{s} \triangleq \vec{r}-\vec{R}$ and where $Y_{lm}$ are the spherical harmonics.
Thus, we can calculate $\braket{\vec{k}+\vec{G} | \tilde{p}_{\vec{R}i}}$
\begin{multline}\label{EqB17}
\braket{\vec{k}+\vec{G} | \tilde{p}_{\vec{R}i}} = \int_0^{s_c} ds \, s^2 \frac{\tilde{\mathcal{P}}_{\vec{R}i}(s)}{s} \\
 \int d \hat{\vec{s}} Y_{l_i m_i}(\hat{\vec{s}})  e^{- i (\vec{k}+\vec{G}) \cdot (\vec{s}+\vec{R})},
\end{multline}
where $s_c$ is the radius of the augmentation regions $\Omega_R$ and $\hat{\vec{s}}$ is the normalized version of $\vec{s}$.
This can be simplified, using the identity
\begin{equation}
e^{i \vec{k} \cdot \vec{s}} = 4 \pi \sum_{l=0}^\infty i^l j_l(ks) \sum_{m=-l}^l Y_{lm}(\hat{\vec{s}}) Y_{lm}(\hat{\vec{q}}),
\end{equation}
where $j_l(ks)$ are spherical Bessel functions. Eq.~\eqref{EqB17} thus becomes 
\begin{align}
\braket{\vec{K} | \tilde{p}_{\vec{R}i}} &= 4 \pi i^{l_i} e^{- i \vec{K} \cdot \vec{R}} Y_{l_i m_i}(\hat{\vec{K}}) \int_0^{s_c} ds \, s \, \tilde{\mathcal{P}}_{\vec{R}i}(s) j_{l_i}(Ks), \label{ptildes} \\
& \triangleq 4 \pi i^{l_i} e^{- i \vec{K} \cdot \vec{R}} Y_{l_i m_i}(\hat{\vec{K}}) \tilde{P}_{\vec{R}i}(K), \label{ptildeK} \\
& \triangleq 4 \pi i^{l_i} e^{- i \vec{K} \cdot \vec{R}} \bar P_{\vec{R}i}(\vec{K}), \label{pbar}
\end{align}
where $\vec{K} \triangleq \vec{k}+\vec{G}$.
Substituting Eq.~\eqref{pbar} into Eq.~\eqref{UGG} gives
\begin{align}
\bra{\vec{G}} \hat D_{\vec{k}} \ket{\vec{G}'} 
& \begin{multlined}[t]
= \sum_{\vec{R}ij}  4 \pi i^{l_i} \bar P_{\vec{R}i}(\vec{K}) e^{- i (\cancel{\vec{k}}+\vec{G}) \cdot \vec{R}} 
D_{\vec{R}ij} \\
 e^{i (\cancel{\vec{k}}+\vec{G}') \cdot \vec{R}} \bar P^*_{\vec{R}j}(\vec{K}') 4 \pi (i^{l_j})^*, \nonumber
\end{multlined} \\
&= \sum_{\vec{R}ij} \braket{\vec{K} | \bar p_{\vec{R}i}} D_{\vec{R}ij} \braket{\bar p_{\vec{R}j} | \vec{K}'}, \label{UGG-pbar}
\end{align}
where we have defined 
\begin{equation}
\label{PAW-pbar}
\braket{\vec{K} | \bar p_{\vec{R}i}} \triangleq 4 \pi i^{l_i} \bar P_{\vec{R}i}(\vec{K}) e^{- i \vec{G} \cdot \vec{R}}.
\end{equation}
Together, Eqs.~\eqref{ptildes}-\eqref{UGG-pbar} form the starting point for the calculation of the derivatives of the non-local part of the Hamiltonian $\hat D_{\vec{k}}$ in Section~\ref{nl}.

\section{Derivation of Eq.~\eqref{eps2c} from Eq.~\eqref{eps2a}}
\label{PAW-alg}

Substituting Eq.~\eqref{psi1_dnd} in Eq.~\eqref{eps2a}, using Eq.~\eqref{Pdeg}, simplifying using Eq.~\eqref{PAW-up}, using Eq.~\eqref{eps1}, and invoking the degeneracy at 1st order $(\varepsilon^\alpha_{nn' \vec{k}} = \varepsilon_{\{d\} \vec{k}}^{\alpha} \delta_{nn'})$ yields
\begin{multline}
\label{eps2b}
\varepsilon_{nn' \vec{k}}^{\alpha\beta} = 
  \bra{\tilde{u}_{n' \vec{k}}} \hat{\tilde{H}}_{\vec{k}}^{\alpha\beta} - \frac{\varepsilon_{n \vec{k}} + \varepsilon_{n' \vec{k}}}{2} \hat{\tilde{1}}_{\vec{k}}^{\alpha\beta} \ket{\tilde{u}_{n\vec{k}}} \\
  + \Big ( \bra{\hat{\tilde{Q}}_{\vec{k}} \tilde{u}_{n' \vec{k}}^\alpha} 
           \hat{\tilde{H}}_{\vec{k}}^\beta - \frac{\varepsilon_{n \vec{k}} + \varepsilon_{n' \vec{k}}}{2} \hat{\tilde{1}}_{\vec{k}}^\beta 
           \ket{\tilde{u}_{n\vec{k}}} \\ 
  +        \bra{\hat{\tilde{Q}}_{\vec{k}} \tilde{u}_{n' \vec{k}}^\alpha} \hat{\tilde{H}}_{\vec{k}} - \frac{\varepsilon_{n \vec{k}} + \varepsilon_{n' \vec{k}}}{2} \hat{\tilde{1}}_{\vec{k}} \ket{\hat{\tilde{Q}}_{\vec{k}} \tilde{u}_{n\vec{k}}^\beta} \\
  +        \bra{\tilde{u}_{n' \vec{k}}} \hat{\tilde{H}}_{\vec{k}}^\alpha - \frac{\varepsilon_{n \vec{k}} + \varepsilon_{n' \vec{k}}}{2} \hat{\tilde{1}}_{\vec{k}}^\alpha \ket{\hat{\tilde{Q}}_{\vec{k}} \tilde{u}_{n\vec{k}}^\beta} \\
  + \bra{\tilde{u}_{n' \vec{k}}^\alpha} \hat{\tilde{1}}_{\vec{k}} \ket{\tilde{u}_{n\vec{k}}} \varepsilon^\beta_{\{d\} \vec{k}}
  + \varepsilon^\alpha_{\{d\} \vec{k}} \bra{\tilde{u}_{n'\vec{k}}} \hat{\tilde{1}}_{\vec{k}} \ket{\tilde{u}_{n\vec{k}}^\beta} \Big) \\
  + \alpha \leftrightarrow \beta .
\end{multline}

It is more convenient to reformulate the second line from the end of this expression so that it can be merged with the second and fourth lines. 
To do so, we rearrange the terms included in $\alpha \leftrightarrow \beta$ and invoke the degeneracy to first order $\varepsilon^\beta_{n \vec{k}} = \varepsilon^\beta_{n'\vec{k}} \ \forall \ n,n' \in \{d\}$, so that Eq.~\eqref{orthonorm1} can be used, and obtain
\begin{align}
\label{eps2bc}
& \Big( \bra{\tilde{u}_{n' \vec{k}}^\alpha} \hat{\tilde{1}}_{\vec{k}} \ket{\tilde{u}_{n\vec{k}}} \varepsilon^\beta_{n \vec{k}}
      + \varepsilon^\alpha_{n' \vec{k}} \bra{\tilde{u}_{n'\vec{k}}} \hat{\tilde{1}}_{\vec{k}} \ket{\tilde{u}_{n\vec{k}}^\beta} \Big) 
  + \alpha \leftrightarrow \beta \nonumber \\
& \begin{multlined}
= -\frac{1}{2} \Big( \bra{\tilde{u}_{n'\vec{k}}} \hat{\tilde{1}}_{\vec{k}}^\alpha \ket{\tilde{u}_{n\vec{k}}} \varepsilon^\beta_{n\vec{k}}
                   + \varepsilon^\alpha_{n'\vec{k}} \bra{\tilde{u}_{n'\vec{k}}} \hat{\tilde{1}}_{\vec{k}}^\beta \ket{\tilde{u}_{n\vec{k}}} \Big) \\
+ \alpha \leftrightarrow \beta, 
\end{multlined} \nonumber \\
& \begin{multlined}
= -\frac{1}{2}\Big(
     \bra{\delta \tilde{u}_{n'\vec{k}}^\alpha}
     \hat{\tilde{H}}_{\vec{k}}^\beta - \frac{\varepsilon_{n \vec{k}} + \varepsilon_{n' \vec{k}}}{2} \hat{\tilde{1}}_{\vec{k}}^\beta
     \ket{\tilde{u}_{n \vec{k}}} \\
   + \bra{\tilde{u}_{n' \vec{k}}} 
     \hat{\tilde{H}}_{\vec{k}}^\alpha - \frac{\varepsilon_{n \vec{k}} + \varepsilon_{n' \vec{k}}}{2} \hat{\tilde{1}}_{\vec{k}}^\alpha 
     \ket{\delta \tilde{u}_{n\vec{k}}^\beta} \Big) \\
 +\alpha \leftrightarrow \beta,
\end{multlined}
\end{align}
where we have defined 
\begin{equation}
\label{dpsi1a-A}
 \ket{\delta \tilde{u}_{n\vec{k}}^\alpha} \triangleq \sum_{n' \in \{d\}} \ket{\tilde{u}_{n'\vec{k}}} \bra{\tilde{u}_{n'\vec{k}}} \hat{\tilde{1}}_{\vec{k}}^\alpha \ket{\tilde{u}_{n\vec{k}}},  
\end{equation}
and where we have used $\varepsilon^\alpha_{nn' \vec{k}} = \varepsilon_{\{d\} \vec{k}}^{\alpha} \delta_{nn'}$, Eq.~\eqref{eps1}, and Eq.~\eqref{Pdeg} for the last equality of Eq.~\eqref{eps2bc}.
Substituting this result in Eq.~\eqref{eps2b} and using the fact that we can add any component within the degenerate subspace to the wavefunctions on the third line of this equation (since they don't contribute to the final result as per Eq.~\eqref{PAW-up}), we obtain
\begin{widetext}
\begin{multline}
\label{eps2c-A}
\varepsilon_{nn' \vec{k}}^{\alpha\beta}
= \bra{\tilde{u}_{n' \vec{k}}} \hat{\tilde{H}}_{\vec{k}}^{\alpha\beta} - \frac{\varepsilon_{n \vec{k}} + \varepsilon_{n' \vec{k}}}{2} \hat{\tilde{1}}_{\vec{k}}^{\alpha\beta} \ket{\tilde{u}_{n\vec{k}}} \\
  + \Big ( \bra{\hat{\tilde{Q}}_{\vec{k}} \tilde{u}_{n' \vec{k}}^\alpha - \frac{1}{2} \delta \tilde{u}_{n\vec{k}}^\alpha}   
           \hat{\tilde{H}}_{\vec{k}}^\beta - \frac{\varepsilon_{n \vec{k}} + \varepsilon_{n' \vec{k}}}{2} \hat{\tilde{1}}_{\vec{k}}^\beta 
           \ket{\tilde{u}_{n\vec{k}}}  
  +        \bra{\tilde{u}_{n' \vec{k}}} 
           \hat{\tilde{H}}_{\vec{k}}^\alpha - \frac{\varepsilon_{n \vec{k}} + \varepsilon_{n' \vec{k}}}{2} \hat{\tilde{1}}_{\vec{k}}^\alpha    
           \ket{\hat{\tilde{Q}}_{\vec{k}} \tilde{u}_{n\vec{k}}^\beta - \frac{1}{2} \delta \tilde{u}_{n\vec{k}}^\beta} \\
  +        \bra{\hat{\tilde{Q}}_{\vec{k}} \tilde{u}_{n' \vec{k}}^\alpha - \frac{1}{2} \delta \tilde{u}_{n\vec{k}}^\alpha} 
           \hat{\tilde{H}}_{\vec{k}}    
         - \frac{\varepsilon_{n \vec{k}} + \varepsilon_{n' \vec{k}}}{2} \hat{\tilde{1}}_{\vec{k}} 
           \ket{\hat{\tilde{Q}}_{\vec{k}} \tilde{u}_{n\vec{k}}^\beta - \frac{1}{2} \delta \tilde{u}_{n\vec{k}}^\beta} \Big )   
  + \alpha \leftrightarrow \beta.
\end{multline}
\end{widetext}

Eq.~\eqref{eps2c-A} is the intermediate point (see Eq.~\eqref{eps2c}) from which the final expression for $\varepsilon_{nn' \vec{k}}^{\alpha\beta}$ (Eq.~\eqref{eps2}) is obtained in Section~\ref{d_eps}.

\section{`Transport equivalent effective mass tensor' $\bar{\matr{M}}_{n\vec{k}}$ from band curvature $f_{n\vec{k}}(\theta,\phi)$}
\label{Meqv}

We summarize in this appendix the idea of Ref.~\onlinecite{Mecholsky14}, which associates to $f_{n\vec{k}}(\theta,\phi)$ a `transport equivalent mass tensors' $\bar{\matr{M}}_{n\vec{k}}$ that generates the same contribution to transport properties. 
The association holds within the relaxation time approximation to Boltzmann's transport equation~\cite{AshcroftMermin,Marder} with an energy dependent relaxation time of the form $\matr \tau_{n\vec{k}}(\varepsilon) = \matr U_n \tau_{n\vec{k}}(\varepsilon)$, i.e. where the energy dependence can be factored out of the tensor.
It also requires that $\bar{\matr{M}}_{n\vec{k}}$ be calculated at a band extrema ($\varepsilon_{nn' \vec{k}}^{\alpha} = 0$).
In the current appendix, we generalize the original demonstration to an energy dependent relaxation time $\matr \tau_{n\vec{k}}(\varepsilon)$ (where the energy dependence may not be factored out of the tensor) but specialize to the case of conductivity $\matr \sigma$ for concision. 

With these assumptions, Boltzmann's transport equation becomes
\begin{multline}
g_{n\vec{k}} 
= f(T,\varepsilon_{n\vec{k}}-\mu) 
- \frac{\partial f}{\partial \varepsilon}\bigg|_{\varepsilon_{n\vec{k}}}
\Big( \matr{\tau}_{n\vec{k}}(\varepsilon_{n\vec{k}}) \vec{v}_{n\vec{k}} \Big) \\
\cdot \left( -e \vec{E} - \vec{\nabla} \mu - \frac{\varepsilon_{n\vec{k}} - \mu}{T} \vec{\nabla} T \right),
\end{multline}
with $T$ the temperature, $\mu$ the chemical potential, $f(T,\varepsilon-\mu)$ the Fermi-Dirac distribution, $\vec{v}_{n\vec{k}}$ the electronic velocity, $-e$ the electronic charge, $\vec{E}$ the electric field, and $g_{n\vec{k}}$ the (out of equilibrium) occupation numbers of the electrons.
We can then calculate the resulting current density 
\begin{equation}
\vec{j} 
= -e \, 2 \sum_n \int \frac{d\vec{k}}{8\pi^3} \vec{v}_{n\vec{k}} g_{n\vec{k}},
\end{equation}
then deduce the conductivity
\begin{align}
\matr{\sigma} 
& = \frac{\partial \vec{j}}{\partial \vec{E}} \nonumber \\
& = - e^2 \sum_n \int \frac{d\vec{k}}{4\pi^3} \frac{\partial f}{\partial \varepsilon}\bigg|_{\varepsilon_{n\vec{k}}} \vec{v}_{n\vec{k}} \vec{v}_{n\vec{k}}^T \matr{\tau}_{n\vec{k}}^T(\varepsilon_{n\vec{k}}).
\end{align}
Since we are at a band extrema (located at $\vec{k}$), we have a dispersion of the form 
\begin{equation}
\varepsilon_{n\vec{k}+\vec{q}} 
= \varepsilon_{n\vec{k}} + f_{n\vec{k}}(\theta,\phi) \frac{q^2}{2}. 
\end{equation}
We can now obtain $\vec{v}_{n\vec{k}}$ from the band curvature $f_{n\vec{k}}(\theta,\phi)$
\begin{equation}
\vec{v}_{n\vec{k}} = \frac{\partial \varepsilon_{n\vec{k}}}{\partial \vec{k}} = \frac{q}{2} \bar{\vec{v}}_{n\vec{k}}(\hat{\vec{q}}),
\end{equation}
where $\hat{\vec{q}}$ is the unit vector along the direction $\theta, \phi$ in spherical coordinates and where the quantity $\bar{\vec{v}}_{n\vec{k}}(\hat{\vec{q}})$ takes the following form in Cartesian coordinates
\begin{widetext}
\begin{equation}
\label{speed-nu}
\bar{\vec{v}}_{n\vec{k}}(\hat{\vec{q}}) \triangleq 
\begin{pmatrix}
2 f_{n\vec{k}}(\theta,\phi) \sin(\theta) \cos(\phi) + \frac{\partial f_{n\vec{k}}}{\partial \theta} \cos(\theta) \cos(\phi) - \frac{\partial f_{n\vec{k}}}{\partial \phi} \frac{\sin(\phi)}{\sin(\theta)} \\
2 f_{n\vec{k}}(\theta,\phi) \sin(\theta) \sin(\phi) + \frac{\partial f_{n\vec{k}}}{\partial \theta} \cos(\theta) \sin(\phi) + \frac{\partial f_{n\vec{k}}}{\partial \phi} \frac{\cos(\phi)}{\sin(\theta)} \\
2 f_{n\vec{k}}(\theta,\phi) \cos(\theta)            - \frac{\partial f_{n\vec{k}}}{\partial \theta} \sin(\theta) 
\end{pmatrix}.
\end{equation}
\end{widetext}
Supposing that the parabolic dispersion of the bands holds wherever $\frac{\partial f}{\partial \varepsilon}$ is non-negligible, the conductivity $\matr{\sigma}$ takes the following form
\begin{equation}
\matr{\sigma} 
= - e^2 \sum_n \int \frac{d\vec{k}}{4\pi^3} \frac{\partial f}{\partial \varepsilon}\bigg|_{\varepsilon_{n\vec{k}}} \frac{q}{2} \bar{\vec{v}}_{n\vec{k}}(\hat{\vec{q}}) \frac{q}{2} \bar{\vec{v}}_{n\vec{k}}^T(\hat{\vec{q}}) \matr{\tau}_{n\vec{k}}^T(\varepsilon_{n\vec{k}}),
\end{equation}
which, using the substitution 
\begin{equation}
\varepsilon \triangleq \varepsilon_{n\vec{k}} + {\rm sign}(\varepsilon_{n\vec{k}}-\mu) |f_{n\vec{k}}(\theta,\phi)| \frac{q^2}{2},
\end{equation}
can be split into a product
\begin{equation}
\label{sigma}
\matr{\sigma} = \sum_n \sum_{\vec{k}} \matr{C}_{n\vec{k}} \matr{K}_{n\vec{k}},
\end{equation}
where the sum over $\vec{k}$ runs over the different extrema of band $n$. 
This product distinguishes the integral over the energy $\varepsilon$
\begin{multline}
\label{Kn}
\matr{K}_{n\vec{k}} \triangleq \frac{-e^2}{2^{3/2}\pi^3} {\rm sign}(\varepsilon_{n\vec{k}}-\mu) \int_{\varepsilon_{n\vec{k}}}^{{\rm sign}(\varepsilon_{n\vec{k}}-\mu) \infty} \\
d\varepsilon |\varepsilon - \varepsilon_{n\vec{k}}|^{3/2} \frac{\partial f}{\partial \varepsilon} \matr{\tau}_{n\vec{k}}^T(\varepsilon)
\end{multline}
and the integral over the spherical angles $(\theta,\phi)$
\begin{equation}
\label{Cn}
\matr{C}_{n\vec{k}} \triangleq \int_0^{2\pi} d\phi \int_0^\pi d\theta \sin(\theta) \frac{\bar{\vec{v}}_{n\vec{k}}(\theta,\phi) \bar{\vec{v}}_{n\vec{k}}^T(\theta,\phi)}{2 |f_{n\vec{k}}(\theta,\phi)|^{5/2}}.
\end{equation}

When $f_{n\vec{k}}(\theta,\phi)$ is an ellipsoid, i.e. when the band dispersion can be described by an effective mass tensor $\matr{M}_{n\vec{k}}$, and if we choose the Cartesian axes to be along the ellipsoid principal axes
\begin{equation}
\matr{M}_{n\vec{k}} =
\begin{pmatrix}
m_{n\vec{k} x} &   0   &   0   \\
  0   & m_{n\vec{k} y} &   0   \\
  0   &   0   & m_{n\vec{k} z} 
\end{pmatrix},
\end{equation}
then we have the relation 
\begin{equation}
f_{n\vec{k}}(\theta,\phi) = \hat{\vec{q}}^T(\theta,\phi) \matr{M}_{n\vec{k}} \hat{\vec{q}}(\theta,\phi).
\end{equation}
Calculating $f_{n\vec{k}}(\theta,\phi)$ in terms of $m_{n\vec{k} x}, m_{n\vec{k} y}, m_{n\vec{k} z}, \theta,$ and $\phi$, substituting in $\bar{\vec{v}}_{n\vec{k}}(\theta,\phi)$ (Eq.~\eqref{speed-nu}), then into $\matr{C}_{n\vec{k}}$ (Eq.~\eqref{Cn}) and finally carrying out analytically the integration over $\theta,\phi$ yields
\begin{equation}
\label{Cn-mni}
[\matr{C}_{n\vec{k}}]_{ij} = \frac{8\pi}{3} \frac{\sqrt{m_{n\vec{k} x} m_{n\vec{k} y} m_{n\vec{k} z}}}{m_{n\vec{k} i}} \delta_{ij},
\end{equation}
which allows to deduce $m_{n\vec{k} i}$ from $\matr{C}_{n\vec{k}}$
\begin{equation}
\label{m_ni}
m_{n\vec{k} i} = \frac{[\matr{C}_{n\vec{k}}]_{jj} [\matr{C}_{n\vec{k}}]_{kk}}{(8\pi/3)^2}; \quad i \neq j \neq k \neq i.
\end{equation}

\section{`Transport equivalent effective mass tensor' $\bar{\matr{M}}_{n\vec{k}}$ from band curvature $f_{n\vec{k}}(\theta,\phi)$ in 2D}
\label{Meqv-2D}

When $m_{n\vec{k} z} \to \infty$, we observe from Eq.~\eqref{Cn-mni} that the $x$ and $y$ matrix elements of $\matr{C}_{n\vec{k}}$ diverge. 
Thus, the procedure do find $\matr{\bar{M}}_{n\vec{k}}$ described in Appendix~\ref{Meqv} becomes numerically unstable for 2D systems.
To solve this issue, we adapt the formalism of Appendix~\ref{Meqv} to the 2D context.

The conductivity $\matr{\sigma}$ then becomes
\begin{equation}
\label{sigma-2D}
\matr{\sigma}^{\mathrm{2D}} = \sum_n \sum_{\vec{k}} \matr{C}^{\mathrm{2D}}_{n\vec{k}} \matr{K}^{\mathrm{2D}}_{n\vec{k}},
\end{equation}
with the energy part 
\begin{equation}
\label{Kn-2D}
\matr{K}^{\mathrm{2D}}_{n\vec{k}} \triangleq \frac{-e^2}{2\pi^2} \int_{\varepsilon_{n\vec{k}}}^{\pm \infty} \pm d\varepsilon |\varepsilon - \varepsilon_{n\vec{k}}| \frac{\partial f}{\partial \varepsilon} \matr{\tau}^{\mathrm{2D}\,T}_{n\vec{k}}(\varepsilon)
\end{equation}
and the angular part
\begin{equation}
\label{Cn-2D}
\matr{C}^{\mathrm{2D}}_{n\vec{k}} \triangleq \int_0^{2\pi} d\phi \frac{\bar{\vec{v}}_{n\vec{k}}(\phi) \bar{\vec{v}}_{n\vec{k}}^T(\phi)}{2 |f_{n\vec{k}}(\phi)|^{2}},
\end{equation}
where
\begin{equation}
\label{vbar-2D}
\bar{\vec{v}}_{n\vec{k}}(\phi) \triangleq 
\begin{pmatrix}
2 f_{n\vec{k}}(\phi) \cos(\phi) - \frac{\partial f_{n\vec{k}}}{\partial \phi} \sin(\phi) \\
2 f_{n\vec{k}}(\phi) \sin(\phi) + \frac{\partial f_{n\vec{k}}}{\partial \phi} \cos(\phi) 
\end{pmatrix},
\end{equation}
and where $\matr{\sigma}^{\mathrm{2D}}$ and $\matr{\tau}^{\mathrm{2D}}_{n\vec{k}}(\varepsilon)$ are 2D tensors.

When $f_{n\vec{k}}(\phi)$ is an ellipse, i.e. when the band dispersion can be described by an effective mass tensor $\matr{M}^{\mathrm{2D}}_{n\vec{k}}$, and if we choose the Cartesian axes to be along the principal axes
\begin{equation}
\label{Mbar-2D}
\matr{M}^{\mathrm{2D}}_{n\vec{k}} =
\begin{pmatrix}
m_{n\vec{k} x} &   0   \\
  0   & m_{n\vec{k} y} 
\end{pmatrix},
\end{equation}
then we have the relation 
\begin{equation}
\label{fn-Mbar-2D}
f_{n\vec{k}}(\phi) = \hat{\vec{q}}^T(\phi) \matr{M}^{\mathrm{2D}}_{n\vec{k}} \hat{\vec{q}}(\phi).
\end{equation}
We deduce from Eqs.~\eqref{Cn-2D} and \eqref{vbar-2D} the tensor $\matr{C}^{\mathrm{2D}}_{n\vec{k}}$ resulting from Eqs.~\eqref{Mbar-2D} and \eqref{fn-Mbar-2D}
\begin{equation}
\label{Cn-mni-2D}
[\matr{C}^{\mathrm{2D}}_{n\vec{k}}]_{ij} = 2\pi \frac{\sqrt{m_{n\vec{k} x} m_{n\vec{k} y}}}{m_{n\vec{k} i}} \delta_{ij}.
\end{equation}
A distinct feature of Eq.~\eqref{Cn-mni-2D} with respect to the 3D case (Eq.~\eqref{Cn-mni}) is that $\matr{C}^{\mathrm{2D}}_{n\vec{k}}$ is determined from the ratio of $m_{n\vec{k} x}$ and $m_{n\vec{k} y}$ only and is not influenced by their magnitude. 

It is easier to get an intuitive understanding of this fact if we consider a one-band system with a minimum at $\vec{\Gamma}$ with $\matr{M} = m^* \matr{1}$, $\matr{\tau} = \tau \matr{1}$ $\Rightarrow$ $\matr{\sigma} = \sigma \matr{1}$, and work at a constant Fermi energy with respect to the extremum (minimum or maximum).
We then have 
\begin{equation}
\label{sigma-scalar}
\sigma = \frac{n e^2 \tau}{m^*},
\end{equation}
with
\begin{equation}
n \triangleq - \int \frac{d\vec{k}}{(2\pi)^D} \frac{\partial f}{\partial \varepsilon} \propto (m^*)^{D/2},
\end{equation}
with $D$ the dimensionality of the system considered. 
We see that for the specific case of 2D systems, a cancellation occurs between the carrier density $n$ and the effective mass $m^*$ in Eq.~\eqref{sigma-scalar}. 
Therefore, $m^*$ does not influence the conductivity $\sigma$ in 2D or, in the more general case of Eq.~\eqref{Cn-2D}, rescaling $f_{n\vec{k}}(\phi)$ (or, equivalently, $\matr{\bar{M}}^{\mathrm{2D}}_{n\vec{k}}$) does not influence $\matr{C}^{\mathrm{2D}}_{n\vec{k}}$. 
Thus, the scale of $\matr{\bar{M}}^{\mathrm{2D}}_{n\vec{k}}$ can be set arbitrarily. 

Reciprocally, $\matr{\bar{M}}^{\mathrm{2D}}_{n\vec{k}}$ does not influence the scale of $\matr{C}^{\mathrm{2D}}_{n\vec{k}}$, as per Eq.~\eqref{Cn-mni-2D}. 
This feature of 2D tensorial effective masses does not hold true for general (i.e. warped) $f_{n\vec{k}}(\phi)$.
There is therefore one degree of freedom of $\matr{C}^{\mathrm{2D}}_{n\vec{k}}$ (its scale) that $\matr{\bar{M}}^{\mathrm{2D}}_{n\vec{k}}$ fails to determine. 
Care must therefore be taken when one computes transport quantities from $\matr{\bar{M}}^{\mathrm{2D}}_{n\vec{k}}$ in 2D. 
Once $\matr{C}^{\mathrm{2D}}_{n\vec{k}}$ has been diagonalized, one must extract its scaling $c_{n\vec{k}}$
\begin{align}
\matr{U}^{\mathrm{2D}\,T}_{n\vec{k}} \matr{C}^{\mathrm{2D}}_{n\vec{k}} \matr{U}^{\mathrm{2D}}_{n\vec{k}} &= 
\begin{pmatrix}
C_{n\vec{k} x} & 0 \\
0 & C_{n\vec{k} y}
\end{pmatrix} \nonumber \\
&= 2 \pi c_{n\vec{k}}
\begin{pmatrix}
\sqrt{\frac{C_{n\vec{k} x}}{C_{n\vec{k} y}}} & 0 \\
0 & \sqrt{\frac{C_{n\vec{k} y}}{C_{n\vec{k} x}}}
\end{pmatrix} \label{scaled_C}
\end{align}
and preserve this information. 
Then, substituting $c_{n\vec{k}} \to 1$ into Eq.~\eqref{scaled_C} gives a form of $\matr{C}^{\mathrm{2D}}_{n\vec{k}}$ compatible with a tensorial effective mass, which allows direct comparison with Eq.\eqref{Cn-mni-2D}
\begin{equation}
\label{ratio_mC}
\sqrt{\frac{C_{n\vec{k} y}}{C_{n\vec{k} x}}} =
\sqrt{\frac{m_{n\vec{k} x}}{m_{n\vec{k} y}}}, 
\end{equation}
which still leaves the scale of $\matr{\bar{M}}^{\mathrm{2D}}_{n\vec{k}}$ undetermined. 

Within this implementation, we choose to set the average curvature associated with $\matr{\bar{M}}^{\mathrm{2D}}_{n\vec{k}}$ (through Eq.~\eqref{fn-Mbar-2D}) to the average curvature of the associated band extrema $\bar{f}^{\mathrm{2D}}_{n\vec{k}}$
\begin{equation}
\frac{1}{2} \bigg( \frac{1}{m_{n\vec{k} x}} + \frac{1}{m_{n\vec{k} y}} \bigg) = \bar{f}^{\mathrm{2D}}_{n\vec{k}} \triangleq \frac{1}{2\pi} \int_0^{2\pi} d\phi f_{n\vec{k}}(\phi),
\end{equation}
so that we recover $\matr{\bar{M}}^{\mathrm{2D}}_{n\vec{k}} = \matr{M}^{\mathrm{2D}}_{n\vec{k}}$ when there is no warping (i.e. when the effective mass can be described by a tensor).
This allows to set specific values for $m_{n\vec{k} x}$ and $m_{n\vec{k} y}$
\begin{align}
m_{n\vec{k} x} &= \frac{1}{2 \bar{f}^{\mathrm{2D}}_{n\vec{k}}} \bigg( 1 + \frac{C_{n\vec{k} y}}{C_{n\vec{k} x}} \bigg); \\
m_{n\vec{k} y} &= m_{n\vec{k} x} \frac{C_{n\vec{k} x}}{C_{n\vec{k} y}}.
\end{align}
As discussed above Eq.~\eqref{scaled_C}, when these values are used to obtain a transport quantity, one must remember to multiply the final result by the scaling factor $c_{n\vec{k}}$ obtained in Eq.~\eqref{scaled_C}
\begin{equation}
\label{scaling_factor}
c_{n\vec{k}} \triangleq 
\frac{\sqrt{C_{n\vec{k} x} C_{n\vec{k} y}}}{2 \pi},
\end{equation}
since tensorial $\matr{\bar{M}}^{\mathrm{2D}}_{n\vec{k}}$ are unable to account for it. 
For instance, rather than using Eq.~\eqref{Cn-mni-2D}, which applies only to tensorial effective masses (i.e. to non-warped bands), one should substitute Eq.~\eqref{ratio_mC} into \eqref{scaled_C} 
\begin{equation}
\matr{U}^{\mathrm{2D}\,T}_{n\vec{k}} \matr{C}^{\mathrm{2D}}_{n\vec{k}} \matr{U}^{\mathrm{2D}}_{n\vec{k}} = 2 \pi c_{n\vec{k}}
\begin{pmatrix}
\sqrt{\frac{m_{n\vec{k} y}}{m_{n\vec{k} x}}} & 0 \\
0 & \sqrt{\frac{m_{n\vec{k} x}}{m_{n\vec{k} y}}}
\end{pmatrix}. 
\end{equation}

\bibliography{article.bib}

\end{document}